\documentclass{aa}  

\usepackage{graphicx}
\usepackage{txfonts}
\newcommand{\um}{\,$\mu$m}
\newcommand{\kms}{km\,s$^{-1}$}
\newcommand{\nii}{[N\,{\sc ii}]}
\newcommand{\cii}{[C\,{\sc ii}]}
\newcommand{\oiii}{[O\,{\sc iii}]}

\newcommand{\ci}{[C\,{\sc i}]}
\newcommand{\oi}{[O\,{\sc i}]}
\newcommand{\transl}{$^3$P$_1$-$^3$P$_0$}
\newcommand{\transu}{$^3$P$_2$-$^3$P$_1$}
\newcommand{\hii}{H\,{\sc ii}}
\newcommand{\hi}{H\,{\sc i}}
\newcommand{\hh}{H$_2$}
\newcommand{\cc}{cm$^{-3}$}

\newcommand{\thco}{$^{13}$CO}
\newcommand{\ceio}{C$^{18}$O}
\newcommand{\cplus}{C$^+$}
\newcommand{\czero}{C$^0$}
\newcommand{\tex}{$T_\mathrm{ex}$}
\begin{document}

   \title{Velocity profiles of \cii, \ci, CO, and \oi\ and physical conditions in four star-forming regions in the Large Magellanic Cloud}

   \titlerunning{Velocity profiles of \cii, \ci, CO, and \oi\ and physical conditions in the LMC}

   \subtitle{}

   \author{Yoko Okada \inst{1}
          \and
          Rolf G\"{u}sten \inst{2}
          \and
          Miguel Angel Requena-Torres \inst{3,2}
          \and
          Markus R\"{o}llig \inst{1}
          \and
          J\"{u}rgen Stutzki \inst{1}
          \and
          Urs U. Graf \inst{1}
          \and
          Annie Hughes \inst{4}
          }

   \institute{I. Physikalisches Institut der Universit\"{a}t zu K\"{o}ln, Z\"{u}lpicher Stra{\ss}e 77, 50937 K\"{o}ln, Germany
              \email{okada@ph1.uni-koeln.de}
              \and
              Max-Planck-Institut f\"{u}r Radioastronomie, Auf dem H\"{u}gel 69, 53121 Bonn, Germany
              \and
              Department of Astronomy, University of Maryland, College Park, MD 20742, USA
              \and
              L'Institut de Recherche en Astrophysique et Plan\'{e}tologie, 14, avenue Edouard Belin, 31400 Toulouse, France
             }

   \date{}

 
  \abstract
   {}
   {The aim of our study is to investigate the physical properties of the star-forming interstellar medium (ISM) in the Large Magellanic Cloud (LMC) by separating the origin of the emission lines spatially and spectrally. The LMC provides a unique local template to bridge studies in the Galaxy and high redshift galaxies because of its low metallicity and proximity, enabling us to study the detailed physics of the ISM in spatially resolved individual star-forming regions. Following Okada\ et~al. (2015, Paper I), we investigate different phases of the ISM traced by carbon-bearing species in four star-forming regions in the LMC, and model the physical properties using the KOSMA-$\tau$\ PDR model.}
   {We mapped 3--13 arcmin$^2$ areas in 30~Dor, N158, N160 and N159 along the molecular ridge of the LMC in \cii\ 158\um\ with GREAT on board SOFIA. We also observed the same area with CO(2-1) to (6-5), \thco(2-1) and (3-2), \ci\ \transl\ and \transu\ with APEX. For selected positions in N159 and 30~Dor, we observed \oi\ 145\um\ and \oi\ 63\um\ with upGREAT. All spectra are velocity resolved.}
   {In all four star-forming regions, the line profiles of CO, \thco, and \ci\ emission are similar, being reproduced by a combination of Gaussian profiles defined by CO(3-2), whereas \cii\ typically shows wider line profiles or an additional velocity component. At several positions in N159 and 30~Dor, we observed the velocity-resolved \oi\ 145\um\ and 63\um\ lines for the first time. At some positions, the \oi\ line profiles match those of CO, at other positions they are more similar to the \cii\ profiles. We interpret the different line profiles of CO, \cii\ and \oi\ as contributions from spatially separated clouds and/or clouds in different physical phases, which give different line ratios depending on their physical properties. We model the emission from the CO, \ci, \cii, and \oi\ lines and the far-infrared continuum emission using the latest KOSMA-$\tau$ PDR model, which treats the dust-related physics consistently and computes the dust continuum SED together with the line emission of the chemical species. We find that the line and continuum emissions are not well-reproduced by a single clump ensemble. Toward the CO peak at N159~W, we propose a scenario that the CO, \cii, and \oi\ 63\um\ emission are weaker than expected because of mutual shielding among clumps.}
   {}


   \maketitle
%

\section{Introduction}\label{sec:intro}

The life cycle of the interstellar medium (ISM) is a central part of understanding star formation and galaxy evolution. The ISM not only influences present and future star formation but it also interacts with recently formed stars and continues evolving. A photon-dominated region \citep[PDR,][]{TH85I,Sternberg1995} is one of the places with such interactions, where the ultra-violet (UV) radiation from stars dominates the physical and chemical conditions of the surrounding ISM. In the Galaxy, detailed analyses of PDRs that take account of the source geometry have been performed \citep[e.g.][]{Andree-Labsch2017}. On the other hand, some of the dominant emission lines such as \cii\ 158\um\ or \oi\ 63\um\ are often used to probe star formation in distant galaxies where individual star-forming regions are rarely resolved. In order to bridge our understanding in Galactic regions to high redshift work, studies in nearby galaxies with low (or different from solar) metallicity are necessary. The best local templates are the Large Magellanic Cloud (LMC) and the Small Magellanic Cloud (SMC) since they have significantly sub-solar metallicity ($0.5$\,$Z_\odot$ and $0.2$\,$Z_\odot$ for carbon \citep{Garnett1999}, where $Z_\odot$ is the solar abundance by \citet{Sofia2004}). They are also the nearest gas-rich systems (50~kpc and 63~kpc, respectively, where we quote the median distance from the NASA/IPAAC Extragalactic Database), enabling excellent spatial resolution compared to other extragalactic targets.

\subsection{\cplus/\czero/CO in low metallicity}

Carbon is the second most abundant metal in the ISM. Dominant carbon-bearing species in PDRs are \cplus, \czero, and CO. The \cii/CO ratio in environments with different metallicities has been studied observationally and theoretically \citep{Mochizuki1994,Poglitsch1995,Israel1996,Madden1997,Madden2011,Bolatto1999,Roellig2006}, with nearly all studies finding that ratio increases with decreasing metallicity. This can be understood as a thicker \cplus\ layer in regions where the UV radiation penetrates deeper due to lower dust extinction and, to a lesser extent, less self-shielding. By contrast, the transition H$^0$/\hh\ does not shift as much as that of \cplus/\czero\ or \cplus/CO at lower metallicity because \hh\ self-shielding is the dominant factor controlling the H$^0$/\hh\ transition and is metallicity independent. An ISM phase where hydrogen is in molecular form but CO is photo-dissociated is called the CO-dark molecular gas. Understanding the mass and spatial distribution of this phase is important for estimates of the total gas mass using CO emission \citep{Bolatto2013}. However, it is not clear whether a similar explanation applies to the \ci/CO ratio. \citet{Bolatto2000a} show that the gradient of \ci/CO against metallicity is slightly shallower than the case A model by \citet{Bolatto1999}, which assumes that the extent of the \czero\ layer in a clump scales linearly with the inverse of the metallicity (1/$Z$) and uses a mean column density of \cplus, \czero, and CO over a clump size distribution. The latest KOSMA-$\tau$ PDR model (see Sect. \ref{sec:PDRintro}) show different dependencies of \cplus/CO and \czero/CO on metallicity; the difference in \czero/CO between the Galactic model and LMC model is much smaller than for \cplus/CO, for reasons that are not yet well-understood.

\subsection{Line profiles of CO, \ci, \cii, and \oi}

The velocity profile of spectral line emission represents the turbulent motions of emitting material. In our observing beam, different phases of the ISM co-exist on spatial scales smaller than the beam and along the line-of-sight. When their motions are significantly different along the line-of-sight, we can distinguish them as separate velocity components in the line profile. This is especially useful for \cii\ emission, which can originate from ionized gas, atomic gas, and molecular gas phases \citep{Fahrion2017,Roellig2016,Carlhoff2013,PerezBeaupuits2012}. \citet[][hereafter Paper I]{Okada2015} show that the \cii\ emission line profile is substantially wider than that of CO and \ci\ in the N159 star-forming region in the LMC. The fraction of the \cii\ integrated intensity that cannot be fitted by the CO-defined line profile varies across the map from 20\% to 50\%.  The significant difference in the velocity profile of the \cii\ emission observed in many sources is interpreted as an indication of a substantial fraction of the \cii\ emitting material being accelerated relative to the quiescent material, e.g. it undergoes ablation \citep{Dedes2010,Mookerjea2012,Okada2012,Schneider2012,Simon2012,Pilleri2012b}, or photoevaporated \citep{Sandell2015}. Although an order of a few \kms\ displacement is consistent with modeled photoevaporation in globules \citep{Lefloch1994}, only a few studies to compare the observed line profiles of different emission lines with simulation incorporating full PDR chemistry have been conducted \citep{Bisbas2018}. An interesting question is whether the velocity profile difference is observed only in \cii\ or also in \oi, because the \oi\ intensities constrain physical conditions of the ablated material. Since 2014 the German REceiver for Astronomy at Terahertz Frequencies (GREAT) on board the Stratospheric Observatory for Infrared Astronomy (SOFIA) enables observations of velocity-resolved \oi\ 63\um\ emission. These observations often show strong absorption features \citep{Ossenkopf2015,Leurini2015}. Since the \oi\ 145\um\ emission is typically optically thin, the velocity profile of this line gives unambiguous information about the gas dynamics with minimal opacity effects on the line profile.

\subsection{Observed regions}

\begin{figure*}
\centering
\includegraphics{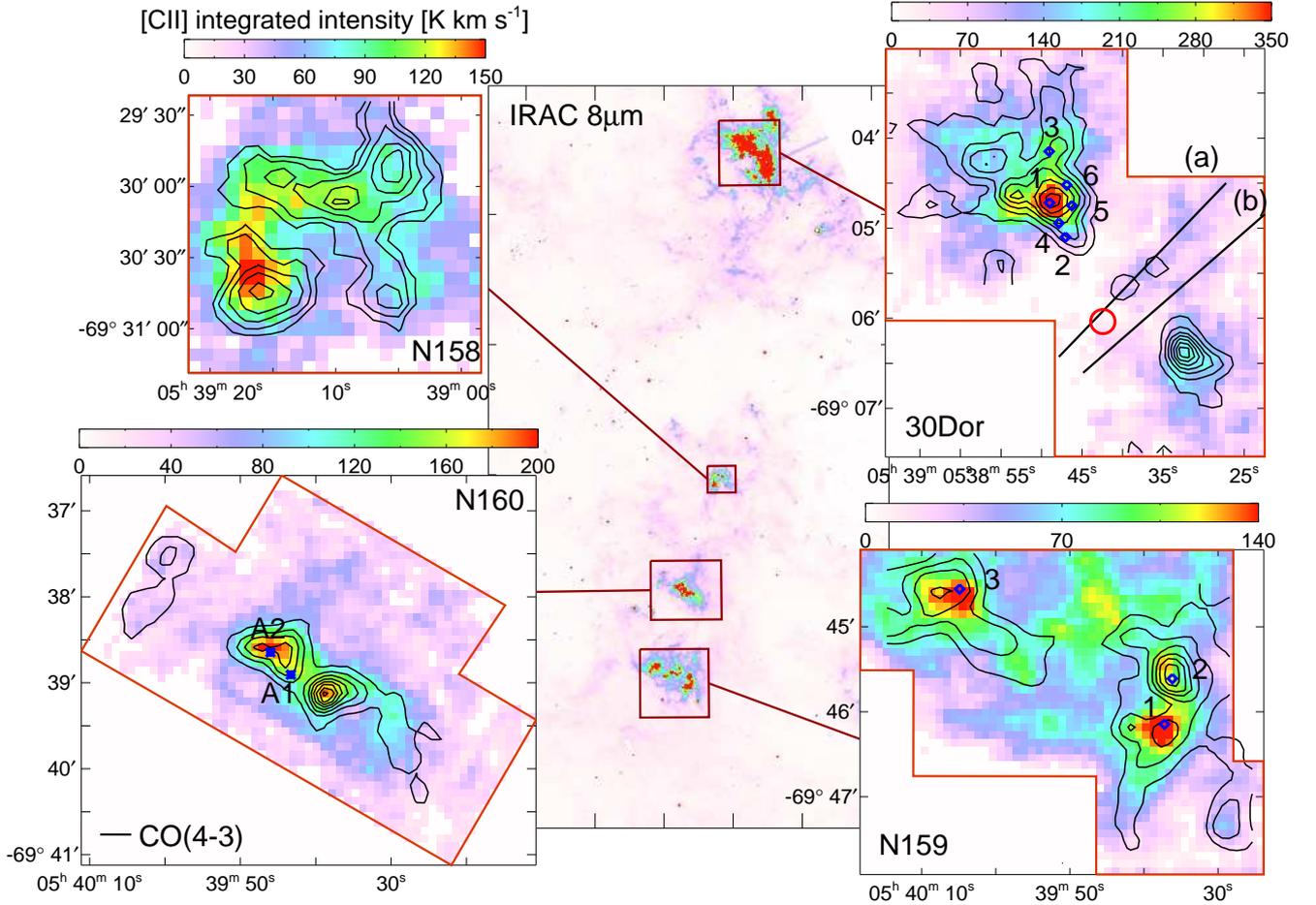}
\caption{\emph{Center:} IRAC 8\um\ map with boxes showing the areas we observed. \emph{Four inset color maps with contours:} Integrated intensity map of \cii\ (color) overlaid with contours of CO(4-3) integrated intensity at 16\arcsec\ resolution. The red lines outline the area observed in \cii. Contour levels are at intervals of  5~K\,\kms\ for N158, 10~K\,\kms\ for N160, and 15~K\,\kms\ for N159 and 30~Dor. The unit of color bar is K\,\kms. In the maps of N159 and 30~Dor, the positions of the \oi\ observations are marked following the numbering in Table~\ref{table:oi_detected}. In 30~Dor, the position of R136 is shown as a red circle, and the two cuts to make position-velocity diagrams (Fig.~\ref{figure:pvdiagram_30Dor}) are shown as black lines. In N160, two \hii\ regions N160A1 and A2 \citep{MartinHernandez2005} are marked.}
\label{figure:cii_co_integ}
\end{figure*}

As mentioned above, the LMC provides a good low metallicity template to bridge Galactic and high redshift observations, since it is close enough that star-forming regions are spatially well-resolved, and independent resolution elements do not show significant blending of different velocity components. In the LMC, 30~Doradus (30~Dor) and other high-mass star-forming regions in the molecular ridge south of 30~Dor (see Fig.~\ref{figure:cii_co_integ}), provide a good opportunity to study star formation and the ISM in a low metallicity environment. 

30~Dor is a well-studied, prominent star-forming region excited by the super star cluster R136, where 117 massive stars provide a total ionizing flux of $4\times 10^{51}$ Ly\,photons\,s$^{-1}$ \citep{CrowtherDessart1998}. Many observations suggest that the ISM must be highly clumped; \citet{Poglitsch1995} show that the integrated intensity maps of the \cii\ and CO(1-0) emission are spatially strongly correlated, suggesting that the molecular and the photodissociated gas appear coextensive over $\sim 30$\,pc. \citet{Kawada2011} show that the \oiii\ emission is widely extended spanning more than 150~pc from R136. \citet{Indebetouw2013} resolved clumps with the Atacama Large millimeter/submillimeter Array (ALMA) and show that most of the flux in the CO map is contained within an area of 5\% to 10\%. They also show that relative velocities of individual clumps, typically with a line width of 2--3\,\kms, dominate the velocity structure, with an additional overall east-west gradient. \citet{Yeh2015} suggest that the H$_2$ rotational-vibrational transition in 30~Dor is due to fluorescence without evidence of shock excitation. \citet{Chevance2016} model FIR fine-structure lines and determine the three-dimensional structure of the gas by comparing the modeled UV radiation field strength with the emitted radiation by stars.

N158 is located south of 30~Dor with an elongated \hii\ region \citep{Henize1956}. The southern part excited by the OB association L101 \citep{LuckeHodge1970} is identified as N158C or NGC2074. LH101 contains two populations of massive stars \citep[$\leq 2$~Myr and 3--6\,Myr;][]{TestorNiemela1998}, creating elongated \hii\ regions in a north-south direction with an extension to the west seen in H$\alpha$ and the optical \oiii\ emission \citep{Fleener2010,Galametz2013}. \citet{Fleener2010} found that the majority of the YSOs and central stars of ultracompact \hii\ regions in NGC2074 are not earlier than the late O-type, whereas there are several evolved early O-type stars, indicating that their formation may have started at a similar time, a few $10^5$ years ago.

N160 is the hottest among the star-forming regions in the molecular ridge in terms of the dust temperature \citep{Galametz2013} and the CO excitation \citep{Bolatto2000b,Heikkila1999}. N160A is the brightest cloud in the radio continuum which has three compact \hii\ regions \citep{HeydariMalayeriTestor1986,Indebetouw2004}. In the southwest of N160A, there is another radio continuum source N160D, where the ionized gas has a shell-like structure with a WR star as well as several OB stars at the center \citep{HeydariMalayeriTestor1986,MartinHernandez2005,Farina2009}, and two compact \hii\ regions are located at the eastern edge of the shell.

N159 is the most southern region among the four regions that we study here. Three CO cores are identified as N159~W, E, and S by \citet{Johansson1998}. Recent ALMA observations reveal detailed filamentary structures and converging flows at N159~W and E \citep{Fukui2015,Saigo2017}. \citet{Lee2016} report that the PDR properties that produce a good fit to the \oi, \cii\, and \ci\ emission predict too low CO intensities compared to the observations, suggesting that the CO emission may be excited by low-velocity C-type shocks.

A global age gradient from 30~Dor southwards to N159 has been suggested by various studies. \citet{Nakajima2005} found an age gradient of the YSOs from N160 to N159 toward the tip of N159~S and suggest a propagation of triggered star formation from north to south along the molecular ridge. \citet{Israel1996} measure a lower CO luminosity compared to \cii\ and far-infrared (FIR) in 30~Dor compared to N159 and N160, presumably reflecting that 30~Dor is the most evolved region. On the other hand, since both N160 and N159 contain young objects as well as evolved objects, \citet{Farina2009} suggest a common time for the origin of recent star formation in the N159/N160 complex as a whole, while sequential star formation at different rates is probably present in several subregions. The analysis by \citet{Gordon2017} indicated that N159 is indeed younger than N160, but these authors did not conclude whether there is an overall north-south age gradient. \citet{deBoer1998} proposed a scenario in which star formation is triggered at the leading edge due to the bow-shock of the LMC with its motion through the halo of the Milky Way; in this scenario, an age gradient is expected because of the rotation of LMC.

We performed velocity-resolved mapping observations of CO, \thco, \ci\ \transl\ and \transu, and \cii\ in N159, N160, N158, and 30Dor, and velocity-resolved pointed observations of \oi\ 145\um\ and 63\um\ at selected positions in N159 and 30~Dor. In Sect.~\ref{sec:obs} we present the data that we use in this study. A brief summary of the KOSMA-$\tau$ PDR model is given in Sect.~\ref{sec:PDRintro}. Sect.~\ref{sec:results} presents all the results and discussion. We start by presenting a line profile analysis of the \oi\ emission (Sect.~\ref{subsec:result_oi_profile}) and other emission lines (Sect.~\ref{subsec:result_lineprofile}). Next we estimate the contribution of the atomic gas (Sect.~\ref{subsec:hi_cii}) and ionized gas (Sect.~\ref{subsec:ionized_to_cii}) to the \cii\ emission. An overview of the line profile analysis is given in Sect.~\ref{subsec:discuss_lineprofile}. In Sect.~\ref{subsec:spatial_dist} we present details of the spatial distribution of detected emission lines in individual regions, and discuss the column density of CO, \czero, and \cplus\ in the low metallicity LMC environments in Sect.~\ref{subsec:columndensity}. The results of the PDR model fitting are given in Sect.~\ref{subsec:pdrmodeling}. Our main findings are summarized in Sect.~\ref{sec:summary}.

\section{Observations and data reduction} \label{sec:obs}

\subsection{\cii\ 158\um\ observations with SOFIA/GREAT}

\begin{table*}
\caption{Summary of \cii\ observations.}
\label{table:obssummary}
\centering
\begin{tabular}{cclcc}
\hline\hline
Region & map size & OFF position (J2000) & $t_{\textrm{source}}^\textrm{a}$ [sec] & $\sigma_{\textrm{rms}}^\textrm{b}$\\
\hline
N159 & 4\arcmin$\times$(3\arcmin--4\arcmin) & 05:38:53.3\ \ -69:46:29.0 & 59 & 0.18\\
N160 & 5\arcmin$\times$(2\arcmin--3\arcmin) & 05:41:35.3\ \ -69:39:01.1 & 28 & 0.39\\
N158 & 2\arcmin$\times$2\arcmin & 05:41:04.2\ \ -69:30:20.0 & 7 & 0.46\\
30~Dor & 3\arcmin$\times$3\arcmin$+$2\arcmin$\times$3\arcmin &  05:39:33.7\ \ -69:04:00.0 & 50 & 0.18\\
&& 05:40:34.1\ \ -69:08:35.0 $^\textrm{c}$ &&\\
\hline\hline
\end{tabular}
\begin{list}{}{\setlength{\itemsep}{0ex}}
\item[$^\textrm{a}$]Effective integration time for one pixel in a 30\arcsec\ resolution map.\ \ \ $^\textrm{b}$ Median of the baseline noise in a 30\arcsec\ resolution map.
\item[$^\textrm{c}$]The farther OFF position, which was used to correct the contaminated primary OFF position (see text).
\end{list}\end{table*}

\begin{table*}
\caption{Summary of the pointed observations for the \oi\ lines. Only positions where at least one of the \oi\ line is clearly detected and used for discussion are shown.}
\label{table:oi_detected}
\centering
\begin{tabular}{cccccc}
\hline\hline
Region & ID$^\textrm{a}$ & position$^\textrm{b}$ (J2000) & pixels (lines) & $t_{\textrm{ON}}$ [min] & $\sigma_{\textrm{rms}}$\\
\hline
N159 &  1 &05:39:37.2\ \ -69:46:08.7 & LFAH/V\_PX00 (\oi\ 145\um) & 7.7 & 0.20\\
&&& HFAV\_PX00 (\oi\ 63\um) & 3.8 & 0.16\\
& 2 & 05:39:36.2\ \  -69:45:36.6 & LFAV\_PX03 (\oi\ 145\um) & 3.8 & 0.49\\
& 3 & 05:40:04.9\ \  -69:44:33.3 & LFAH/V\_PX00 (\oi\ 145\um) & 12.8 & 0.16\\
&&& HFAV\_PX00 (\oi\ 63\um) & 7.7 & 0.14\\
30~Dor & 1&  05:38:49.0\ \  -69:04:43.1 & LFAH/V\_PX00 (\oi\ 145\um) & 3.4 & 0.36\\
&&& HFAV\_PX00 (\oi\ 63\um) & 3.4/1.7/1.7/1.7$^\textrm{c}$ & 0.7/1.1/1.0/0.8\\
& 2 & 05:38:47.0\ \  -69:05:05.9 & LFAH\_PX01 (\oi\ 145\um) & 1.7 & 0.45\\
& 3 & 05:38:49.1\ \  -69:04:08.8 & LFAV\_PX04 (\oi\ 145\um) & 1.7 & 0.45 \\
& 4 & 05:38:47.9\ \  -69:04:56.6 & HFAV\_PX01 (\oi\ 63\um) & 3.4/1.7/1.7/1.7$^\textrm{c}$ & 0.42/0.57/0.46/0.45\\
& 5 & 05:38:46.2\ \  -69:04:44.9 & HFAV\_PX02 (\oi\ 63\um) & 3.4/1.7/1.7/1.7$^\textrm{c}$ & 0.42/0.68/0.64/0.52\\
& 6 & 05:38:46.9\ \  -69:04:31.2 & HFAV\_PX03 (\oi\ 63\um) & 3.4/1.7/1.7/1.7$^\textrm{c}$ & 0.50/0.58/0.50/0.57\\
\hline\hline
\end{tabular}
\begin{list}{}{\setlength{\itemsep}{0ex}}
\item[$^\textrm{a}$] ID numbers used in Figs.~\ref{figure:oi_spec_N159} and \ref{figure:oi_spec_30Dor}.
\item[$^\textrm{b}$] Positions are determined as follows: when both LFAH and LFAV observations are available, we use the average of these. When only one of the LFA observations is available, we took its position. When 4 raster observations of the HFAV are available, we take the average of the center and the north positions. These positions are used when extracting spectra from the mapping observations (e.g. \cii, CO(4-3)).
\item[$^\textrm{c}$] 4 position raster.
\end{list}
\end{table*}

\begin{table*}
\caption{Summary of observed emission lines.}
\label{table:linesummary}
\centering
\begin{tabular}{ccccccccc}
\hline\hline
Line & Frequency$^\mathrm{a}$ [GHz] & Instrument & $\eta_\mathrm{mb}^\mathrm{b}$ & HPBW$^\mathrm{c}$ [\arcsec] & \multicolumn{4}{c}{detection$^\mathrm{d}$}\\
&&&&& N159 & N160 & N158 & 30~Dor\\
\hline
APEX &&&&&&&&\\
\hline
\ceio(2-1) & 219.5603541 & SHeFI APEX-1 & 0.81 & 28.4 & -- & $\times$ & -- & --\\
\thco(2-1) & 220.3986842 & SHeFI APEX-1 & 0.81 & 28.3 & $(\checkmark)^\mathrm{e}$ & $\checkmark$ & -- & --\\
CO(2-1) & 230.5380000 & SHeFI APEX-1 & 0.81 & 27.1 & $(\checkmark)^\mathrm{e}$ & $\checkmark$ & -- & $\checkmark$ \\
\ceio(3-2) & 329.3305525 & FLASH$^+$ & 0.69 & 19.5 & -- & -- & -- & $\times$ \\
\thco(3-2) & 330.5879653 & FLASH$^+$ & 0.69 & 19.0 & $\checkmark$ & $\checkmark$ & $\checkmark$ & $\checkmark$ \\
CO(3-2) & 345.7959899 & FLASH$^+$ & 0.69 & 18.2 & $\checkmark$ & $\checkmark$ & $\checkmark$ & $\checkmark$ \\
CO(4-3) & 461.0407682 & FLASH$^+$ & 0.61 & 13.6 & $\checkmark$ & $\checkmark$ & $\checkmark$ & $\checkmark$ \\
\ci\ \transl & 492.1606510 & FLASH$^+$ & 0.6 & 12.8 & $\checkmark$ & $\checkmark$ & $\checkmark$ & $\checkmark$ \\
CO(6-5) & 691.4730763 & CHAMP$^+$ LFA & 0.42 & 8.8 & $\checkmark$ & $\checkmark$ & -- & $\checkmark$ \\
\ci\ \transu & 809.3419700 & CHAMP$^+$ HFA & 0.38 & 7.7 & $\checkmark$ & $\times$ & -- & $\checkmark$ \\
\hline
SOFIA &&&&&&&&\\
\hline
\nii & 1461.1338000 & GREAT L1 & 0.67 & 18.3 & $\checkmark$ & -- & -- & -- \\
\cii & 1900.5369000 & GREAT L2 & 0.67--0.69 & 14.1 & $\checkmark$ & $\checkmark$ & $\checkmark$ & $\checkmark$ \\
\oi & 2060.0688600 & upGREAT LFA & 0.65-0.73 & 12.9 & $(\checkmark)^\mathrm{e}$ & -- &-- & $(\checkmark)^\mathrm{e}$ \\
\oi & 4744.7774900 & upGREAT HFA & 0.63 & 6.3 & $(\checkmark)^\mathrm{e}$ & -- &-- & $(\checkmark)^\mathrm{e}$ \\
\hline\hline
\end{tabular}
\begin{list}{}{\setlength{\itemsep}{0ex}}
\item[$^\textrm{a}$]The Cologne Database for Molecular Spectroscopy (CDMS)\ \ \ $^\textrm{b}$ Main beam efficiency.\ \ \ $^\textrm{c}$ Half power beam width.
\item[$^\textrm{d}$]$\checkmark$ is detected, $\times$ is not detected, and -- is not observed.\ \ \ $^\textrm{e}$ Only pointed observations.
\end{list}
\end{table*}

\cii\ mapping observations were made in four regions using the German REceiver for Astronomy at Terahertz Frequencies \citep[GREAT\footnote{GREAT is a development by the MPI f\"{u}r Radioastronomie and the KOSMA / Universit\"{a}t zu K\"{o}ln, in cooperation with the MPI f\"{u}r Sonnensystemforschung and the DLR Institut f\"{u}r Planetenforschung};][]{Heyminck2012} on board the Stratospheric Observatory for Infrared Astronomy \citep[SOFIA;][]{Young2012}. We observed N159 and 30~Dor in July 2013, as part of the open time and guaranteed time in cycle 1 observations, and N160 and N158 in July 2015, as part of the guaranteed time in cycle 3 observations. The L2 channel was tuned to \cii, and the backend was XFFTS with 2.5 GHz bandwidth in 2013, FFTS4G with 4 GHz bandwidth in 2015. The \nii\ 205\um\ was observed in parallel only for N159. All mapping observations were made in total-power on-the-fly (OTF) mode with 6\arcsec\ step size. Other parameters of observations are summarized in Table~\ref{table:obssummary}. See also Paper I for details of the observations in N159. For 30~Dor, we found that the primary OFF position is contaminated in the \cii\ emission, and we executed total-power pointed observations of this contaminated OFF position against another more distant OFF position (see Table~\ref{table:obssummary}). We averaged 12 spectra with an integration time of 15 sec each to obtain the OFF spectrum, and add it to the mapping spectra to correct the OFF contamination. The uncertainty of the OFF spectra was estimated by the standard deviation of the 12 OFF spectra, which is 0.2~K as a median of all velocity bins between 220~\kms\ and 280~\kms, and 0.7~K as a maximum.

The data were calibrated by the standard GREAT pipeline \citep{Guan2012}, which converts the observed counts to the main beam temperature ($T_\mathrm{mb}$). One improvement since Paper I is that we interpolate the internal calibration (load) measurement and the OFF measurement in time to better account for the gain drift. We compare this new calibration in N159 with the one used in Paper I and confirmed that the resulting spectra do not have a significant change but we get better signal-to-noise ratio (S/N).  We then subtract second-order polynomial baselines and spectrally resample to 1~\kms\ channel width. We convolved the \cii\ map to different spatial resolutions as required for further analysis. For the discussion of the spatial distribution, we convolved it to 16\arcsec\ resolution (Sect.~\ref{subsec:spatial_dist}). Since we use a simple Gaussian kernel for the convolution, where we define the width of the Gaussian kernel by the difference between the square of the goal beam size and the square of the original beam size, we cannot avoid losing the spatial resolution slightly compared to the original resolution. To compare the line profiles with the \oi\ emissions, we also used the 16\arcsec\ resolution. For the analysis of the line profile and the PDR modeling, we use a spatial resolution of 30\arcsec\ for all regions in order to make use of CO(2-1) data (see \ref{subsec:obs_apex}). This also gives maps with a better S/N.  When comparing the \cii\ emission with the \hi\ emission, we convolved the \cii\ map to 1\arcmin\ resolution.

\subsection{\oi\ 145\um\ and 63\um\ observations with SOFIA/upGREAT}\label{subsec:obs_oi}

We observed \oi\ 145\um\ and 63\um\ at selected positions in N159 and 30~Dor with the upGREAT \citep{Risacher2016} onboard SOFIA in July 2017, as part of the guaranteed time in cycle 5 observations. The two polarizations of the Low Frequency Array (LFAH and LFAV) were tuned to the \oi\ 145\um\ line, and the High Frequency Array (HFA) was tuned to the \oi\ 63\um\ line, and they were operated in parallel. Each array has 7 pixels in a hexagonal configuration and the beam size of the \oi\ 145\um\ and 63\um\ is 13\arcsec\ and 6.3\arcsec, respectively. We selected two positions in each of N159 and 30~Dor, where the line profiles of the CO and \cii\ emission are significantly different, and executed pointed single-phase chopped observations. Since the \oi\ 63\um\ line for the source velocity of the LMC ($\sim 250$~\kms) is located at the edge of the atmospheric window and at a difficult tuning range for the quantum cascade laser (QCL) used in the HFA, we optimized the observations to the \oi\ 145\um\ emission.  We chose the rotation of the arrays so that some of the LFA outer pixels observed other positions of interest in parallel. We focused on the HFA at only one position in the 30~Dor region, where we observed 4 point rasters (central position plus 4\arcsec east, west, and north) with a step of 4\arcsec\ in order to fill a LFA beam, so that we can compare the line profile of the \oi\ 63\um\ with the \oi\ 145\um\ and \cii.

The central pixels of LFAH and LFAV are aligned within 2\arcsec, which is small enough compared to the LFA beam size. The central pixel of HFA is about 3\arcsec\ away from the LFA central pixels, and outer pixels of HFA and LFA are located at different positions because the size of the array is scaled by the beam size.  

Before applying the standard calibration with the GREAT pipeline, we made a frequency correction and a gain correction. The frequency correction is always needed for the HFA, since the QCL locks only to a discrete frequency. We correct the frequency scale scan by scan by using the narrow atmospheric \oi\ line. In addition, we had a synthesizer reference problem for the LFA arrays during flight, which results in an imprecise LO frequency and requires a similar frequency correction for the \oi\ 145\um\ observations. In the observed LFA band, only a wider atmospheric feature was available, thus the frequency uncertainty of the \oi\ 145\um\ in the present observations is $\sim 5$~MHz, which corresponds to $0.7$~\kms. We also applied a gain correction using the total raw count (see Appendix~\ref{app:gain_correction} for details). Then the data were calibrated by the standard GREAT pipeline, as for the mapping observations. In Table~\ref{table:oi_detected}, we list the positions where at least one of the \oi\ lines was detected.

\subsection{CO and \ci\ observations with APEX}\label{subsec:obs_apex}

We carried out complementary observations with the Atacama Pathfinder Experiment \citep[APEX\footnote{APEX is a collaboration between the Max-Planck-Institut fur Radioastronomie, the European Southern Observatory, and the Onsala Space Observatory.};][]{Guesten2006} of CO and \ci\ emission lines. The list of observed lines is shown in Table~\ref{table:linesummary}. CO(2-1) and its isotopes were observed with The Swedish Heterodyne Facility Instrument \citep[SHeFI;][]{Vassilev2008} in 2012--2014. For N159, only pointed observations toward 12 positions for CO(2-1) and one position for \thco(2-1) were made. CO(3-2) and its isotopes, CO(4-3), and \ci\ \transl\ emissions were observed with the MPIfR heterodyne receiver FLASH$^+$ \citep{Klein2014} in 2012--2016, and CO(6-5) and \ci\ \transu\ with the Carbon Heterodyne Array of the MPIfR \citep[CHAMP$^+$;][]{Kasemann2006} in 2012--2014. All observations were made in the OTF mode except for the CO(2-1) and \thco(2-1) in N159. We used the same OFF position as in the GREAT observations. We reduced the spectra in the same way as GREAT; resampled to 1~\kms\ and spatial resolution of 16\arcsec\ or 30\arcsec, depending on the analysis.

\subsection{\hi\ data from ATCA and Parkes}
To investigate the contribution of \hi\ gas to the \cii\ emission, we used the combined \hi\ survey data by Australia Telescope Compact Array (ATCA) and Parkes telescope \citep{Kim2003}. The angular resolution is 1\arcmin, corresponding to a linear scale of $\sim15$\,pc, and the spectral resolution is 1.65\,\kms. The cube covers heliocentric velocities between $-33$ and $627$ \kms, with emission from the LMC arising between $190$ and $386$\,\kms. The rms of the brightness fluctuations in the final cube is $2.4$~K per channel. Integrating over three channels, this corresponds to a column density sensitivity limit of $\sim2.2 \times 10^{19}$\,cm$^{-2}$.

\subsection{\oi\ observations with PACS} \label{subsec:obs_pacs}

We also use the \oi\ 63\um\ and 145\um\ data, observed with the Photoconductor Array Camera and Spectrometer \citep[PACS;][]{Poglitsch2010} on the Herschel Space Observatory\footnote{Herschel is an ESA space observatory with science instruments provided by European-led Principal Investigator consortia and with important participation from NASA.} \citep{Pilbratt2010}, in order to better constrain the PDR modeling. All these data are presented in \citet{Cormier2015}. A detailed analysis of these lines in 30~Dor has been presented by \citet{Chevance2016}, and in N159~W by \citet{Lee2016}. In order to avoid regridding more than once, we downloaded the level 2 data in the original $5\times 5$ spaxels from the Herschel Science Archive and directly extracted the spectra. We assume that the brightness of individual spaxel is uniformly distributed in a $9.4$\arcsec$\times9.4$\arcsec\ square. We defined a geometrical area which corresponds to each pixel in our GREAT and APEX dataset as a circle with diameter of 30\arcsec, and extracted spectra as an average within this circle. As mentioned in \citet{Cormier2015}, some observations in 30~Dor have contaminated OFF spectra, where we re-ran the last step of the pipeline to skip the procedure of the OFF subtraction and simply used only ON spectra because we are not interested in the continuum level.

\subsection{FIR continuum emission with Spitzer and Herschel}\label{subsec:obs_continuum}

We also use the Spitzer MIPS 70\um\ continuum from the SAGE project \citep{Meixner2006} and PACS 100\um, 160\um, SPIRE 250\um\, and 350\um\ by the HERITAGE project \citep{Meixner2013} to construct the dust spectral energy distribution (SED) to be fitted with the PDR model. As for the \oi\ line emission, we took a geometrical average within the circle of a 30\arcsec\ diameter to make the same grid as the GREAT and APEX dataset. We did not use SPIRE 500\um\ because its spatial resolution is worse than 30\arcsec.

\section{PDR model}\label{sec:PDRintro}

We used the KOSMA-$\tau$ PDR model \citep{Stoerzer1996,Roellig2006,Roellig2013a} to investigate the effect of metallicity on the PDR structure and chemistry and to investigate physical properties of the four regions. As presented in \citet{Roellig2013a}, the revised KOSMA-$\tau$ model treats the dust-related physics consistently and computes the dust continuum emission together with the line emission of chemical species. This has been now extended to include all dust size distributions tabulated in \citet{WD2001}, including the LMC and SMC dust models. By applying the elemental abundances for the LMC \citep[$X(\textrm{C})= 7.94\times 10^{-5}$, $X(\textrm{O})= 2.51\times 10^{-4}$;][]{Garnett1999}, the model can be compared with the far-infrared continuum SED and the line intensities in the LMC.

The KOSMA-$\tau$ model assumes either a single clump filling the beam (hereafter 'non-clumpy model') or an ensemble of clumps with a power law mass spectrum $dN/dM \propto M^{-\alpha}$ where $\alpha=1.8$ (hereafter 'clumpy model'). Each clump follows the mass-size relation $M_\textrm{clump}\propto R_\textrm{clump}^\gamma$ with $\gamma=2.3$, and has a power-law radial density profile outside of the core, where the density is constant \citep{Stoerzer1996,Cubick2008}. The line and continuum intensities are calculated by summing the contribution from each clump in a model grid of three independent parameters; total mass ($m$), mean gas density ($n$), and far-ultraviolet (FUV; $h\nu=$6--13.6~eV) flux ($\chi$) in units of the Draine field ($2.7\times 10^{-6}$~W\,m$^{-2}$).

\section{Results and discussion} \label{sec:results}

\begin{figure}
\centering
\includegraphics[width=0.8\hsize]{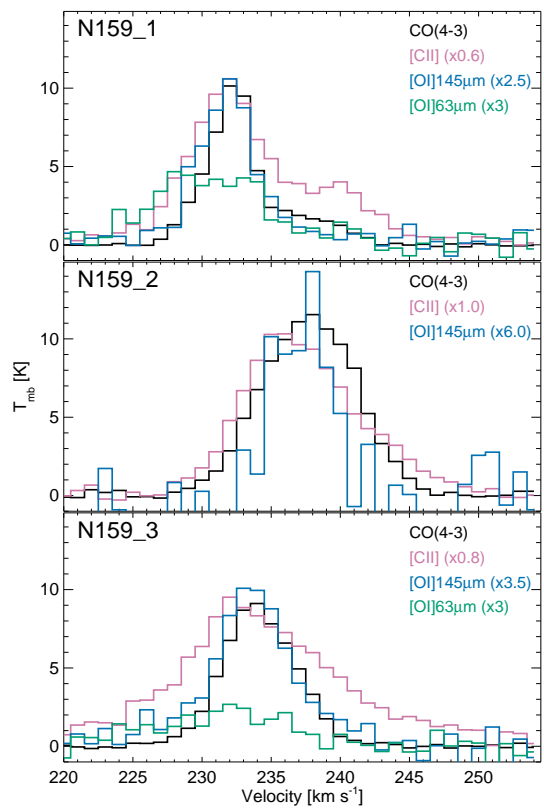}
\caption{\oi\ 145\um\ and 63\um\ spectra in N159 at the positions marked in Fig.~\ref{figure:cii_co_integ}. The CO(4-3) and \cii\ spectra were extracted from the maps with a resolution of 16\arcsec, while the \oi\ observations are shown with the original spatial resolution (13\arcsec\ for the \oi\ 145\um\ and 6.3\arcsec\ for the \oi\ 63\um).}
\label{figure:oi_spec_N159}
\end{figure}

\begin{figure}
\centering
\includegraphics[width=0.9\hsize]{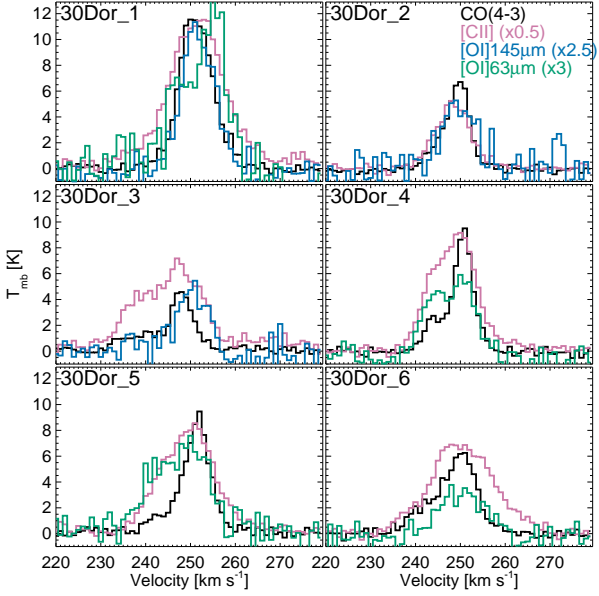}
\caption{Same as Fig.~\ref{figure:oi_spec_N159} but for 30~Dor. Here the \oi\ 63\um\ spectra are the average of the 4 raster positions (see text).}
\label{figure:oi_spec_30Dor}
\end{figure}

\begin{figure}
\centering
\includegraphics[width=0.9\hsize]{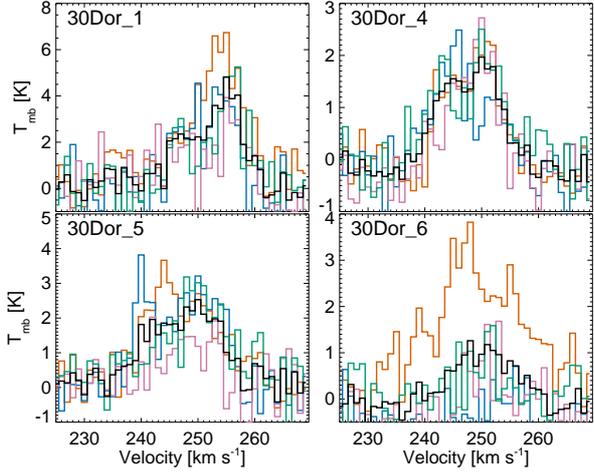}
\caption{\oi\ 63\um\ spectra of 4 raster positions and their average (black histogram) in 30~Dor positions 1, 4, 5, and 6.}
\label{figure:oi_raster_30Dor}
\end{figure}

The properties of the detected emission lines in each region are summarized in Table~\ref{table:linesummary}. Fig.~\ref{figure:cii_co_integ} shows the integrated intensity maps of \cii\ and CO(4-3) with 16\arcsec\ angular resolution, together with an overview of IRAC 8\um\ mission. In Appendix~\ref{app:map_and_spectra}, integrated intensity maps of the detected emissions at 30\arcsec\ resolution and spectra of selected positions are shown for N160, N158, and 30~Dor. For N159 data we refer to Paper I. The \oi\ 145\um\ and 63\um\ lines at selected positions (Table~\ref{table:oi_detected}, and marked in Fig.~\ref{figure:cii_co_integ}) are shown in Figs.~\ref{figure:oi_spec_N159} and \ref{figure:oi_spec_30Dor}.

\subsection{The velocity resolved \oi\ emissions in the LMC}\label{subsec:result_oi_profile}

We observed the velocity resolved \oi\ 63\um\ and 145\um\ emission in N159 and 30~Dor for the first time. In N159, the central pixels of LFAH and LFAV detected the \oi\ 145\um\ at two observed positions. We average the LFAH and LFAV spectra using a weight based on the baseline noise. In addition, we detected the line with the LFAV pixel 3 at the first observed position. We used only the LFAV data for this position because the LFAH pixel 3 performed much worse than the LFAV pixel 3. The \oi\ 145\um\ spectra at these three positions are shown in Fig.~\ref{figure:oi_spec_N159} and the positions are marked in Fig.~\ref{figure:cii_co_integ}. The 63\um\ emission was detected by the HFAV center pixel and is also shown in Fig.~\ref{figure:oi_spec_N159}. A direct comparison of the line profile is difficult because of the different beam sizes. The line profile of the \oi\ 145\um\ has a mixture of the characteristics of the CO(4-3) and \cii\ profiles. At position 1 in Fig.~\ref{figure:oi_spec_N159}, the blue wing of the \oi\ 145\um\ line is identical to that of \cii, while the broad red wing and a second velocity component at $\sim 240$~\kms\ is missing in the \oi\ 145\um. On the other hand, at position 3, the \oi\ 145\um\ profile follows that of CO(4-3), which is narrower than the \cii\ line. The \oi\ 145\um\ profile is difficult to interpret at position 2 due to low S/N, but the \oi\ 145\um\ peak velocity may be between that of \cii\ and CO.

In 30~Dor, we detected \oi\ emission only at the first pointed positions, where the 4 point raster observations were carried out (see Section~\ref{subsec:obs_oi}). The 6 positions listed in Table~\ref{table:oi_detected} are the positions of different pixels. In Fig.~\ref{figure:oi_spec_30Dor}, the 4 raster positions were averaged for the \oi\ 63\um, so that we can compare its line profile with other emission lines. We took a simple mean instead of a weighted mean in order to weight different spatial positions equally. The averaged \oi\ 63\um\ spectra have line profiles that tend to be more similar to the \cii\ profile in terms of the overall line width. At position 1, a dip around $\sim 250$~\kms\ may be self-absorption, because it corresponds to the peak of CO and \oi\ 145\um. On the other hand, the profiles at positions 4 and 5 seem to show two velocity components around $\sim 242$~\kms\ and $\sim 250$~\kms, where CO(4-3) has only a very weak emission for the former component, and \cii\ emission fills the velocity between these two components, although it is not excluded that all of CO(4-3), \cii, and \oi\ 63\um\ have an absorption feature. The \oi\ 145\um\ emission behaves as in N159: a mixture of the characteristics of the CO(4-3) profile and \cii\ profile. At position 1 it is closer to the CO(4-3) profile, at position 2 to the \cii\ profile, and at position 3, it has the blue profile of CO(4-3) and red profile of \cii. Fig.~\ref{figure:oi_raster_30Dor} shows that the \oi\ 63\um\ spectra at individual raster positions are different, indicating that different velocity components often originate from spatially separated components. This is consistent with the ALMA observations by \citet{Indebetouw2013}, showing that the relative velocities of individual clumps dominate the velocity structures.

We compare the \oi\ integrated intensities measured by upGREAT to those by PACS. We extracted PACS spectra following the method described in Sect.~\ref{subsec:obs_pacs} but using a circle with diameter of 13\arcsec. For \oi\ 145\um, the integrated intensities match within 20\% except for the two faintest positions (N159 position 2 and 30~Dor position 3), where they match within 30\%. For \oi\ 63\um, we make comparisons only at positions in 30~Dor where the upGREAT observations consist of the 4 point raster. For the bright positions (30~Dor position 1 and 5), the discrepancy between PACS and upGREAT (raster positions averaged) is within 20\%, which is also consistent with the uncertainties caused by the gain correction (Appendix~\ref{app:gain_correction}). In view of the difficult to assess beam coupling effects for our undersampled data set, and the different beam width of the observations, we consider the agreement encouraging.

\subsection{Line profile analysis}\label{subsec:result_lineprofile}

In Paper I, we show that a combination of Gaussian profiles defined by CO(3-2) typically well reproduces the line profiles of CO, \thco, and \ci\ emissions. \cii, by contrast, has a very different line profile that is typically wider than the CO profile, but is often not symmetric and is complex in N159. The other three regions studied here also show the same trend. Appendix~\ref{app:map_and_spectra} shows spectra of selected positions in each region.

To quantitatively investigate the velocity profiles, we fit one of the CO spectra with Gaussians as a reference, and apply the fitted center velocity and width in the fit to the other emission lines, fitting only their amplitudes. For the mapping observations, we used the CO(3-2) as a reference emission line and performed fitting at each grid position in a 30\arcsec\ resolution map (a detailed description is given in the Paper I). For the pointed observations of \oi\ 63\um\ and 145\um, we used the 16\arcsec\ resolution data and used CO(4-3) as a reference line. We fit the reference CO spectra with two Gaussians, except for eye-selected positions where there are clearly three velocity components identified. In the following, we refer to the fitted center velocity and width of those Gaussians as 'the CO-defined line profile', and investigate how well the (amplitude-scaled) CO-defined line profile describes the other emission lines.

We compare the sum of the fitted Gaussians with the integrated intensity for each emission line. For all detected CO and \thco\ emissions in four regions, the integrated intensity is well reproduced by the sum of Gaussians. As shown in the Paper I, the \thco(3-2) and \ci\transl\ lines in N159 show that the sum of Gaussians yields larger values than the integrated intensity where the emission is strong, which is interpreted as a broadening of the $^{12}$CO lines by the optical depth effect. The \thco(3-2) in N160 shows the same trend, but it is not the case for 30~Dor and N158. For the \ci\transl\ emission, we do not see the trend other than in N159 because of the large scatter due to the lower S/N in other regions.

What is common across all four regions is that the sum of the Gaussians of the \cii\ emission is on average 30\% lower than its integrated intensity, which suggests that the CO-defined line profile can reproduce only a fraction of the \cii\ emission. As in N159, the \cii\ profile is closer to the CO-defined profile toward the CO peaks, and the fraction of the \cii\ emission that cannot be fit by the CO-defined Gaussians increases toward the regions between CO peaks, up to 60\% in some positions in 30~Dor. Possible contributions from the atomic gas and the ionized gas are discussed in the next subsections.

As shown in Sect.~\ref{subsec:result_oi_profile}, the \oi\ emission has a profile between CO(4-3) and \cii. The Gaussian fit result confirms this conclusion: the fraction of the \oi\ emission that cannot be fit by the CO-defined Gaussians are clearly correlated to its fraction for the \cii\ emission, and somewhat smaller than the \cii\ emission. In N159, this fraction for the \oi\ 145\um\ is about 25\% at position 1 and 3, whereas the fraction for the \cii\ is 35-45\%. In 30Dor, the fraction for \oi\ 145\um\ is 0--20\%, \oi\ 63\um\ is 0--30\%, and \cii\ is 15--45\%.

\subsection{Contribution of the atomic gas to the \cii\ emission}\label{subsec:hi_cii}

\begin{figure}
\centering
\includegraphics[width=0.9\hsize]{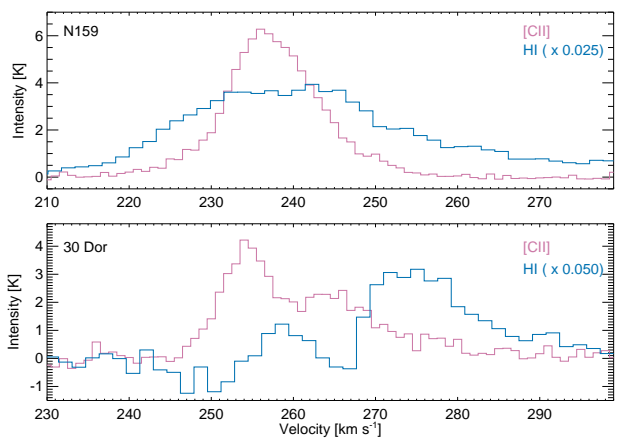}
\caption{\cii\ and \hi\ spectra at a \cii\ blob in N159 (05:39:45.9, -69:44:32.2, upper panel) and 30~Dor (05:38:31.6, -69:04:50.0, lower panel) at 1\arcmin\ spatial resolution.}
\label{figure:cii_hi_spec}
\end{figure}

In order to estimate the contribution of the atomic gas to the \cii\ emission, we compared the \cii\ spectra with the \hi\ spectra at 1\arcmin\ resolution. In general, the \hi\ emission has much broader line profiles than those of \cii\ in N159, N160, N158, and regions not associated with CO peaks in 30~Dor. Towards CO peaks in 30~Dor, the \hi\ shows absorption features. Figure ~\ref{figure:cii_hi_spec} shows the \cii\ and \hi\ spectra at two positions in N159 and 30~Dor, where the \cii\ emission dominates over the CO and \ci\ emission. At the position in 30~Dor, the \cii\ velocity component at $\sim 255$~\kms\ has a counterpart in the CO emission, but the component at $\sim 265$~\kms\ is only seen in \cii\ and not in CO, and the \hi\ emission shows an absorption feature at this velocity. This is consistent with a picture of the \hi\ self-absorption by cold molecular clouds with a mixture of atomic hydrogen \citep{LiGoldsmith2003,Gibson2005,Kavars2005,Klaassen2005,Tang2016}, although we do not resolve individual clouds in the LMC and the absorption feature is not as narrow as specified in their studies.

We estimate the \hi\ gas contribution to the \cii\ emission using the line profile. We scaled the \hi\ spectra to fit the wing of the \cii\ emission and attributed it as a contribution of the \hi\ gas to the \cii\ emission. In most regions in N158, N159 and N160, the estimated fraction of the \hi\ gas contribution to the \cii\ emission is 15\% or less, except for N159~E and the western edge of N160, where the fraction is higher. In 30~Dor, we could not fit the wing well because the velocity profile is more complex than in other regions. Since the \hi\ profile is much broader than the \cii\ line profile, subtracting this contribution does not make the \cii\ line profile significantly narrower, and it still cannot be explained by the CO-defined line profile.

We then estimate the thermal pressure (or density) of the \hi\ gas that is required to emit the fitted wings of the \cii\ intensity assuming a temperature \citep{Pineda2017}. Using the equation of optically thin subthermal \cii\ emission from \citet{Goldsmith2012} and the gas temperature of 100~K, the derived thermal pressure is $4\times 10^3$--$10^5$~K\,cm$^{-3}$ in three regions except for 30~Dor. They are higher than the standard Galactic ISM \citep[e.g. the cold neutral medium with $T_k=100$~K and $n=30$~\cc;][]{Draine2011a}, and consistent with previous studies suggesting higher thermal pressure in the LMC \citep{Welty2016,Pineda2017}. Regions with higher pressure $\ge 3\times 10^4$~K\,cm$^{-3}$ occur around N159~E, the \cii\ peak in N159~W, the N160 peak and its southern and western part. This is consistent with \citet{Welty2016}, who find higher pressure toward complex regions including \hii\ regions, molecular clumps and supernova remnants, where the pressures may be enhanced by energetic activities. The derived pressure is not sensitive to the assumed temperature for $T>80$~K \citep{Pineda2017}.

With the above method, we may underestimate the \cii\ integrated intensity that can be attributed to the atomic phase because we assume that the \hi\ line profile is optically thin.  \citet{Braun2009} and \citet{Braun2012} show that flat-topped \hi\ emission profiles can be modeled by an opacity effect, and the opacity correction of the \hi\ column density can be an order of magnitude in our observed regions. In the above method, we do not use the \hi\ column density itself but use the \hi\ profile to fit the \cii\ emission. A flat-topped profile and steeper line wings of the \hi\ emission lead to an underestimate of the corresponding \cii\ emission, if the \cii\ emission is not as optically thick as the \hi\ emission.

\subsection{Contribution of the ionized gas to the \cii\ emission}\label{subsec:ionized_to_cii}

\begin{figure}
\centering
\includegraphics[bb=45 0 405 350,width=0.53\hsize,clip]{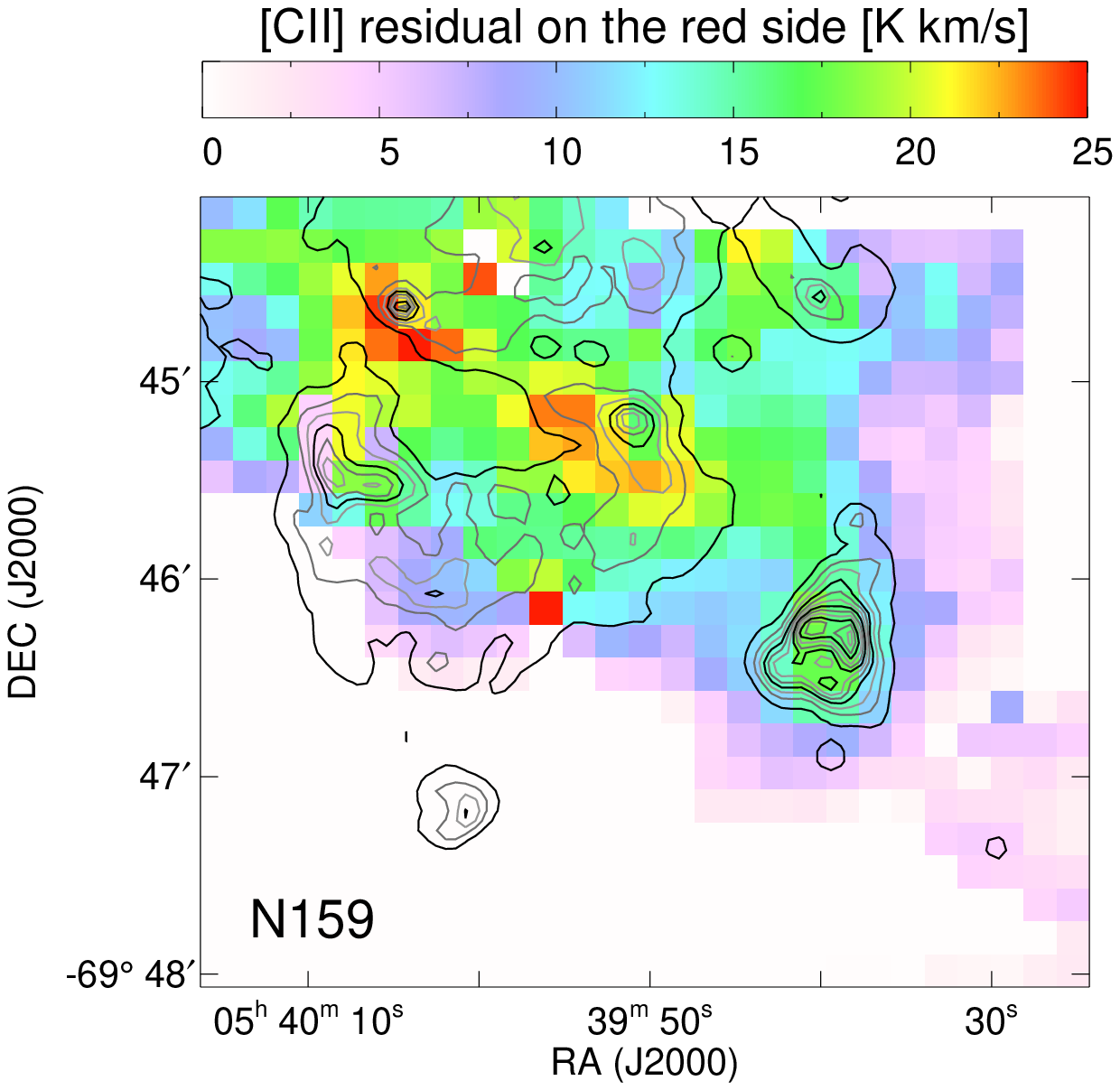}
\includegraphics[bb=100 0 405 350,width=0.45\hsize,clip]{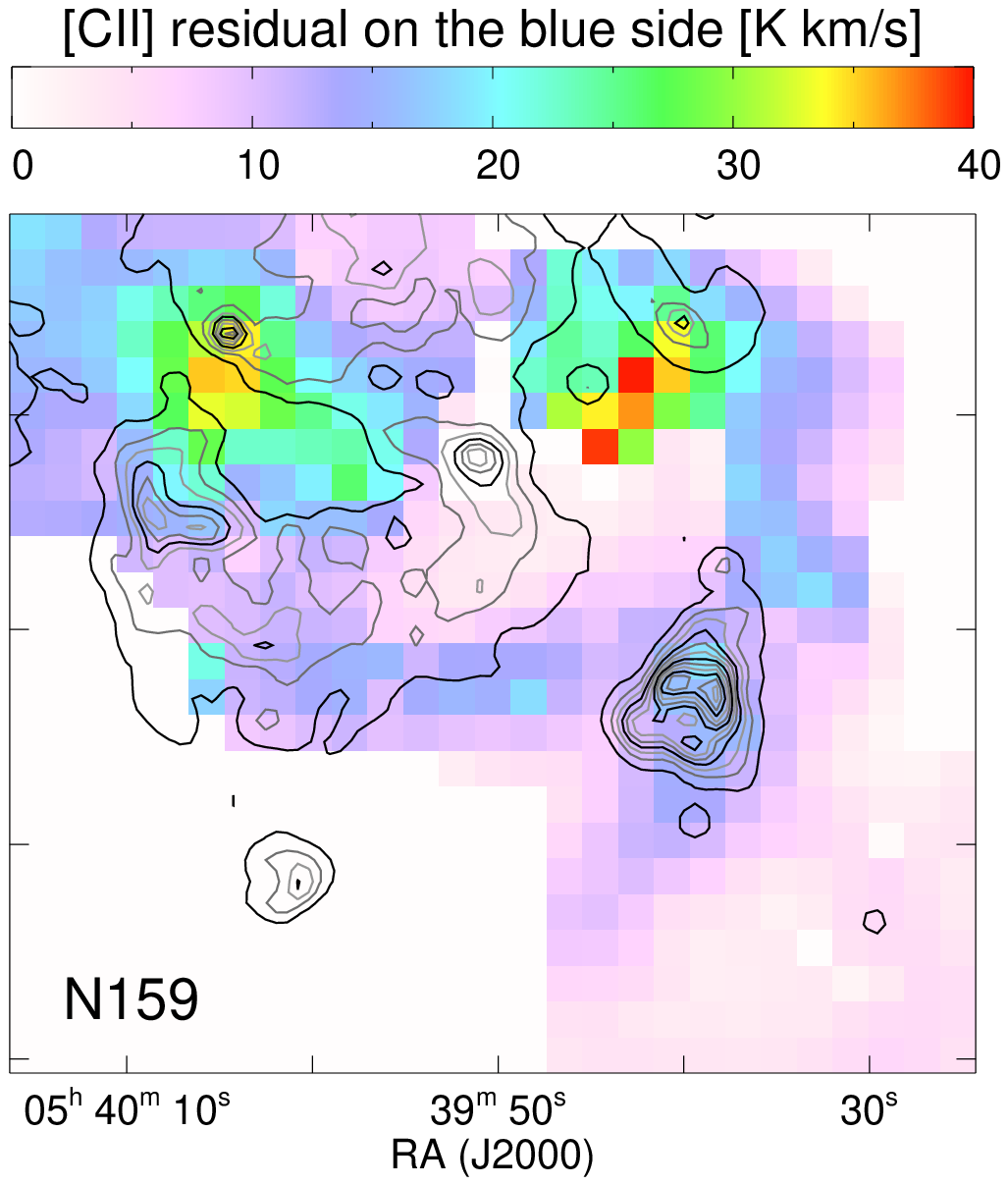}
\caption{The integrated intensity of the red-side (left) and blue-side (right) residual of the \cii\ emission after subtracting the CO-defined line profile (see text) in N159 (color, at 30\arcsec\ resolution) overlaid with contours of H$\alpha$\ \citep{Chen2010}.}
\label{figure:CII_residual_N159}
\end{figure}

\begin{figure*}
\centering
\includegraphics[width=\hsize]{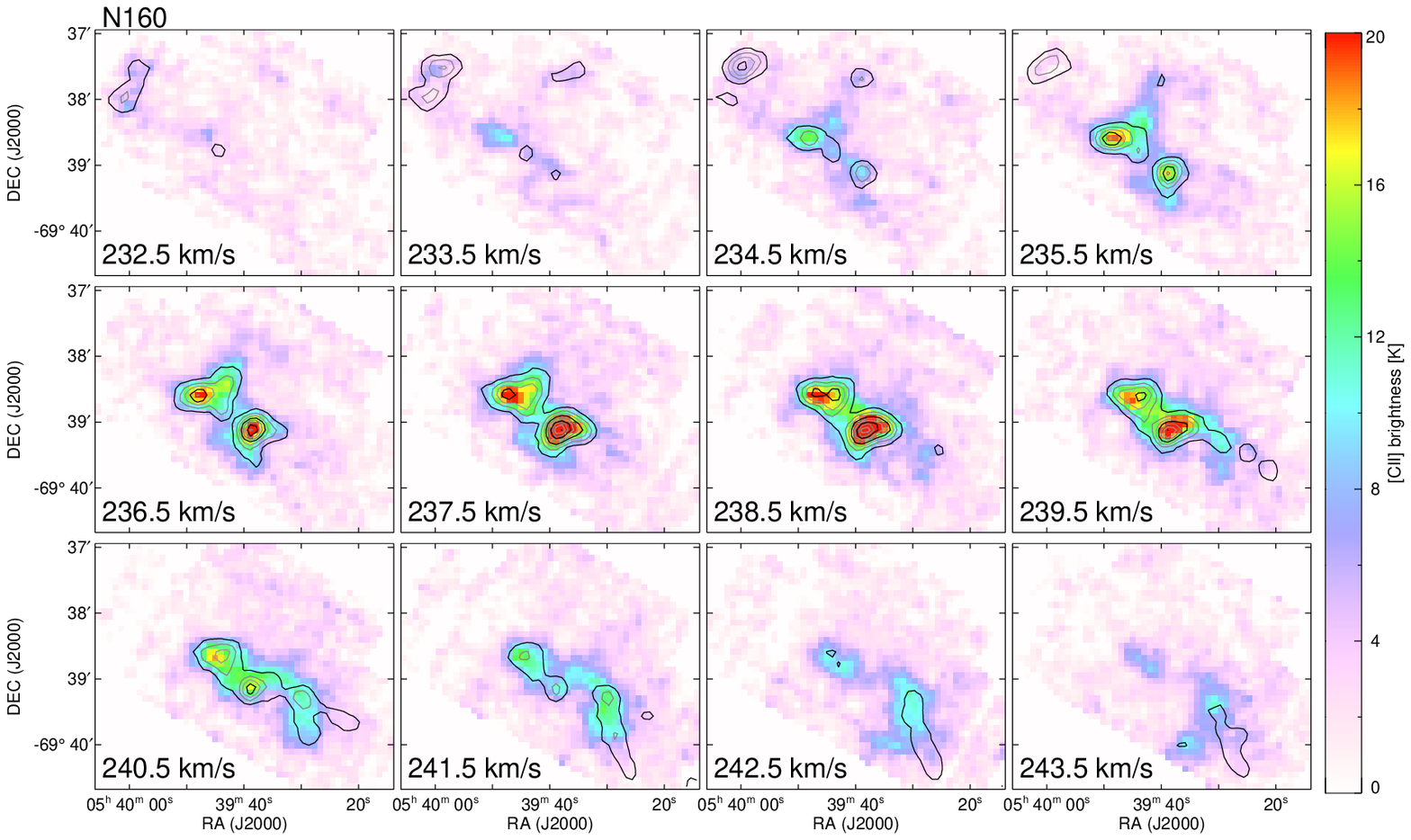}
\caption{1\kms\ wide channel maps of the \cii\ emission (color) overlaid with contours of CO(4-3) emission in N160. The contours start at 1.5~K\,\kms\ with an interval 1.5~K\,\kms. The central velocity is written in each panel.}
\label{figure:N160_channelmap}
\end{figure*}

As shown for the case of N159 in Paper I, the velocity-resolved \nii\ spectra are a useful tool to investigate the contribution of the ionized gas to the \cii\ emission. However, the \nii\ observations were made only in N159. Here we examine a possible ionized gas contribution through the spatial distribution of the residual of the \cii\ emission after subtracting the CO-defined profile. In contrast to Galactic-scale analysis \citep[e.g.][]{Pineda2014}, the contribution of the ionized gas is expected to vary locally in spatially resolved massive star-forming regions. As described in Sect.~\ref{subsec:result_lineprofile}, we fit the \cii\ spectra with Gaussians defined by the CO profile. We refer to the residual at greater velocities than the peak of the Gaussian (if more than one Gaussians are used in the fit, the peak with the largest velocity) as the red-side residual, and the residual at smaller velocities than the bluest Gaussian peak  as the blue-side residual.

The spatial distributions of the integrated intensity of the red-side residual and blue-side residual are clearly different in all regions. The left panel of Fig.~\ref{figure:CII_residual_N159} shows that the integrated intensity of the red-side residual in N159 correlates with the H$\alpha$\ emission, which confirms the result of Paper I, i.e. that the ionized gas makes some contribution to red wings of the \cii\ emission. On the other hand, the blue residual is strong in N159~E and the \cii\ blob (see Paper I). In N159~E, ALMA observations resolved a complex velocity structure consist of a few colliding filaments \citep{Saigo2017}. The \cii\ blue residual in this study may be a gas component at peculiar velocities in this kinematically complex region with physical conditions that emits dominantly in \cii, rather than CO. 

In N160, neither the red-side nor blue-side residual shows a correlation with the H$\alpha$ emission. The red-side residual is strong towards the extended \cii\ emission in the northwest (Fig.~\ref{figure:cii_co_integ}), which may be the ionized gas contribution because the radio continuum emission extends weakly there \citep{Mills1984,Israel1996}. In other regions,  the spatial distribution of the red-side and blue-side residuals do not give a strong indication of an ionized gas contribution. In 30~Dor, the region $\sim 1$\arcmin\ northeast of the \cii\ peak at 30~Dor-10 shows a strong blue-side residual, which may be due to a gas stream (Sect.~\ref{subsec:spatial_dist}).

\subsection{Discussion of the line profile shapes}\label{subsec:discuss_lineprofile}

The pointed observations of \oi\ 145\um\ and 63\um\ at N159 and 30~Dor suggested that neither the CO nor the \cii\ line profiles simply represent the \oi\ profile. Since 16\arcsec\ corresponds to $\sim 4$~pc at the distance of LMC, the observed line profile must be a composite of different cloud components. This interpretation is consistent with the fact that the \oi\ 63\um\ profile varies among the 4 raster positions with a spacing of 4\arcsec\ (Fig.~\ref{figure:oi_raster_30Dor}), and that individual clumps detected by ALMA in 30~Dor typically have linewidths of 2--3\,\kms\ \citep{Indebetouw2013}, which is narrower than the results presented here within a 30\arcsec\ beam. We propose the following interpretation for the observed different line profiles among CO, \cii\ and \oi\ in the LMC. In addition to a possible ionized gas contribution to the \cii\ emission, we consider that our beam includes several PDR components that are spatially separated and/or are in different physical phases, and each component contributes to a certain velocity range in the observed line profiles depending on their dynamics. In this case, line ratios of individual velocity components can vary depending on the physical properties of the corresponding gas component. When a PDR component emits dominantly the \cii\ line and the intensity of the CO emission is below our detection limit (for example, a low density case), we detect it as a \cii\ component which cannot be reproduced by the CO-defined velocity profile. As far as the contribution of the \hi\ gas and ionized gas to the \cii\ emission is excluded, we could call it CO-dark molecular gas. We should, however, keep in mind that the derived fraction depends on the detection limit of the CO emission.

\subsection{Spatial distributions in individual regions} \label{subsec:spatial_dist}

\begin{figure*}
\centering
\includegraphics[width=\hsize]{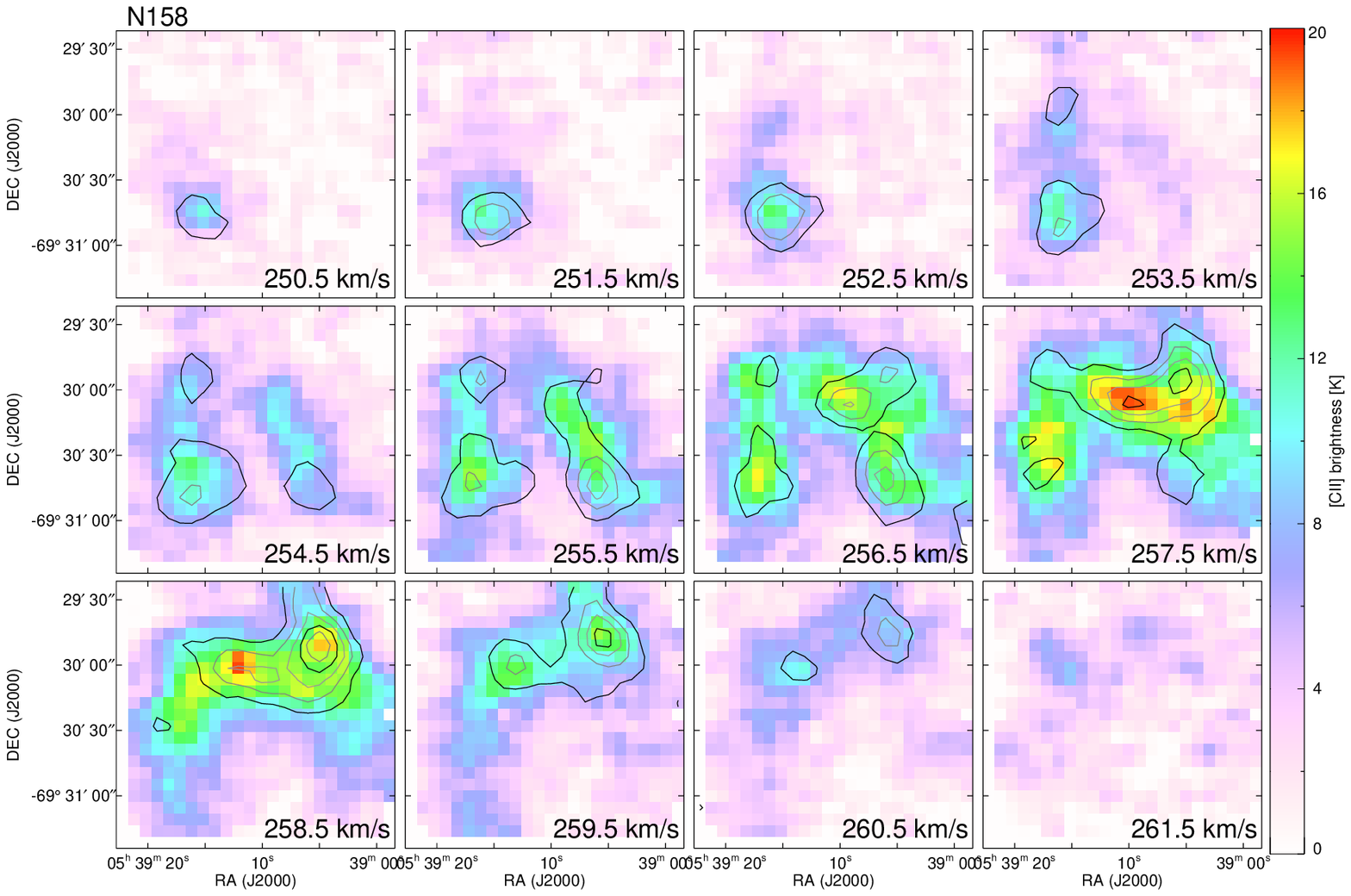}
\caption{Same as Fig.~\ref{figure:N160_channelmap} but for N158.}
\label{figure:NGC2074_channelmap}
\end{figure*}

\begin{figure*}
\centering
\includegraphics[width=\hsize]{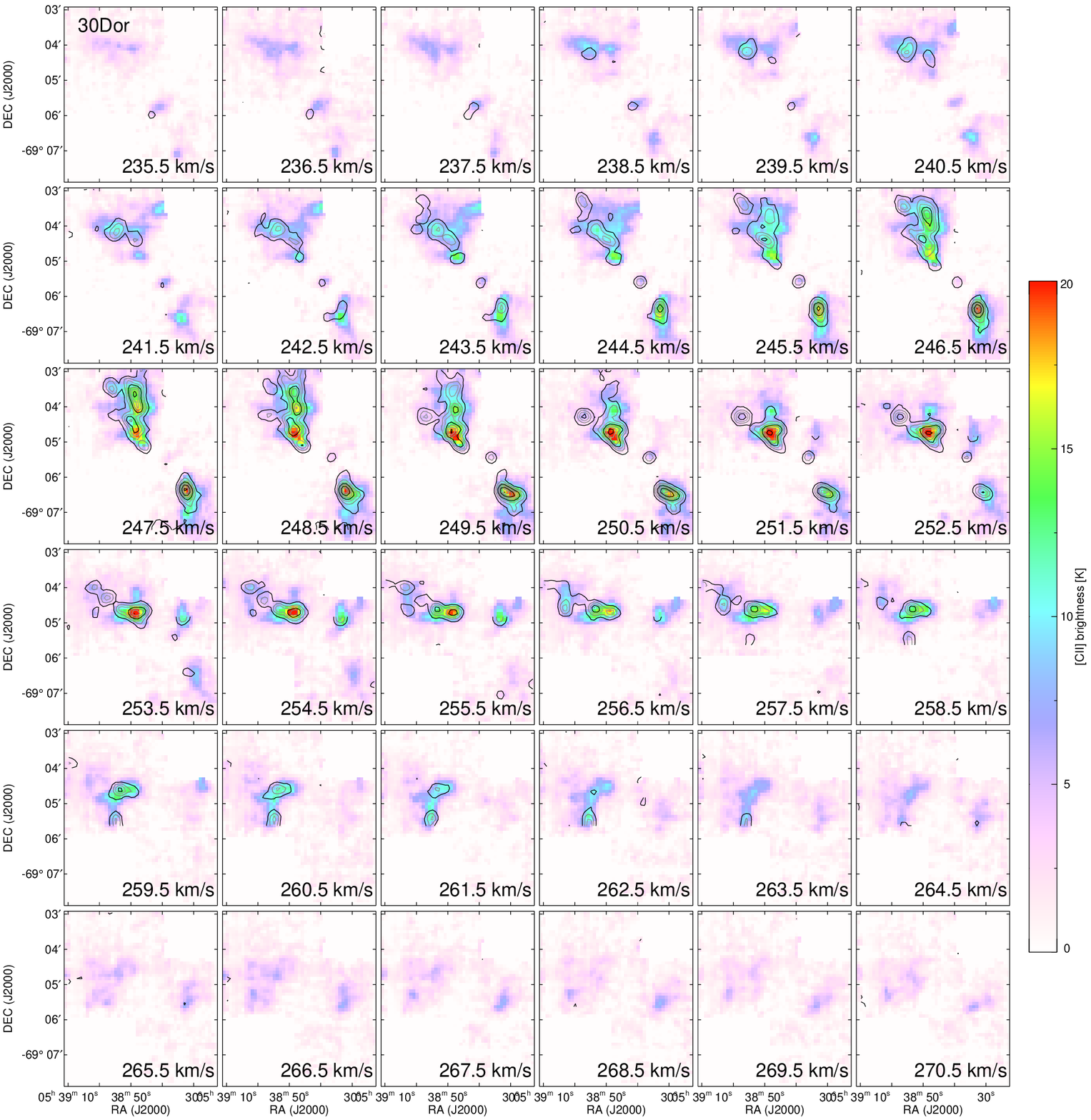}
\caption{Same as Fig.~\ref{figure:N160_channelmap} but for 30~Dor.}
\label{figure:30Dor_channelmap}
\end{figure*}

Figure~\ref{figure:cii_co_integ} shows the integrated intensity map of CO(4-3) and \cii\ in the observed four regions, and Figures~\ref{figure:N160_channelmap}--\ref{figure:30Dor_channelmap} show the channel maps (N159 is not shown). As discussed in Paper I, the spatial distribution of the integrated intensity of \cii\ and CO(4-3) emission is not the same in N159. This is also the case for N158 to some degree, whereas in 30~Dor and N160, they are more similar. The integrated intensity maps of all detected emissions converted to 30\arcsec\ are shown in Appendix~\ref{app:map_and_spectra}. In the following,  we describe the detailed morphology in each region except for N159, for which we refer the reader to Paper I.

\subsubsection{N160}

\begin{figure}
\centering
\includegraphics[width=\hsize]{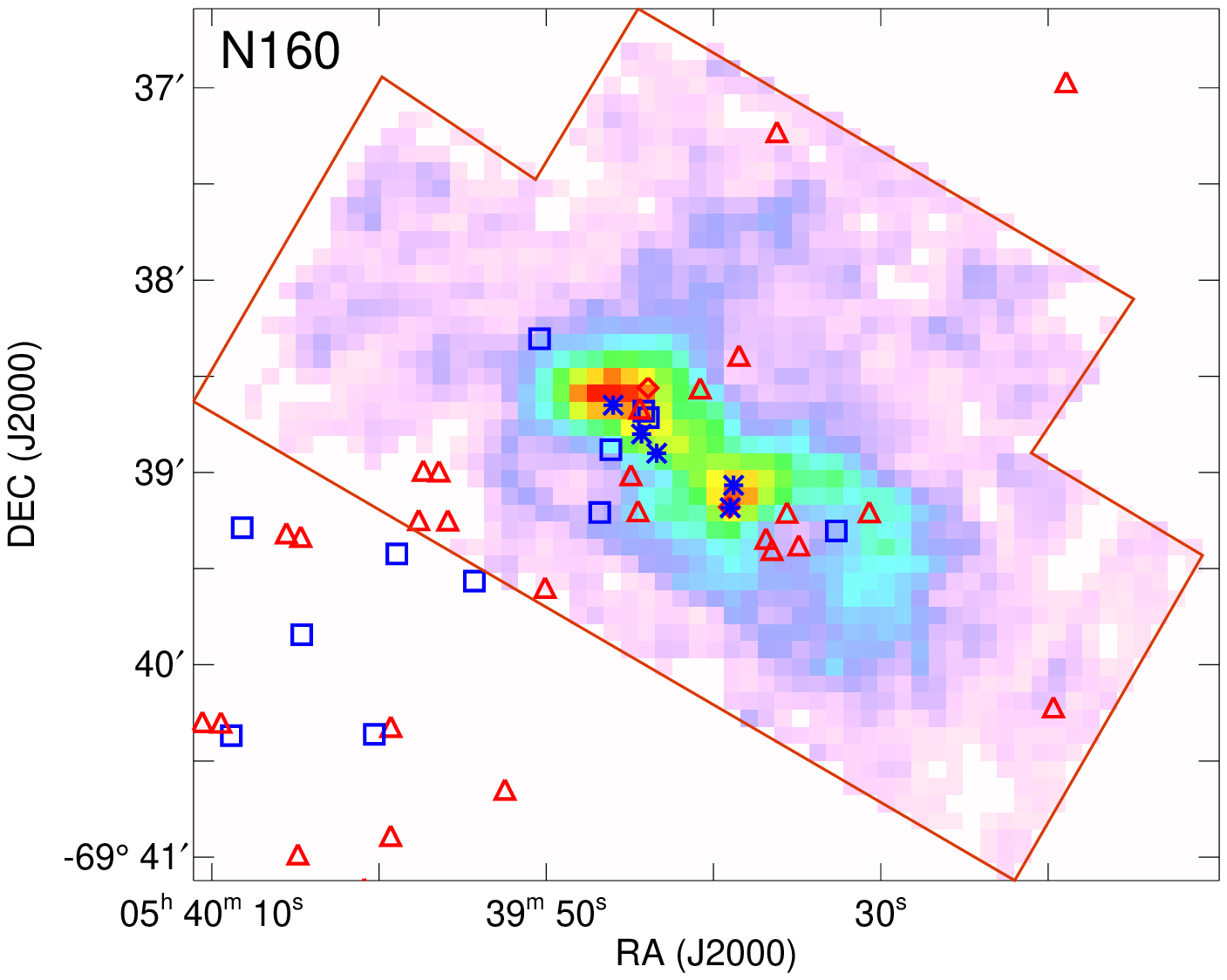}
\caption{The \cii\ integrated intensity map of N160 (color scale is the same in Fig.~\ref{figure:cii_co_integ}). Open blue squares show the position of stars with a spectral type of O7 or earlier, while red triangles indicate stars with spectral type O7 to B1 \citep{Farina2009}. The blue asterisks are compact \hii\ regions \citep{Indebetouw2004}, and the red diamond marks the position of an H$_2$O master \citep{Lazendic2002}.}
\label{figure:N160_cii_stars}
\end{figure}

The peak of the \cii\ integrated emission is located at N160A, which is the brightest region in the radio continuum and H$\alpha$ emission \citep{HeydariMalayeriTestor1986,MartinHernandez2005}. N160A consist of subregions A1, A2, and another cluster in between \citep{HeydariMalayeri2002}. Both CO(4-3) and \cii\ emission show an elongated shape tracing those subregions (Fig.~\ref{figure:cii_co_integ}). In radio continuum emission, another \hii\ region is identified as N160D southwest of N160A, with the ionized gas having a shell-like structure with a WR star as well as several OB stars at the center \citep{HeydariMalayeriTestor1986,MartinHernandez2005,Farina2009}. Two compact \hii\ regions are located at the eastern edge of the shell, where our CO(4-3) has the strongest peak and the \cii\ has a peak as well. It corresponds to N160-4 in \citet{Johansson1998}. The shell structure can be traced in the \cii\ map in the bottom panels of Fig.~\ref{figure:N160_channelmap} and Fig.~\ref{figure:N160_cii_stars}, where positions of OB stars and compact \hii\ regions are marked. OB stars in \citet{Farina2009} are distributed not only in N160A and D, but continue towards the southeast of the observed region toward another radio continuum peak \citep{Mills1984}. The \cii\ emission extends northwards from N160A, which can be also seen in the \cii\ map by \citet{Israel1996}. There is one late O-star to the north of this extended distribution \citep{Farina2009}. To the east of N160A, we detected two clouds in CO(4-3), where the \cii\ emission is very weak \citep[N160-1 and N160-3 in][]{Johansson1998}. Fig.~\ref{figure:integmap_N160} shows that these clouds are bright in \ci\ \transl\ and low-J CO, while CO(6-5) is not detected, indicating a cold and quiescent environment. However they must have exciting UV sources because the \cii\ emission is weakly detected and the IRAC 8\um\ map shows diffuse structure as well as cores.

The \cii\ and CO(4-3) channel maps (Fig.~\ref{figure:N160_channelmap}) show an overall good correlation, with a global velocity gradient from the northeast to the southwest. The shell structure in N160D also follows the gradient, but the \cii\ emission at the eastern edge of the shell extends in velocity up to 244\,\kms\ (it is also seen as a \cii\ wing in the spectrum of position 4 in Fig.~\ref{figure:integmap_N160}). The \cii\ emission to the north of N160A appears at a velocity of 235--236\,\kms, and it connects to clouds at N160A, while the CO(4-3) emission is detected at 233-235\,\kms\ and is somewhat isolated from N160A. 

\subsubsection{N158}

The observed region is N158C. The southeast and the most northwestern CO peaks (Fig.~\ref{figure:cii_co_integ}) correspond to N158-2 and N158-1 respectively in \citet{Johansson1998}. The channel maps (Fig.~\ref{figure:NGC2074_channelmap}) indicate three different cloud components: north-south structures at right ascensions of 05:39:17 and 05:39:07, and an east-west structure with a declination of -69:30:00. The \hii\ region traced by H$\alpha$ is also elongated in the north-south direction \citep[slightly inclined from northeast to southwest;][]{Fleener2010}, and is located between the two north-south structures seen in \cii\ emission. The eastern CO(4-3) peak corresponds to the prominent dust lane in the H$\alpha$ image \citep{Fleener2010}. The strongest peak of the CO(4-3) and \cii\ towards the southeast is shifted, and two compact \hii\ regions identified by \citet{Indebetouw2004} are located between the CO(4-3) peak and the \cii\ peak. The ionized gas extends also in an east-west direction at the northern edge of our observed region. This clearly shows that the \cii\ emission is distributed dominantly outside the ionization front, together with the CO emission. On the other hand, the \cii\ distribution is more or less continuous spatially and spectrally, while the CO(4-3) emission shows more clearly separated clouds.

\subsubsection{30~Dor}

\begin{figure}
\centering
\includegraphics[width=0.95\hsize]{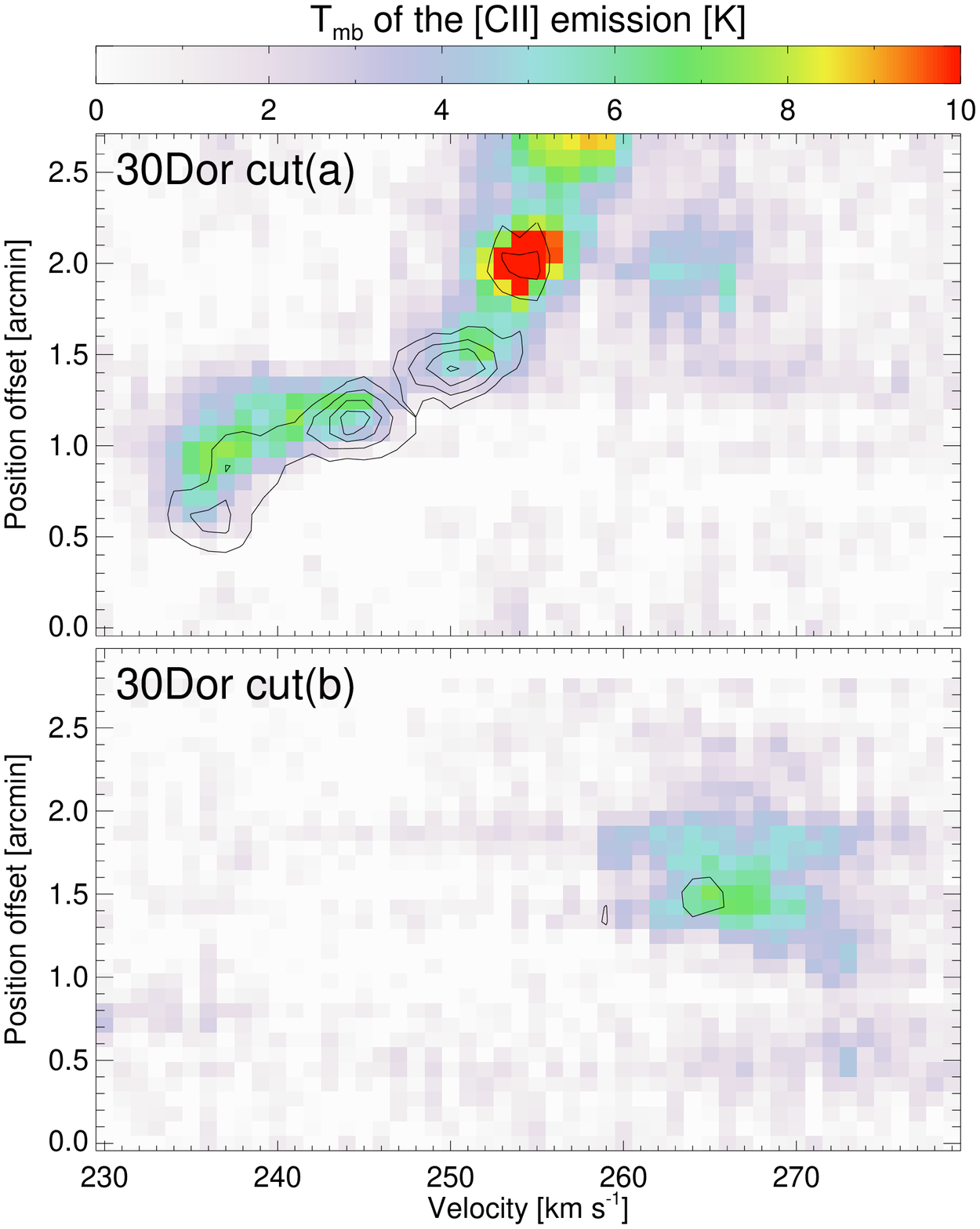}
\caption{Position-velocity diagrams of \cii\ (color) and CO(4-3) (contour) emission along the cuts shown in Fig.~\ref{figure:cii_co_integ}. The contour spacing is 1~K.}
\label{figure:pvdiagram_30Dor}
\end{figure}

There are two main clouds of CO and \cii\ emission in our observed regions (Fig.~\ref{figure:cii_co_integ}): northeast (30~Dor-10) and southwest of R136 \citep[30~Dor-12,][]{Johansson1998}. The brightest part of 30~Dor-10 is closer to R136 with a relatively sharp boundary on the side facing R136, while both CO(4-3) and \cii\ are extended towards the northeast. \citet{Indebetouw2013} identify about one hundred clumps with the ALMA observations at the central part of 30~Dor-10. Their clumps 52 and 72 are the main contributors to our brightest peak and the neighboring peak towards the east.

In the channel maps, the peak velocities of 30~Dor-10 and 30~Dor-12 appear similar, while the emission at $>255$~\kms\ appears only in 30~Dor-10 (Fig.~\ref{figure:30Dor_channelmap}). In 30~Dor-10, we see four different velocity streams, connecting at the CO(4-3) and \cii\ emission peaks: northeast-southwest direction at $\sim 244$\,\kms, north-south direction at $\sim 247$\,\kms, east-west direction at $\sim 256$\,\kms, and south-north direction at $\sim 260$\,\kms. The CO(4-3) and \cii\ emission show basically similar structures, except for the following few differences. At $\sim 243$\,\kms, the \cii\ emission has an additional northwest-southeast stream, which is not seen in CO(4-3). The CO(4-3) has a peak northeast of the brightest peak (see the channel map at $\sim 251$\,\kms), which does not have a corresponding \cii\ peak; \cii\ instead shows a hole there (Fig.~\ref{figure:cii_co_integ}). At $\sim 260$\,\kms, the CO(4-3) shows a cloud south of 30~Dor-10, while the \cii\ emission extends continuously towards the south and it extends further in the velocity domain up to $\sim 266$\,\kms. 

Position-velocity diagrams along two cuts (Fig.~\ref{figure:cii_co_integ}) are shown in Fig.~\ref{figure:pvdiagram_30Dor}. Cut (a) goes through the so-called `stapler nebula', where \citet{Kalari2018} found three molecular clouds with CO(2-1). These authors suggested that we see the tails of pillar-like structures whose ionized heads are pointing toward R136, and that the large observed velocity dispersion can be explained by the velocity difference between the head and tail of a photoevaporating cloud. In our observations, the \cii\ velocity is shifted compared to the CO(4-3) velocity (Fig.~\ref{figure:pvdiagram_30Dor}a, see also position 10 and 11 in Fig.~\ref{figure:selected_spectra_30Dor}), which is consistent with their picture since the contribution from the head should be more dominant in \cii\ emission than the CO(4-3) emission. Cut (b) goes through two weak \cii\ blobs in the integrated intensity map, which appear as a velocity component at $\sim 265$\,\kms. Positions 10 and 11 in Fig.~\ref{figure:selected_spectra_30Dor} also show this velocity component at 30\arcsec\ resolution. The p-v diagrams show that there is no clear connection between different velocity components among these two cuts. As discussed in Sect.~\ref{subsec:hi_cii}, the \hi\ spectra show an absorption feature at $\sim 265$\,\kms. Unfortunately the spatial resolution of the \hi\ observations is insufficient to determine whether the absorption is spatially associated with the \cii\ blobs.

\subsection{Column densities}\label{subsec:columndensity}

\begin{figure}
\centering
\includegraphics[width=\hsize]{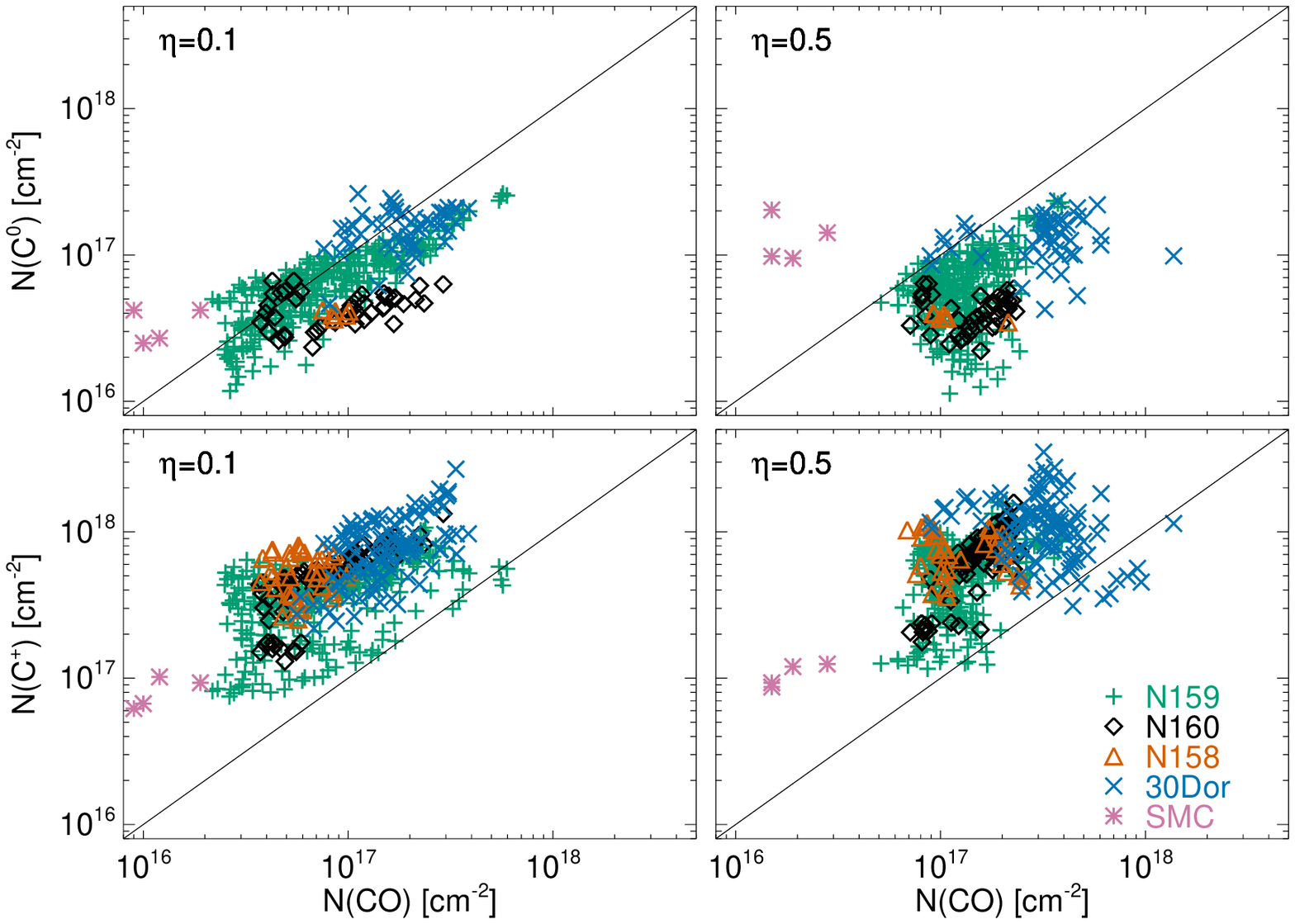}
\caption{The column density $N$(C) (upper panels) or $N$(\cplus) (lower panels) against $N$(CO) for the beam filling factor of $0.1$ (left panels) and $0.5$ (right panels). The SMC data are taken from \citet{Requena-Torres2016}. The black line indicates that the two column densities are the same.}
\label{figure:columndensity}
\end{figure}

We derive column densities of \cplus, C, and CO at each position in our four regions. The detailed derivation is presented in Paper I. Here we describe it briefly. We assume local thermodynamic equilibrium (LTE) for each species, and a uniform excitation temperature (\tex) for different velocity bins at each position. Assuming that CO and \thco\ have the same \tex, beam (area) filling factor ($\eta$), and an isotope ratio $^{12}$C/$^{13}$C of 49 \citep{Wang2009}, we derive the optical depth of \thco(3-2) ($\tau_{13}$) from the intensity ratio CO(3-2)/\thco(3-2). With the absolute intensity of \thco(3-2) and $\tau_{12}$ we can estimate \tex, and then the CO column density for a given beam filling factor ($\eta$). To estimate the \czero\ column density, we defined \tex\ of the \ci\ emission as follows. In N159 and 30~Dor, we derived \tex\ by the ratio of the two \ci\ intensities where both lines are detected and took its median. We used this median \tex\ as a constant \tex\ over the whole map in each region. For the other two regions, where we do not have the \ci\ \transu\ line, we use the \tex\ obtained in N159. To estimate the \cplus\ column density we have to assume \tex\ of \cii. Here we also used a constant \tex\ over the map in each region, which is determined so that it is above the \tex\ lower limit at any position of the map (i.e. for a $\tau\gg 1$ case).

Figure~\ref{figure:columndensity} shows the relation between the column density $N$(CO), $N$(C), and $N$(\cplus) integrated over the whole velocity range. Here we present beam averaged column densities, i.e. the column densities presented in the Paper I multiplied by the beam filling factor. In most of the positions $N$(\cplus) is in the range of 1--10 times $N$(CO), and $N$(\czero) is $0.1$--1 times $N$(CO). There is some hint that N159 has a lower $N$(\cplus)/$N$(CO) ratio compared to other regions, but the trend is weak. N160 appears to have two different populations; for the two clouds in the northeast, where the \cii\ emission is very weak and the \ci\ emission is strong, there are several positions with lower $N$(\cplus) and higher $N$(C). This may be due to a lower radiation field in this region. Toward the main cloud of N160 and in N158, the $N$(C)/$N$(CO) ratio is smaller than in N159 and 30~Dor, while the $N$(\cplus)/$N$(CO) ratio is in a similar range as N159 and 30~Dor. This result is not very sensitive to a different \tex\ assumption for the \ci\ emission in N160 and in N158, for example using \tex\ of CO, since it is above the energy of the \ci\transl\ transition (24~K) at most positions. It is most likely not a radiation field effect since the representative radiation field strength is $\chi=60$--220 in N159 and N160 \citep{Israel1996,Pineda2008,Lee2016} and $10^3$--$10^4$ in 30~Dor \citep{Chevance2016,Pineda2012}. It is also inconsistent with a scenario that a more evolved region has less CO, since 30~Dor is the most evolved region and N159 is likely to be one of the youngest regions in our analysis. A comparison with the SMC data is also puzzling. In Fig.~\ref{figure:columndensity} the data of N66 (plume and ridge), N25, and N88 by \citet{Requena-Torres2016} are plotted. Their $N$(\cplus)/$N$(CO) ratio falls within the range of values obtained for LMC regions, while they find a larger $N$(C)/$N$(CO) ratio. As mentioned in Sect.~\ref{sec:intro}, $N$(\cplus)/$N$(CO) is expected to be large in a lower metallicity environment, while $N$(C)/$N$(CO) is expected to be less dependent on metallicity. We do not see this effect. For N25 and N88, where the radiation field is $\chi=20$--80 \citep{Israel2011}, the high $N$(C)/$N$(CO) may be due to the low radiation field. However the data point with the highest $N$(C)/$N$(CO) comes from N66, where the radiation field is estimated as 80--570 \citep{Israel2011}, which is not very different to the LMC regions. The result here indicates that $N$(\cplus)/$N$(CO) and $N$(C)/$N$(CO) in resolved regions are affected significantly by local conditions and are not a simple function of metallicity.

\subsection{PDR modeling} \label{subsec:pdrmodeling}
\subsubsection{Assumptions and fitting procedure}

We fit the KOSMA-$\tau$ PDR model (Sect.~\ref{sec:PDRintro}) to the absolute intensities of the observed line emission and continuum SED at each position of the 30\arcsec\ resolution maps. We assume uncertainties of emission lines to be 10\% for the APEX and continuum observations \citep{Meixner2013} and 15\% for GREAT and PACS observations \citep[PACS Spectroscopy performance and calibration document, ][]{Cormier2015}. In case the baseline noise is bigger than these uncertainties, we take the baseline noise as an uncertainty. We use the clump ensemble model with a lower mass limit of $10^{-3}M_\sun$ and an upper mass limit of $10^3 M_\sun$. The upper mass limit is consistent with the ALMA result for the LMC \citep{Indebetouw2013}. For the dust model, \citet{WD2001} provide 6 different models for the LMC; two types of environment characterized by different extinction curves (LMC average and LMC 2) with 3 different abundances of very small grains each. LMC 2 is a supergiant shell southeast of 30~Dor that partially overlaps our observed regions \citep{Misselt1999}. We selected the LMC average environment with the abundance of very small grains of $10^5b_c=2.0$ (model 28; numbering by the rows of \citet{WD2001} Tables 1 and 3), which is consistent with the polycyclic aromatic hydrocarbon (PAH) abundance quoted by \citet{Galametz2013}. We used a sum of the Gaussians defined by the CO velocity profiles (Sect.~\ref{subsec:result_lineprofile}) as input intensities except for the \oi\ emission, where we use the integrated intensity from the velocity-unresolved PACS data. In this way, the derived physical properties are the average of different gas components that are detected in CO(3-2). The effects of different model assumptions and input intensities are discussed in Sect.~\ref{subsec:PDR_compare_assumptions}.

We select the emission to be used in the model fit as follows. We consider two cases, where we fit only the line emission or both the line and continuum emission. We consider five continuum data (Sect.~\ref{subsec:obs_continuum}) as independent data points and treat them as if they were individual line data in the $\chi^2$ fit. For the selection of line emission to be included in the fit, we have four cases: using all lines, using all lines except for the \oi\ lines, using all lines except for the \ci\ lines, and using only optically thin lines (\thco, \ci, and \oi\ 145\um). For every selection, we consider the map positions where at least four lines are detected in order to derive three model parameters ($n$, $m$, and $\chi$, see Sect.~\ref{sec:PDRintro} for their definitions). The exception is the case for the optically thin case fitted together with the continuum. In this case, we accept positions with three lines since otherwise most of the map positions do not fulfill the criteria, and it gives still a reasonable fit because the number of the total data points is more than four together with the continuum.

The original model grid covers $\log(n)=2$ to 7 in density, $\log(m)=-3$ to 3 in total mass, and $\log(\chi)=0$ to 6 in the UV field strength, all with a logarithmic step of 1. We first extend the mass grid to $\log(m)=6$. In a clumpy model, no extrapolation is needed to extend the mass range because it is just a scaling of the total intensity. We interpolate the model to a $0.1$ step in the logarithmic scale of $n$, $m$, and $\chi$. Then we calculate the reduced $\chi$-square ($\chi^2$) of all the grid points and take the grid point with the minimum $\chi^2$ as a solution. In this way, we can avoid to fall into a local minimum. 

\subsubsection{Result}\label{subsec:PDRfit_result}

\begin{figure*}
\centering
\includegraphics[width=0.9\hsize]{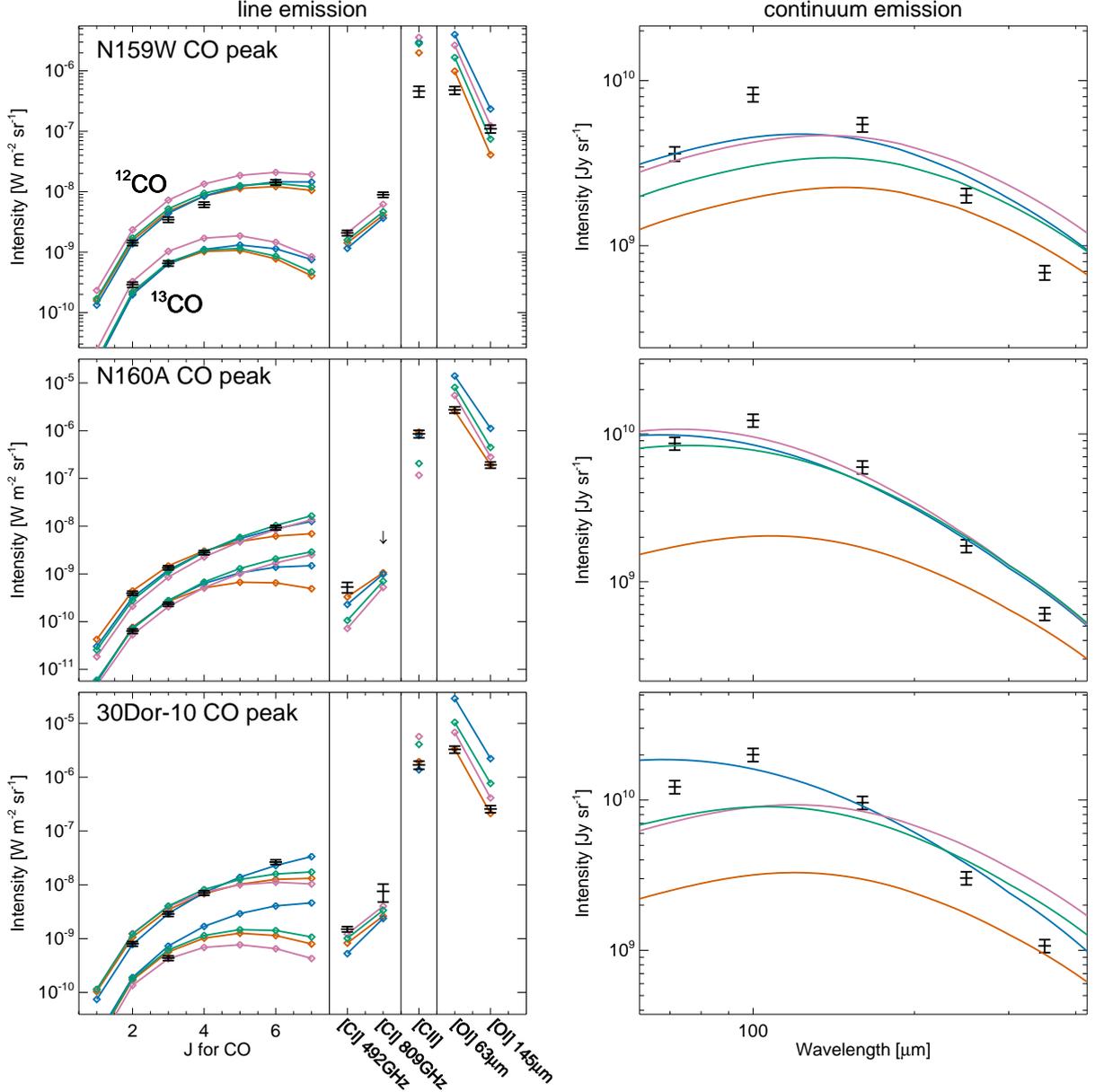}
\caption{Results of fitting the line and continuum emission at the CO peaks in N159~W (upper two panels), in N160A (middle panels), and in 30~Dor-10 (lower panels). Black points are the observed data and colored points represent the models. The left panels are the line SED of CO(2-1) to (6-5), \ci\transl, \ci\transu, \cii, \oi\ 63\um, and \oi\ 145\um. $^{12}$CO and \thco\ are plotted together. For N159, CO(2-1) and \thco(2-1) are not used in the fit because it is a pointed observation and the beam sizes are slightly different. The red model uses all line emissions but no continuum in the fit. The blue model excludes \oi\ from the fit but includes the continuum emission. The green model fits all the lines and the continuum emission. The purple model fits only optically thin lines and the continuum. Input line intensities are the sum of Gaussians except for the \oi\ lines, where the integrated intensities are used (see Table~\ref{table:PDR_input}).}
\label{figure:PDRmodel_result_SED}
\end{figure*}

\begin{table*}
\caption{Input intensities for the fit example in Fig.~\ref{figure:PDRmodel_result_SED}. They are the sum of Gaussians except for the \oi\ lines, where the integrated intensities are used.}
\label{table:PDR_input}
\centering
\begin{tabular}{cccc}
\hline\hline
& N159~W CO peak$^{\textrm{a}}$ & N160~A CO peak$^{\textrm{b}}$ & 30~Dor-10 CO peak$^{\textrm{c}}$ \\
\hline
Lines [W\,m$^{-2}$\,sr$^{-1}$] &&&\\
\hline
CO(2-1)         & $(1.4 \pm 0.1)\times 10^{-9}$ $^{\textrm{d}}$   & $(4.0 \pm 0.4)\times 10^{-10}$  & $(8.0 \pm 0.8)\times 10^{-10}$ \\
CO(3-2)         & $(3.5 \pm 0.3)\times 10^{-9}$   & $(1.4 \pm 0.1)\times 10^{-9}$   & $(2.9 \pm 0.3)\times 10^{-9}$ \\
CO(4-3)         & $(6.1 \pm 0.6)\times 10^{-9}$   & $(2.8 \pm 0.3)\times 10^{-9}$   & $(7.1 \pm 0.7)\times 10^{-9}$ \\
CO(6-5)         & $(1.4 \pm 0.1)\times 10^{-8}$   & $(9.3 \pm 1.0)\times 10^{-9}$   & $(2.7 \pm 0.3)\times 10^{-8}$ \\
\thco(2-1)      & $(2.9 \pm 0.3)\times 10^{-10}$ $^{\textrm{d}}$  & $(6.4 \pm 0.7)\times 10^{-11}$  & $-$                           \\
\thco(3-2)      & $(6.5 \pm 0.7)\times 10^{-10}$  & $(2.3 \pm 0.2)\times 10^{-10}$  & $(4.4 \pm 0.4)\times 10^{-10}$\\
\ci\ \transl    & $(2.1 \pm 0.2)\times 10^{-9}$   & $(5.3 \pm 1.3)\times 10^{-10}$  & $(1.5 \pm 0.2)\times 10^{-9}$ \\
\ci\ \transu    & $(8.9 \pm 0.9)\times 10^{-9}$   & $ <8.0\times 10^{-9}$           & $(7.6 \pm 2.8)\times 10^{-9}$ \\
\cii            & $(4.6 \pm 0.9)\times 10^{-7}$   & $(8.6 \pm 1.6)\times 10^{-7}$   & $(1.7 \pm 0.3)\times 10^{-6}$ \\
\oi\ 63\um      & $(4.8 \pm 0.7)\times 10^{-7}$   & $(2.8 \pm 0.4)\times 10^{-6}$   & $(3.3 \pm 0.5)\times 10^{-6}$ \\
\oi \ 145\um    & $(1.1 \pm 0.2)\times 10^{-7}$   & $(1.9 \pm 0.3)\times 10^{-7}$   & $(2.6 \pm 0.4)\times 10^{-7}$ \\
\hline
Continuum [Jy\,sr$^{-1}$]&&&\\
\hline
70\um   & $(3.6 \pm 0.4)\times 10^{9}$  & $(8.6 \pm 0.9)\times 10^{9}$   & $(1.2 \pm 0.1)\times 10^{10}$\\
100\um  & $(8.3 \pm 0.8)\times 10^{9}$  & $(1.2 \pm 0.1)\times 10^{10}$  & $(2.0 \pm 0.2)\times 10^{10}$\\
160\um  & $(5.4 \pm 0.5)\times 10^{9}$  & $(6.0 \pm 0.6)\times 10^{9}$   & $(9.6 \pm 1.0)\times 10^{9}$ \\
250\um  & $(2.0 \pm 0.2)\times 10^{9}$  & $(1.7 \pm 0.2)\times 10^{9}$   & $(3.0 \pm 0.3)\times 10^{9}$ \\
350\um  & $(6.9 \pm 0.7)\times 10^{8}$  & $(6.1 \pm 0.6)\times 10^{8}$   & $(1.1 \pm 0.1)\times 10^{9}$ \\
\hline\hline
\end{tabular}
\begin{list}{}{\setlength{\itemsep}{0ex}}
\item[$^\textrm{a}$] 05:39:36.8, -69:45:28.9 \qquad $^\textrm{b}$ 05:39:45.9, -69:38:34.6 \qquad $^\textrm{c}$ 05:38:48.8, -69:04:42.1 (J2000)
\item[$^\textrm{d}$] Not used in the fit because of a slightly different beam size.
\end{list}
\end{table*}

\begin{table*}
\caption{Derived physical properties in the fit in Fig.~\ref{figure:PDRmodel_result_SED}. The derived physical properties of all positions for the green model are shown in Figs.~\ref{figure:nmuv_N159}--\ref{figure:nmuv_30Dor}.}
\label{table:PDRproperties}
\centering
\begin{tabular}{lllllr}
\hline\hline
Position &  model in Fig.~\ref{figure:PDRmodel_result_SED} & $\log(n)$ & $\log(m)$ & $\log(UV)$ & Area filling factor \\
\hline
N159~W &       red (w/o cont.)    & $ 3.5  $ & $ 4.9  $ & $ 1.3$ & $  42.5$ \\
   &            blue (w/o \oi)    & $ 3.7  $ & $ 4.9  $ & $ 1.8$ & $  31.3$\\
   &          purple (thin)    & $ 3.6  $ & $ 5.1  $ & $ 1.5$ & $  57.8$\\
   &           green (all)    & $ 3.5  $ & $ 5.0  $ & $ 1.4$ & $  53.6$\\
\hline
  N160~A &       red (w/o cont.)    & $ 3.9  $ & $ 4.3  $ & $ 2.2$ & $   5.8$\\
   &            blue (w/o \oi)    & $ 4.7  $ & $ 4.2  $ & $ 3.7$ & $   1.3$\\
   &          purple (thin)    & $ 6.2  $ & $ 4.1  $ & $ 4.8$ & $   0.1$\\
   &           green (all)    & $ 5.9  $ & $ 4.2  $ & $ 4.4$ & $   0.2$\\
\hline
 30~Dor-10 &       red (w/o cont.)    & $ 3.8  $ & $ 4.7  $ & $ 1.9$ & $  16.9$\\
  &            blue (w/o \oi)    & $ 4.9  $ & $ 4.5  $ & $ 3.8$ & $   2.0$\\
  &          purple (thin)    & $ 3.5  $ & $ 5.1  $ & $ 1.8$ & $  67.4$\\
  &           green (all)    & $ 3.8  $ & $ 4.9  $ & $ 2.2$ & $  26.8$\\
\hline\hline
\end{tabular}
\end{table*}

\begin{figure}
\centering
\includegraphics[width=0.9\hsize]{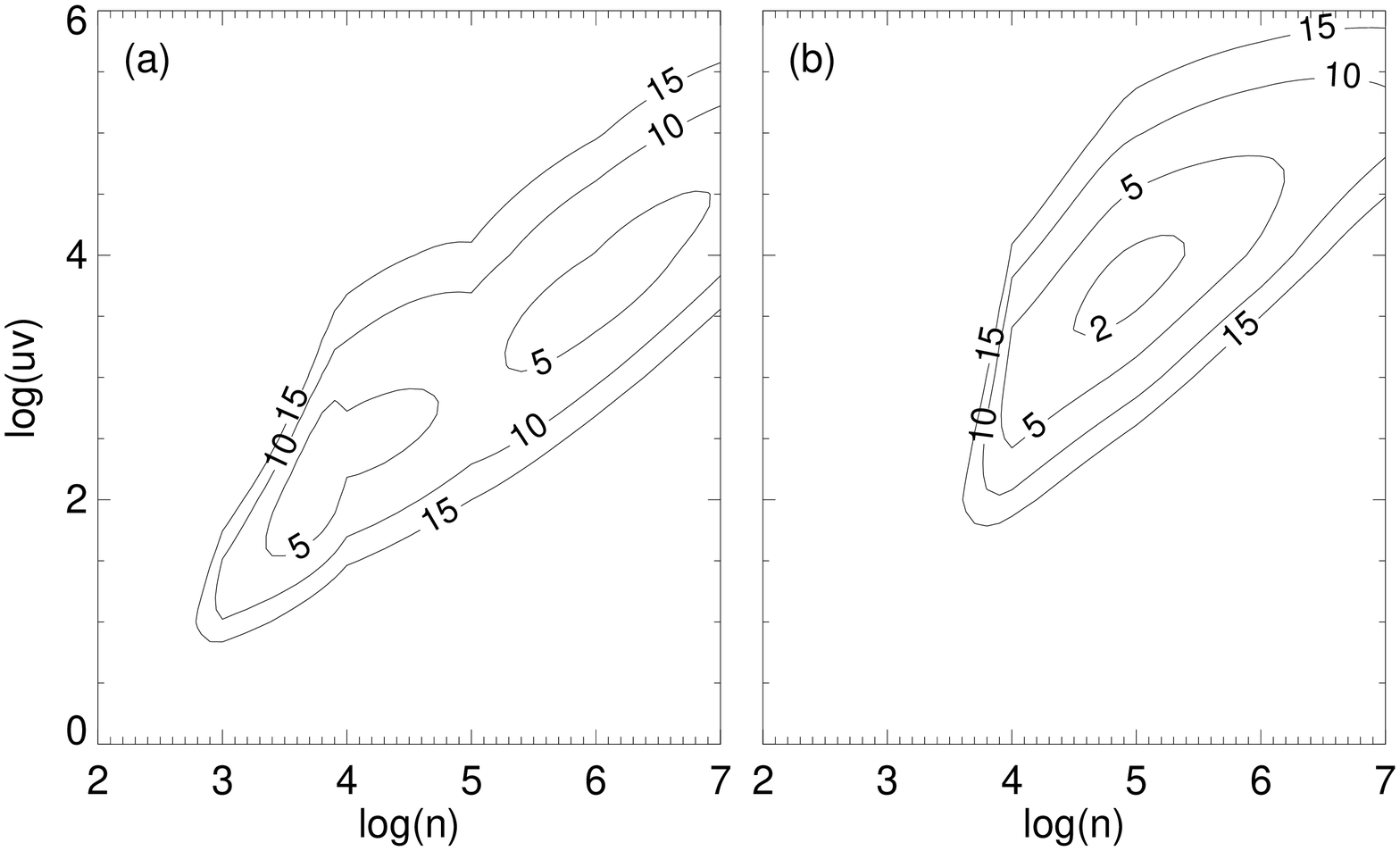}
\caption{The contours of $\chi^2$ at the 30~Dor-10 CO peak position for (a) the green model (fit all line and continuum emission) and (b) the blue model (excluding \oi) in the lower panel of Fig.~\ref{figure:PDRmodel_result_SED}.}
\label{figure:chi2contour}
\end{figure}

We calculate the best-fitting model at each position of the 30\arcsec\ resolution maps of the four star-forming regions. Figure~\ref{figure:PDRmodel_result_SED} and Table~\ref{table:PDRproperties} show an example fit (the coordinates and input intensities are listed in Table~\ref{table:PDR_input}). In most of the observed positions within the four star-forming regions (except for the CO peak at N159~W), we find the following overall trend. When fitting with only the line emission (red lines in Fig.~\ref{figure:PDRmodel_result_SED}), the lines are relatively well fit, with some cases the model underestimating the \ci\ emission. However the continuum emission is clearly underestimated especially at shorter wavelengths, indicating that the dust temperature in the best-fitting model is too low. It is consistent with the fact that the CO-ladder starts to be flat already around $J=5$ or 6, resulting in an underestimate of CO(6-5). When we exclude the \oi\ lines and include the continuum in the fit (blue lines), the estimated UV field is much stronger. The continuum SED as well as the rising trend from CO(4-3) to CO(6-5) are relatively well reproduced, but the \oi\ lines are heavily overestimated and the \ci\ emission is even more severely underestimated than in the first model. When we fit everything (green lines), the fit of \oi\ improves with compromising other parts. In most cases, fitting only optically thin lines (purple lines) or fitting without \ci\ lines (not shown) does not change the result much, or makes the fit worse.

The fundamental problem for fitting the line emission is that the CO and \ci\ have two solutions at low $n$ -- low $\chi$ and at high $n$ -- high $\chi$ range for their given intensities (Fig.~\ref{figure:chi2contour}). The \cii\ intensity contours are also aligned from low $n$ -- low $\chi$ to high $n$ -- high $\chi$. In principle the \oi\ emission lines have a different dependency on $n$ and $\chi$, and hence including the \oi\ emission in the fit should provide a good constraint.  However, for most positions that we study here, the solutions of $n$ and $\chi$ indicated by the \oi\ emission do not overlap with those that other emissions suggest (Fig.~\ref{figure:chi2contour}a). This is why it looks like that the fit excluding \oi\ (blue lines in Fig.~\ref{figure:PDRmodel_result_SED}) look the most reasonable for lines other than the \oi. However, in this case the low $n$ -- low $\chi$ solution and high $n$ -- high $\chi$ solution are degenerate (Fig.~\ref{figure:chi2contour}b), and very different physical conditions are obtained for spatially adjacent pixels. It is also not justified to simply ignore the \oi\ emission, because it is difficult to explain the overestimate of both \oi\ lines. For these reasons, we consider that fitting with all the lines and the continuum (green model) gives the most stable solution for representative physical properties in the sense that it excludes physical properties that would yield completely different line or continuum intensities compared to what is observed.

Since we fit the absolute intensities, the model results include the area filling factor as well (Table~\ref{table:PDRproperties}). However it is sensitive to the derived density and it ranges from the order of 0.01 to 100 in four regions even with the same model assumption, which is again an effect of the degeneracy of the low $n$ -- low $\chi$ solution and high $n$ -- high $\chi$ solution.

The CO peak of N159~W shows a different result (Fig.~\ref{figure:PDRmodel_result_SED} top panel). There all models fail to reproduce the \cii\ emission and the \oi\ 63\um\ to 145\um\ ratio \citep{Lee2016}. When we fit only the optically thin lines, the model overestimates all the CO, \cii, and \oi\ 63\um\ line intensities, which is qualitatively consistent with an argument that it is an optical depth effect among clumps. Quantitatively, we assume that the $N_\textrm{cloud}$ clouds are aligned along the line of sight, with each having a line intensity of $I_\textrm{line}$ and the optical depth of $\tau_\textrm{line}$. The model predicts the line intensity of $I_\textrm{line}\times N_\textrm{cloud}$, while the observed line intensity is $I_\textrm{line}\times \sum_{i=0}^{N_\textrm{cloud}-1} e^{-i\tau_\textrm{line}}$ \citep{Okada2003}. To obtain one order of magnitude difference between the model and observations as seen in Fig.~\ref{figure:PDRmodel_result_SED}, we need $N_\textrm{cloud}=10$ for a $\tau_\textrm{line}=\infty$ case and $N_\textrm{cloud}=15$ for a $\tau_\textrm{line}=1$ case.  The fitted clump ensemble model when using only optically thin lines gives an area filling factor of $>20$ for any combination of model assumptions, which supports the idea of heavily overlapping clumps, although the filling factor is sensitive to the derived density as mentioned above.

\begin{figure}
\centering
\includegraphics[width=0.9\hsize]{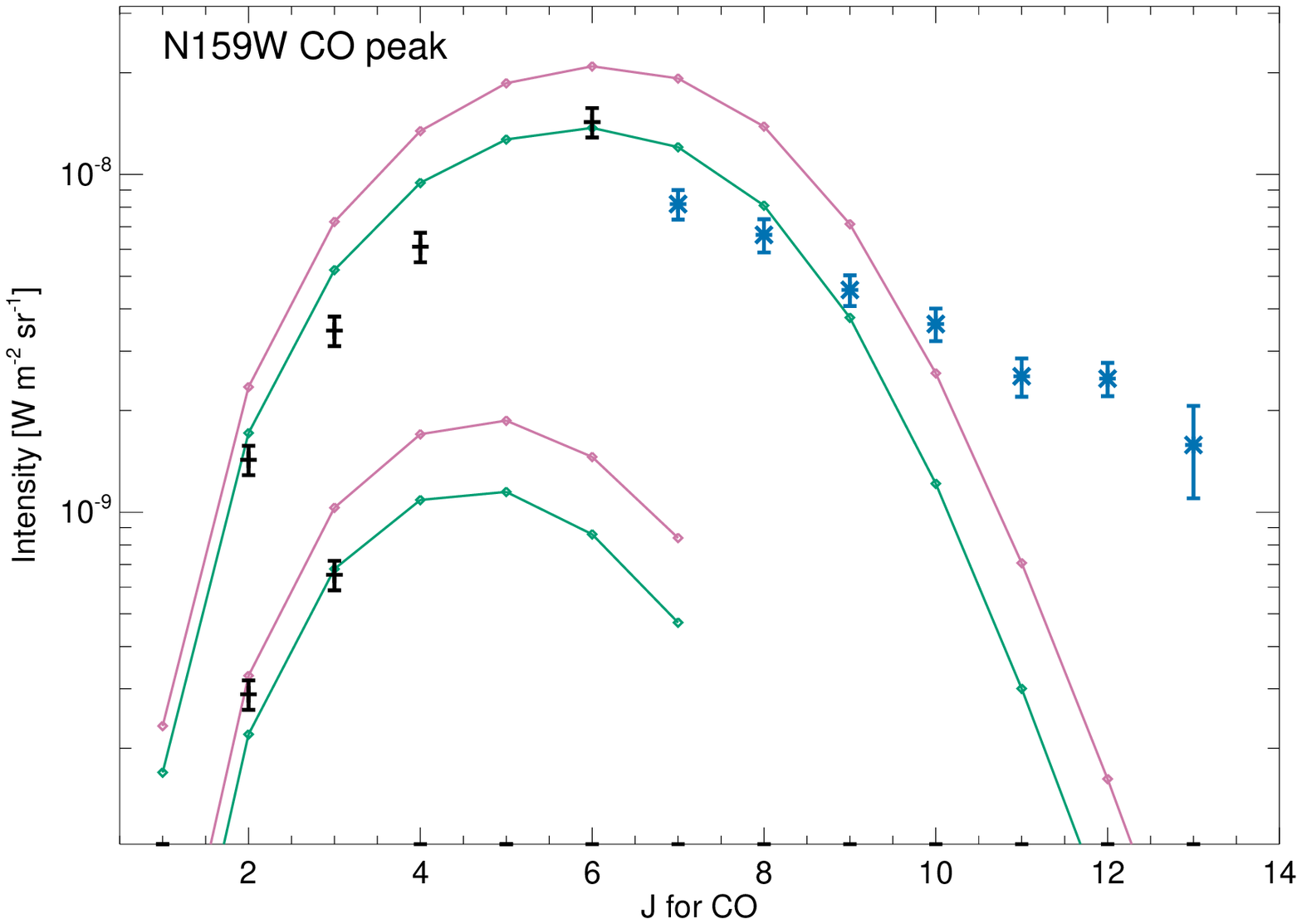}
\caption{The CO ladder including the high-J CO data by \citet{Lee2016} at the CO peak in N159~W. Black data points and the model lines are the same as in the upper panel of Fig.~\ref{figure:PDRmodel_result_SED}. Blue data points are Herschel/SPIRE observations at 42\arcsec\ resolution \citep{Lee2016}.}
\label{figure:PDRmodel_result_SED_N159_CO}
\end{figure}

\citet{Lee2016} modeled \oi\ 145\um, \ci\transu, \cii\ and FIR with the Meudon PDR code and derived P/k=$10^6$ K\cc\ and $G_0=100$. Their model underestimates the CO emission, especially high-J CO, and they concluded that CO is heated by low-velocity shocks. When we use a non-clumpy model of the KOSMA-$\tau$ model and exclude the CO emissions from the fit, we reproduced qualitatively their results. In Fig.~\ref{figure:PDRmodel_result_SED_N159_CO}, we compare our clumpy PDR model with the high-J CO intensities observed with the Spectral and Photometric Imaging REceiver (SPIRE) on Herschel provided by \citet{Lee2016}. The gap between CO(6-5) and CO(7-6) is partially because the SPIRE data is convolved to the 42\arcsec\ beam size. The original beam size of CO(7-6) is 33\arcsec\ \citep{Lee2016}, close to our 30\arcsec, and its value is 15\% higher. The figure shows that the clumpy model provides a reasonable fit to the CO up to around $J=10$. The very shallow slope of the higher-J CO-ladder is consistent with the shock contribution, while low- and mid-J CO can be also explained by a PDR model when we introduce clumpiness. In fact the very similar line profiles of the CO and \ci\ is consistent with the picture that they are both emitted by a PDR gas, although the proposed shock in \citet{Lee2016} has a velocity of only $\sim 10$\,\kms, which may be hard to see in an emission line with a width of $\sim 8$\,\kms\ (Paper I). 

For the continuum SED, the model prediction of the SED is typically broader than the observed SED, independent of which model we adopt. It is typically characterized by a strong drop of 70\um\ continuum (e.g. N159 in Fig.~\ref{figure:PDRmodel_result_SED}). When we choose a dust model with fewer small grains (see Sect.~\ref{subsec:PDR_compare_assumptions}), the model SED is slightly narrower because of weaker 70\um\ emission, but the fit to the observed SED is not significantly better. This is consistent with the model SED by \citet{Bron2014} where the contribution of PAHs to the 70\um\ emission is a few percent for $\chi=1$--$1000$. At the column density peak, it is possible that the optically thin assumption for the 70\um\ continuum emission does not fully hold; we estimated the total hydrogen column density from $N$(CO), $N$(\czero), and $N$(\cplus) in Sect.~\ref{subsec:columndensity} with the carbon abundance of $7.9\times 10^{-5}$ \citep{Garnett1999} and converted it to a dust opacity using the extinction cross section \citep{WD2001}, which is at maximum 0.15 at the peak of 30~Dor-10 in case of $\eta=0.1$. This adds $\sim10$\% uncertainty to a few positions around the CO peak, but does not have a significant impact on the continuum fit, although it is not excluded that the beam filling factor is less than 0.1 and the 70\um\ continuum is marginally optically thick. \citet{Chevance2016} suggest that the FIR continuum has a significant contribution from the ionized gas outside of the molecular clouds in 30~Dor by analyzing its correlation with PAH and \oiii\ emission. This may cause an overestimate of the UV field in the fit with the continuum SED, but it does not explain the observed narrow peak in the continuum SED and its contribution should be higher outside of molecular clouds. 

\begin{figure*}
\centering
\includegraphics[bb=40 0 404 355,width=0.26\hsize,clip]{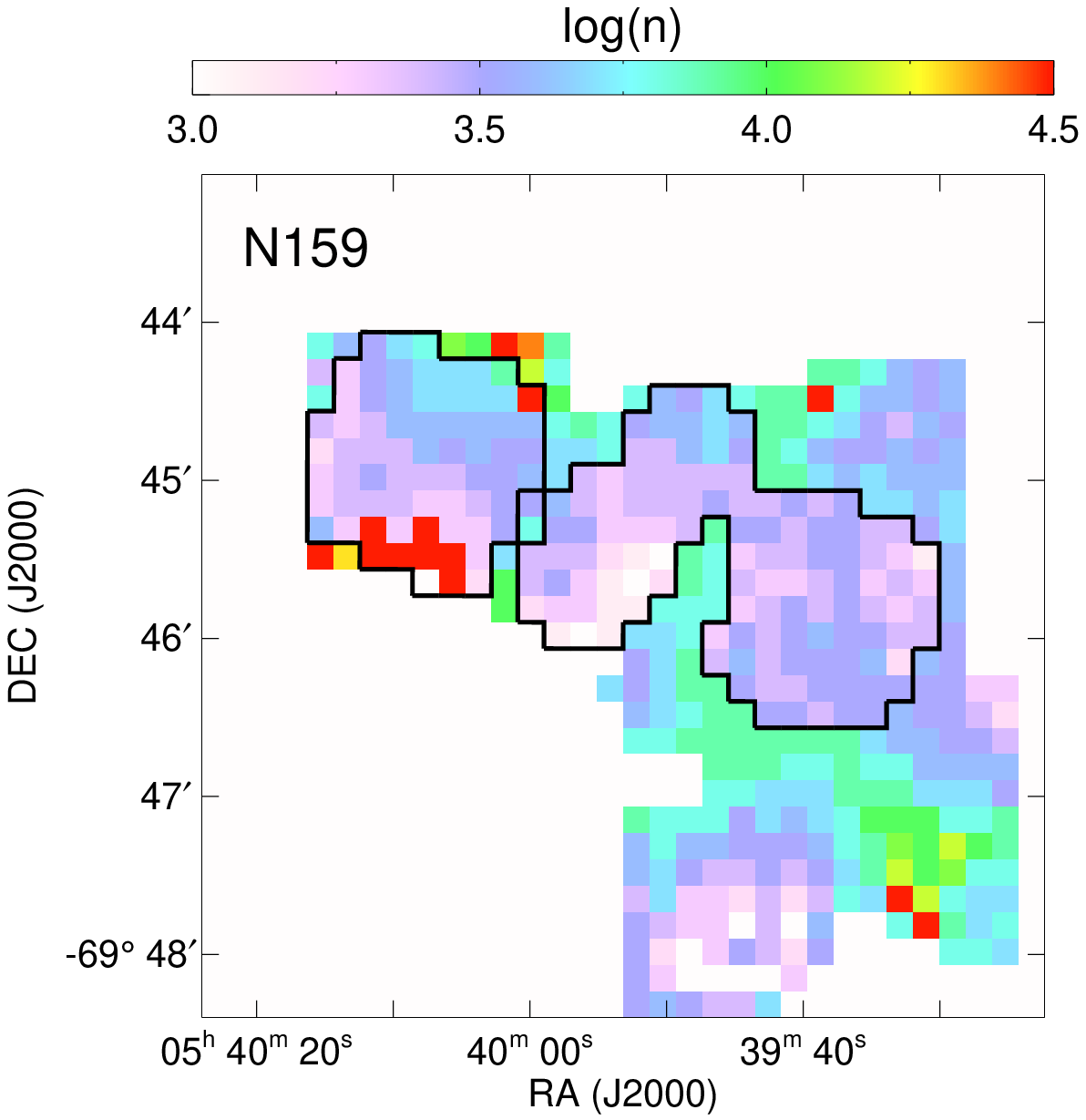}
\includegraphics[bb=80 0 404 355,width=0.23\hsize,clip]{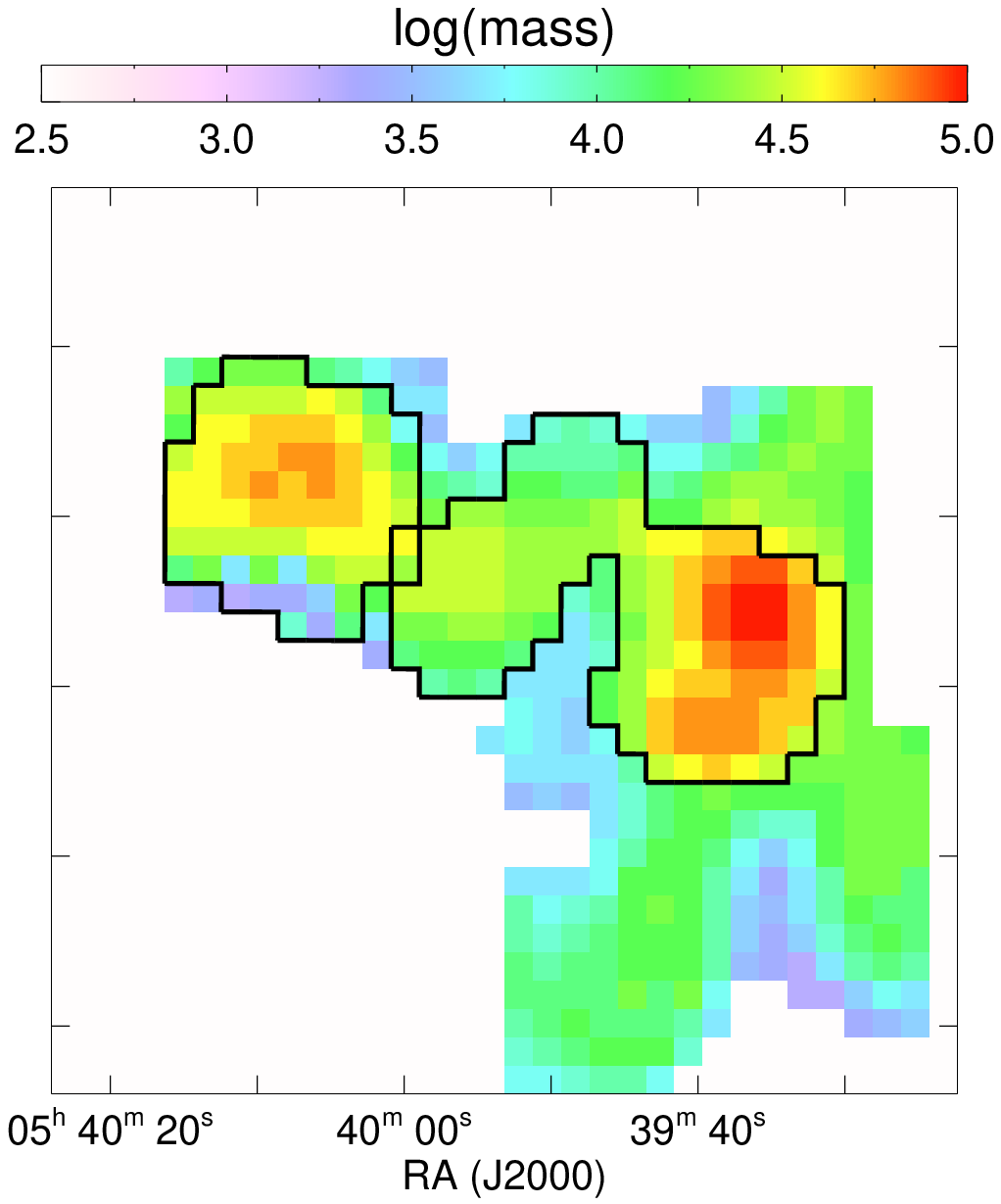}
\includegraphics[bb=80 0 404 355,width=0.23\hsize,clip]{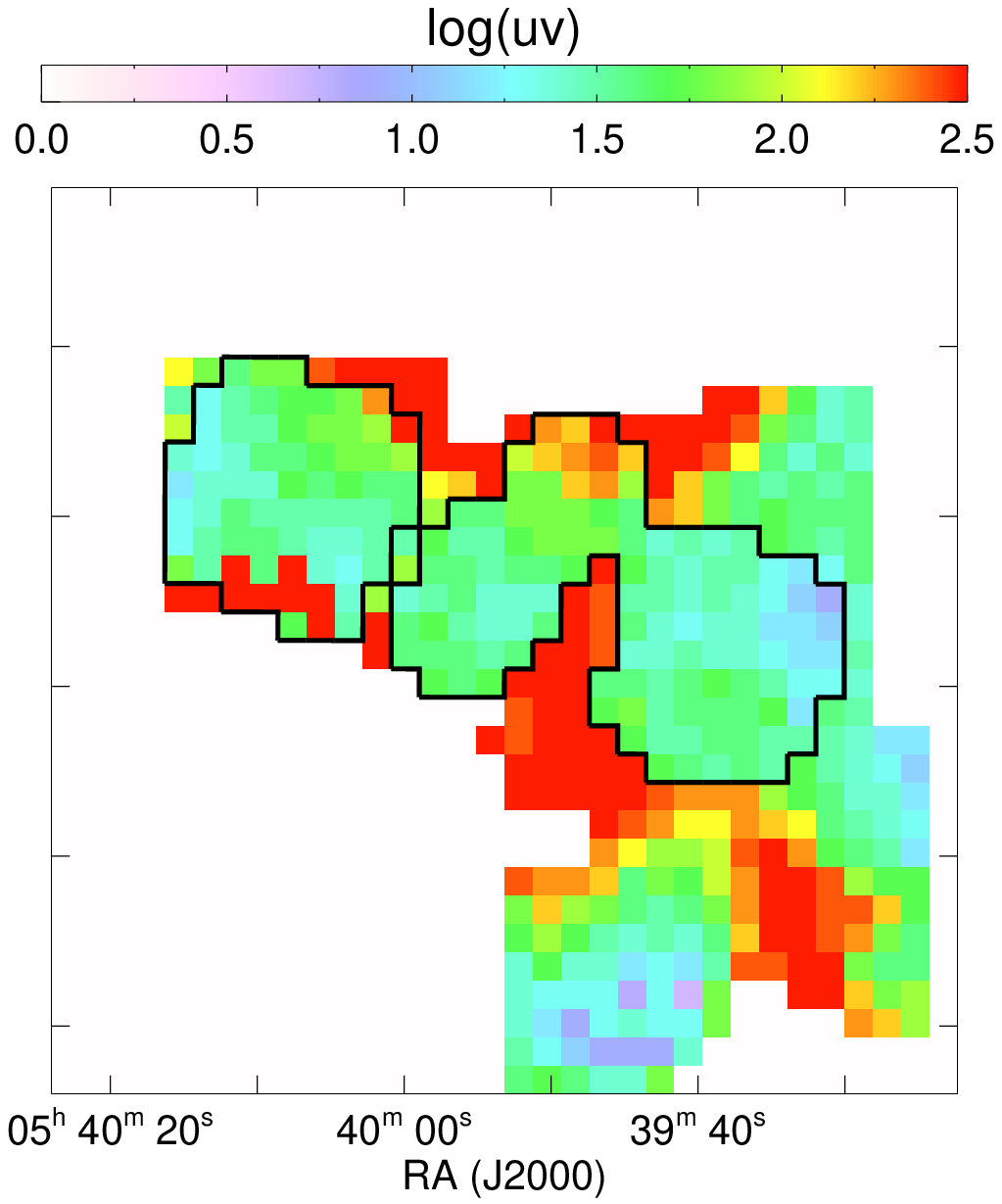}
\includegraphics[bb=80 0 404 355,width=0.23\hsize,clip]{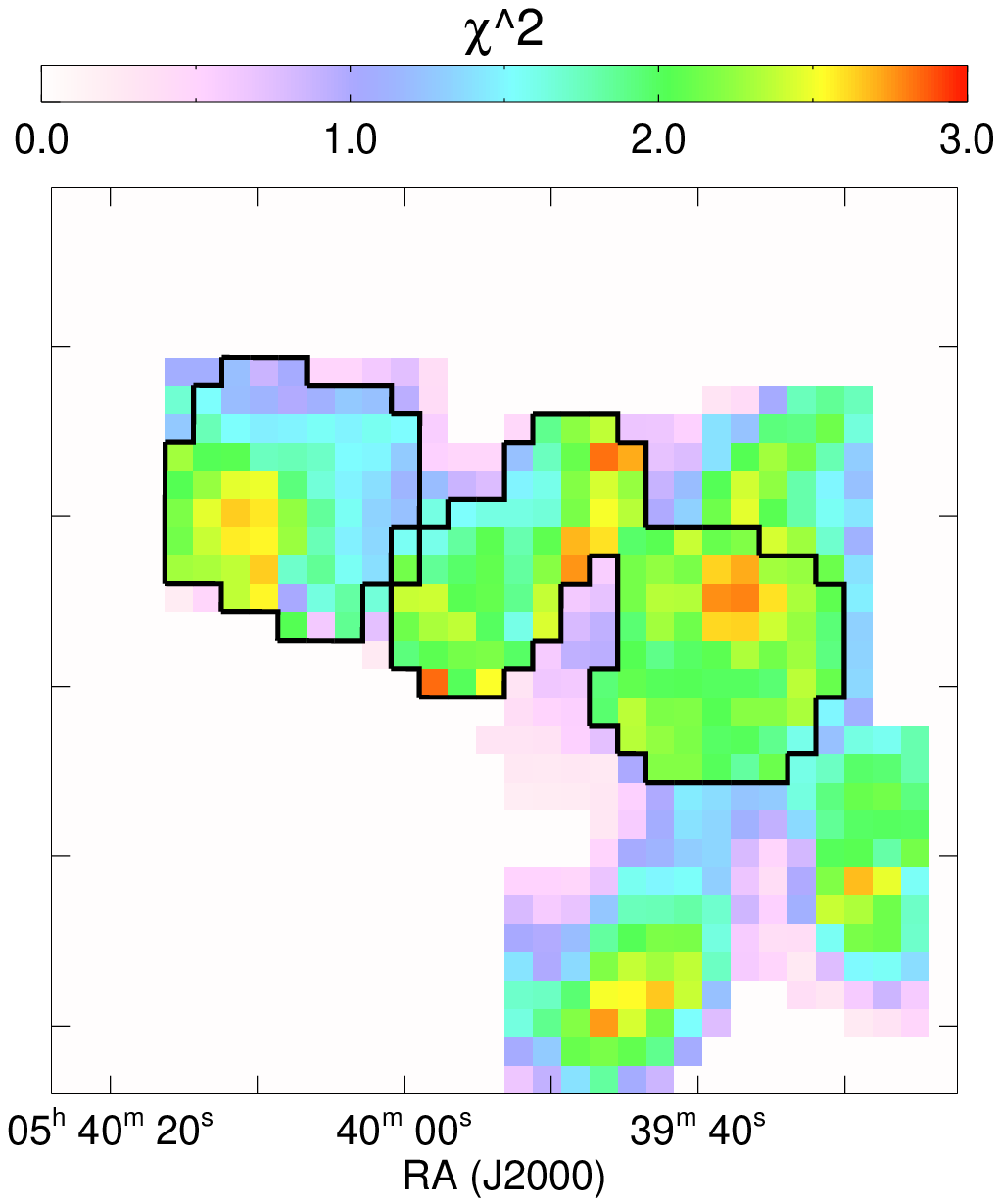}
\caption{Derived density, mass, UV field, and the $\chi^2$ of the fit (from left to right) for the N159 region. The results are obtained using a fit to all emission lines and the continuum (green lines in Fig.~\ref{figure:PDRmodel_result_SED}). Areas surrounded by black lines are where at least one of the \oi\ lines is detected.} 
\label{figure:nmuv_N159}
\end{figure*}

\begin{figure*}
\centering
\includegraphics[bb=0 0 424 355,width=0.26\hsize,clip]{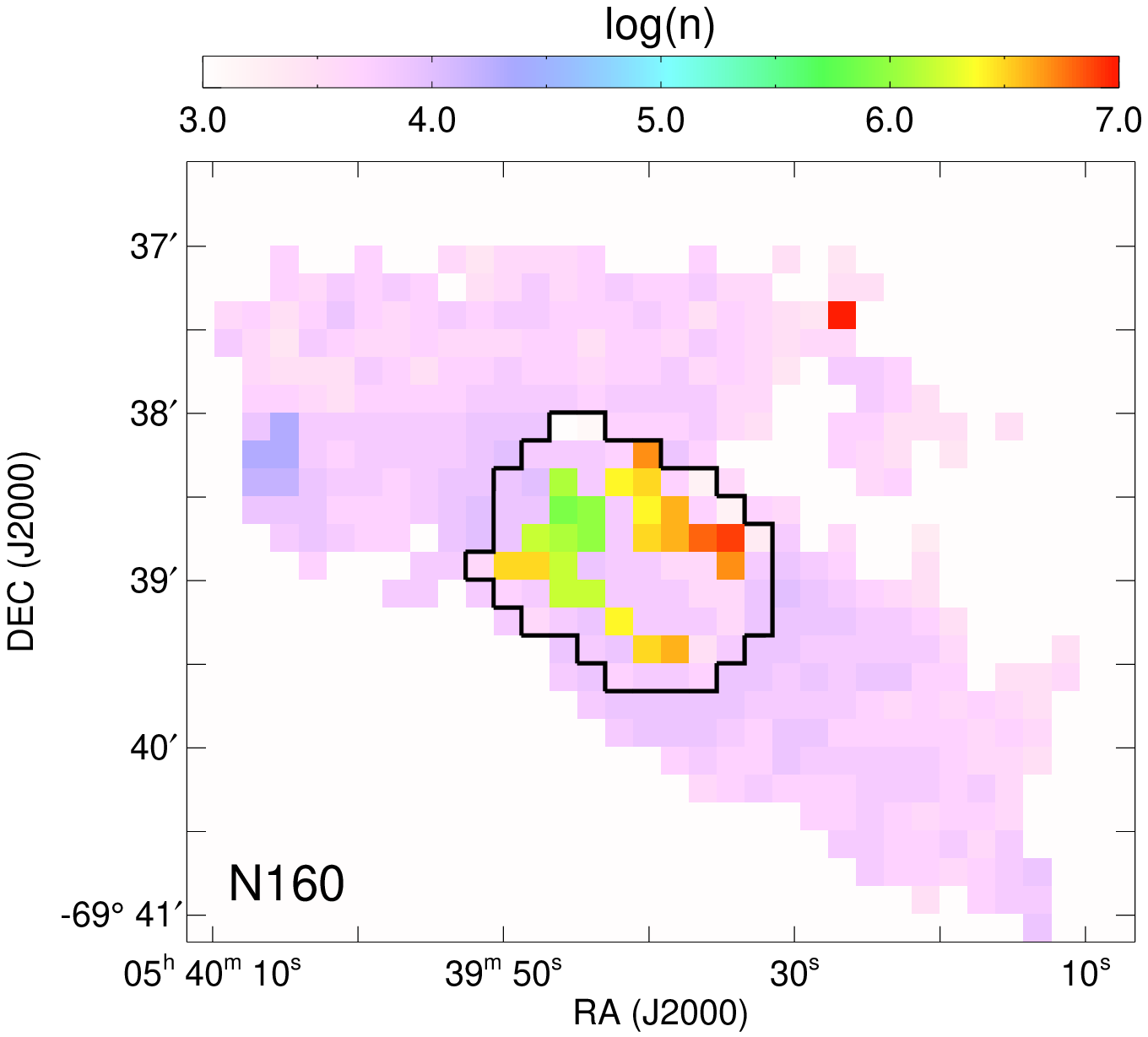}
\includegraphics[bb=40 0 414 355,width=0.23\hsize,clip]{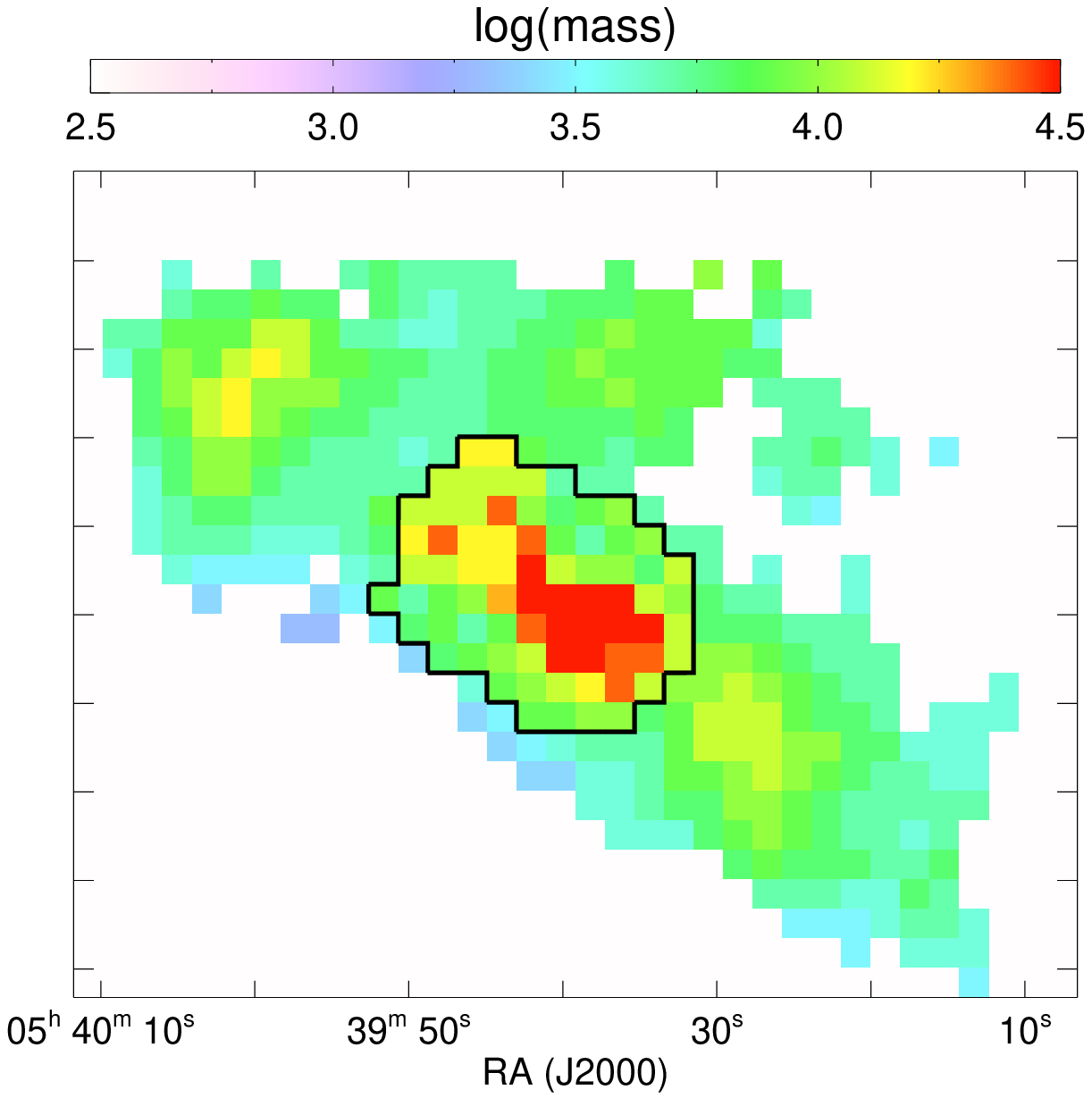}
\includegraphics[bb=40 0 414 355,width=0.23\hsize,clip]{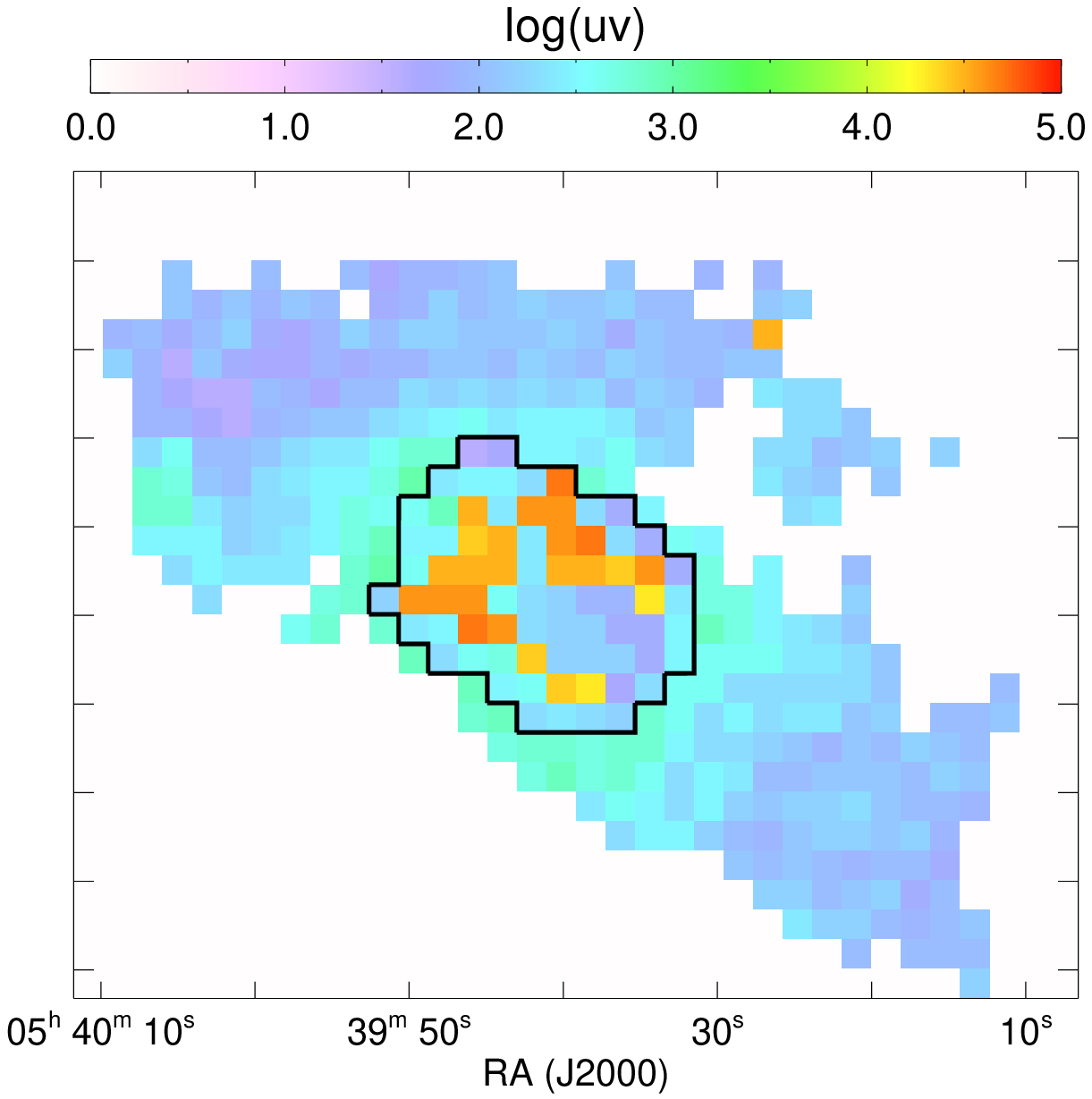}
\includegraphics[bb=40 0 414 355,width=0.23\hsize,clip]{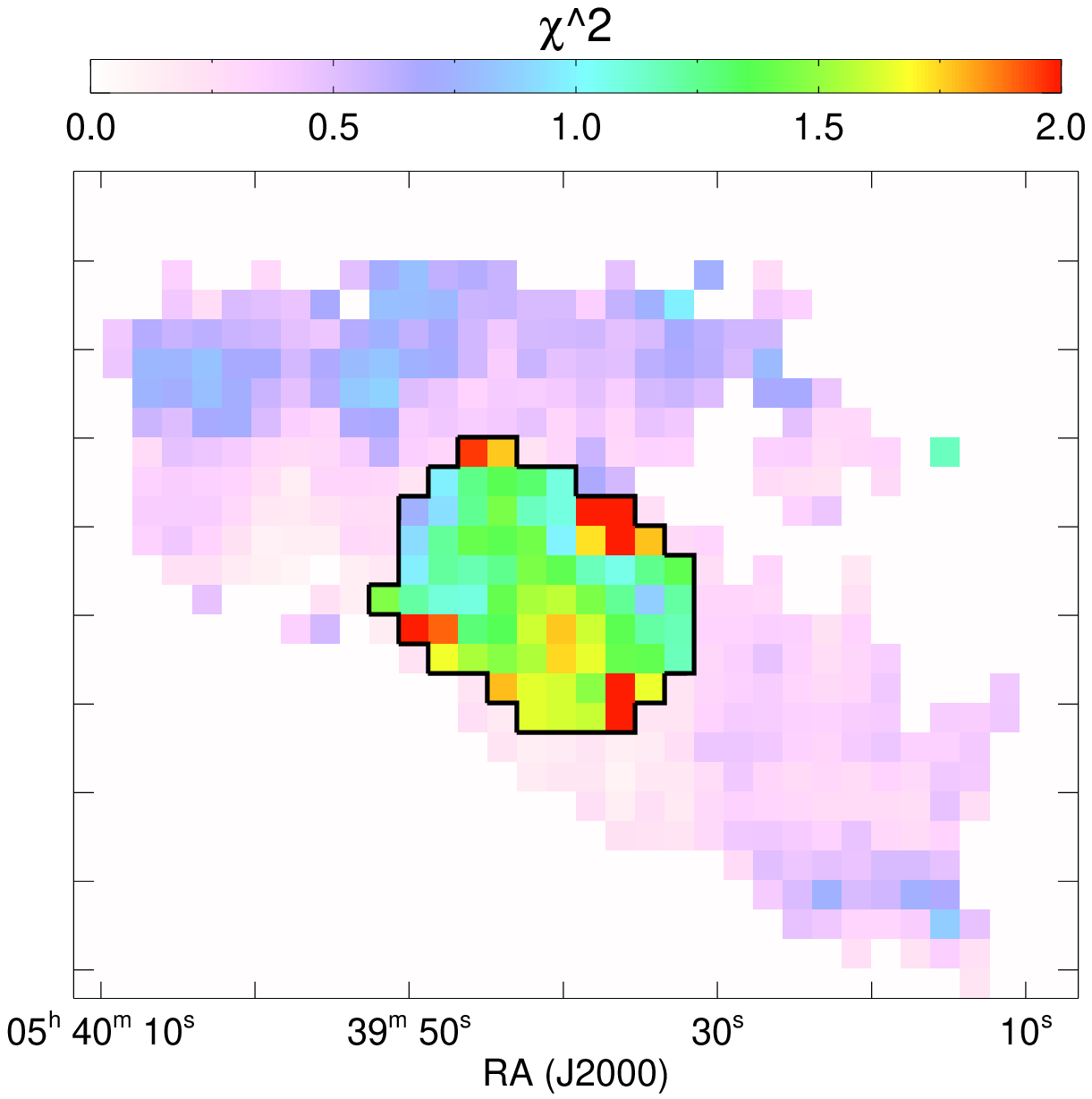}
\caption{Same as Fig.~\ref{figure:nmuv_N159} but for N160.}
\label{figure:nmuv_N160}
\end{figure*}

\begin{figure*}
\centering
\includegraphics[bb=20 0 394 355,width=0.265\hsize,clip]{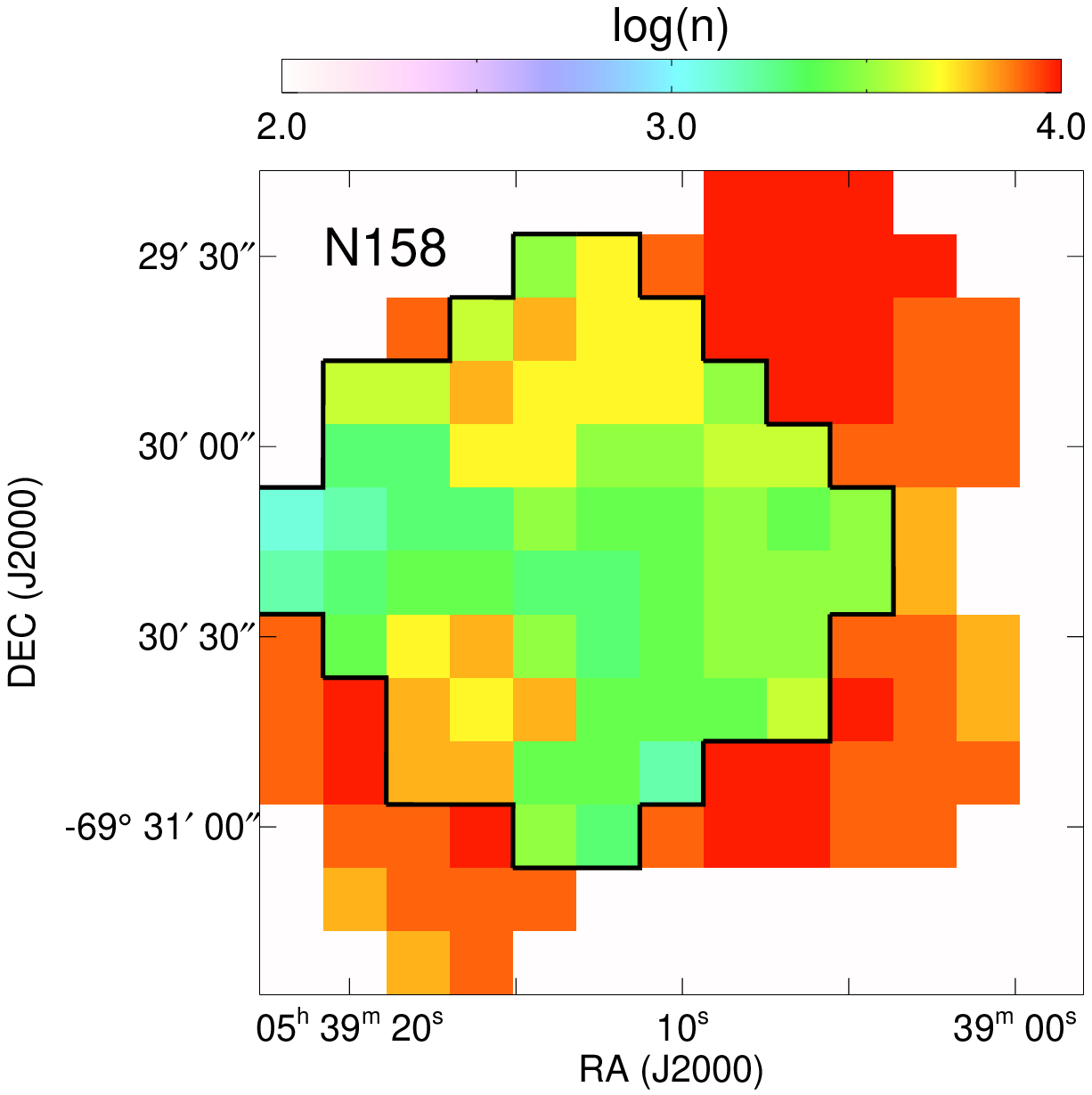}
\includegraphics[bb=80 0 394 355,width=0.22\hsize,clip]{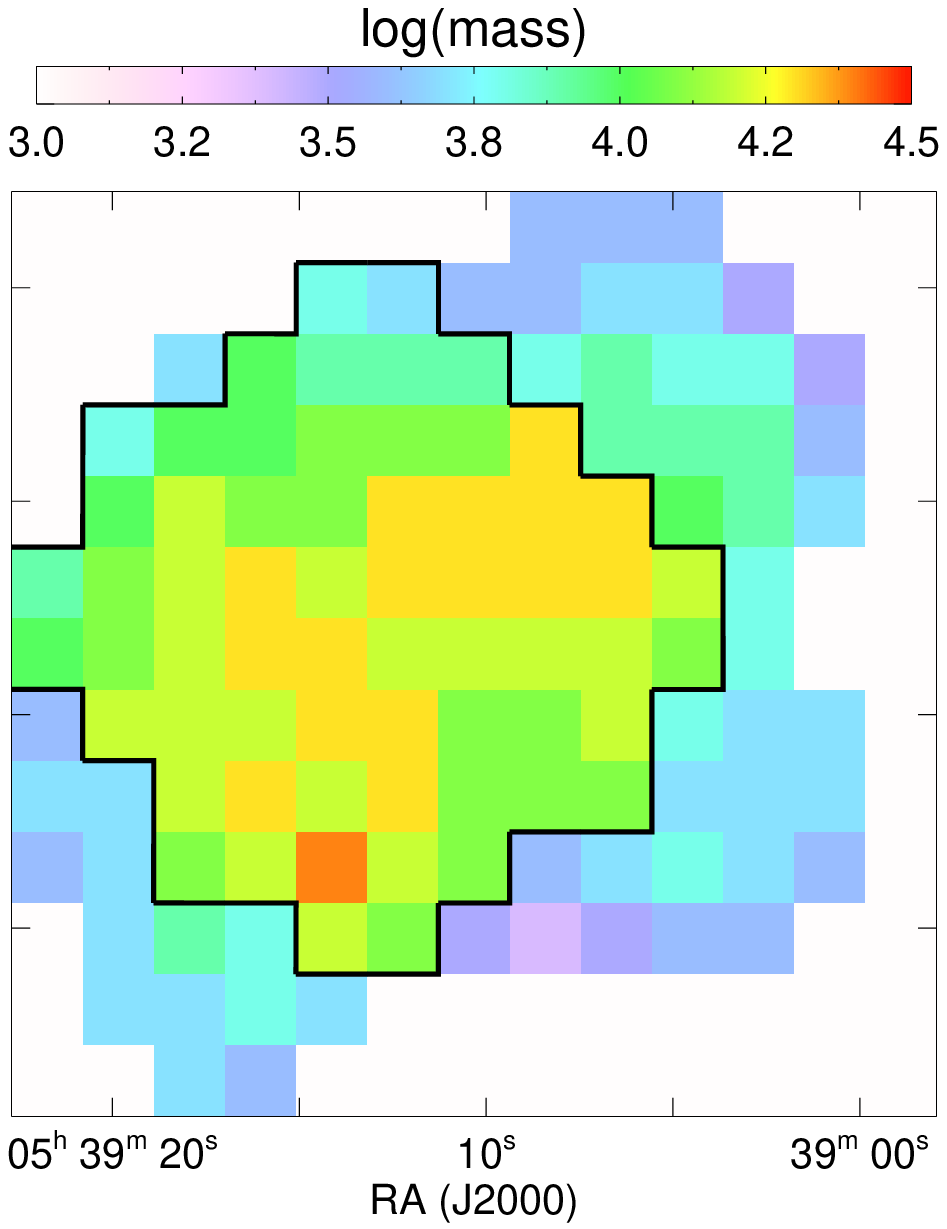}
\includegraphics[bb=80 0 394 355,width=0.22\hsize,clip]{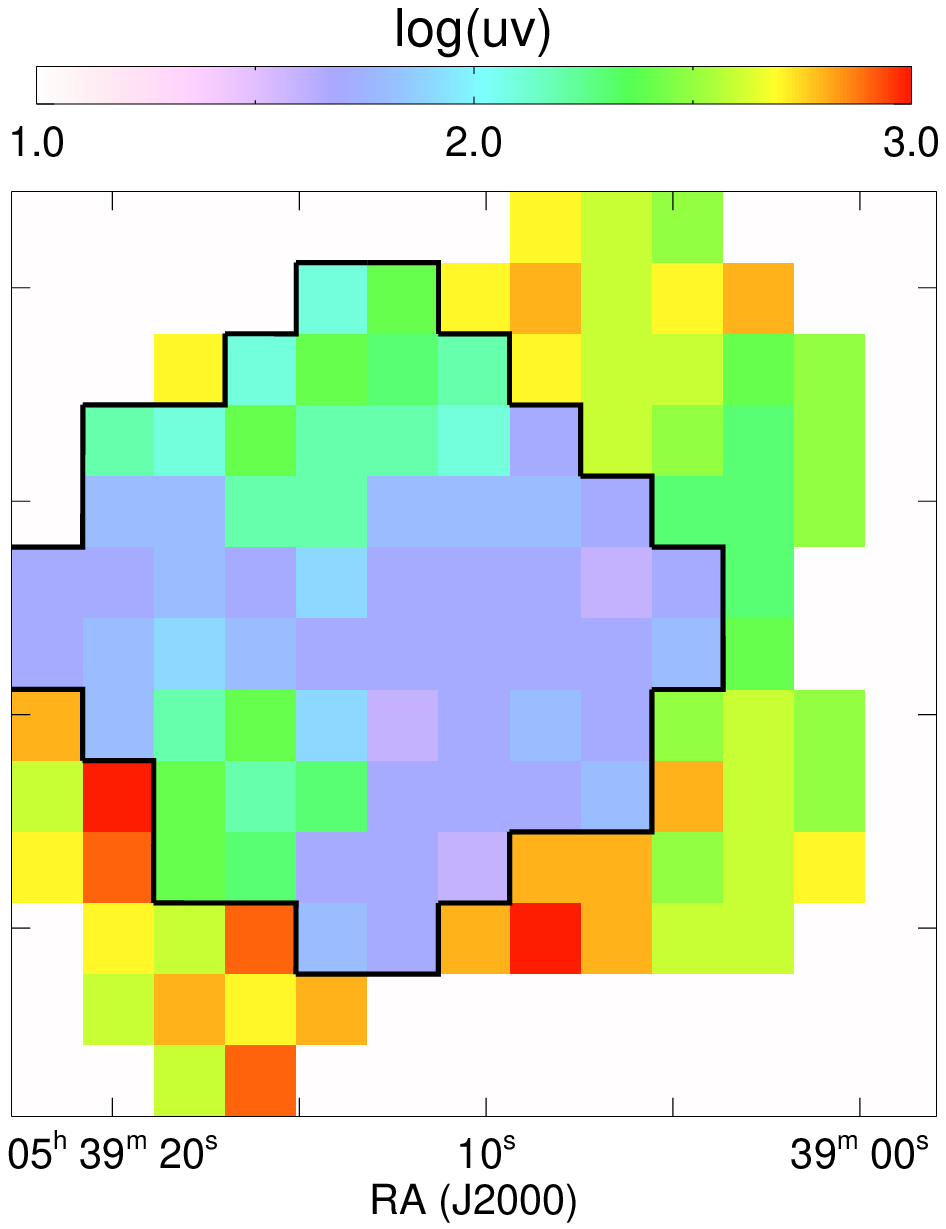}
\includegraphics[bb=80 0 394 355,width=0.22\hsize,clip]{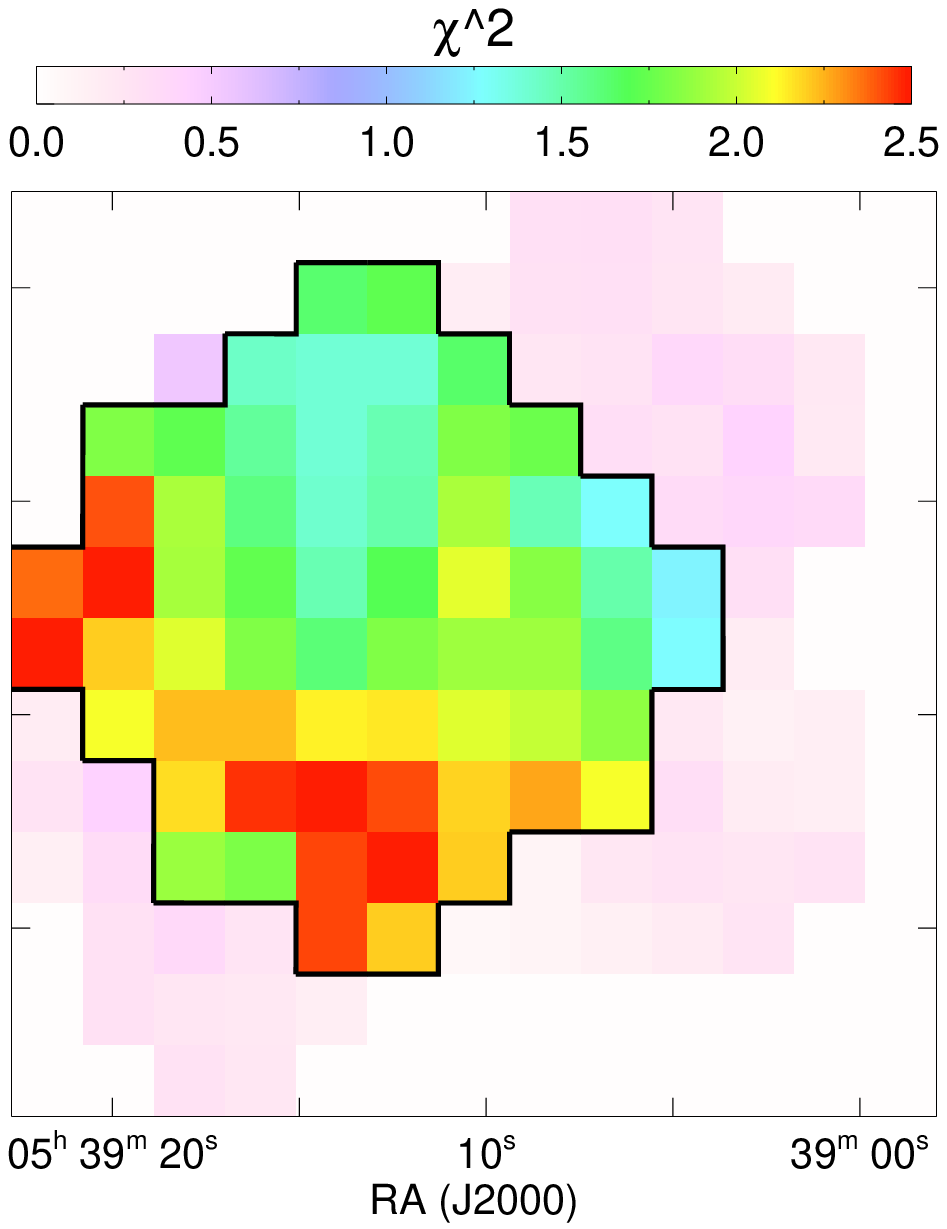}
\caption{Same as Fig.~\ref{figure:nmuv_N159} but for N158.}
\label{figure:nmuv_N158}
\end{figure*}

\begin{figure*}
\centering
\includegraphics[bb=40 0 394 355,width=0.28\hsize,clip]{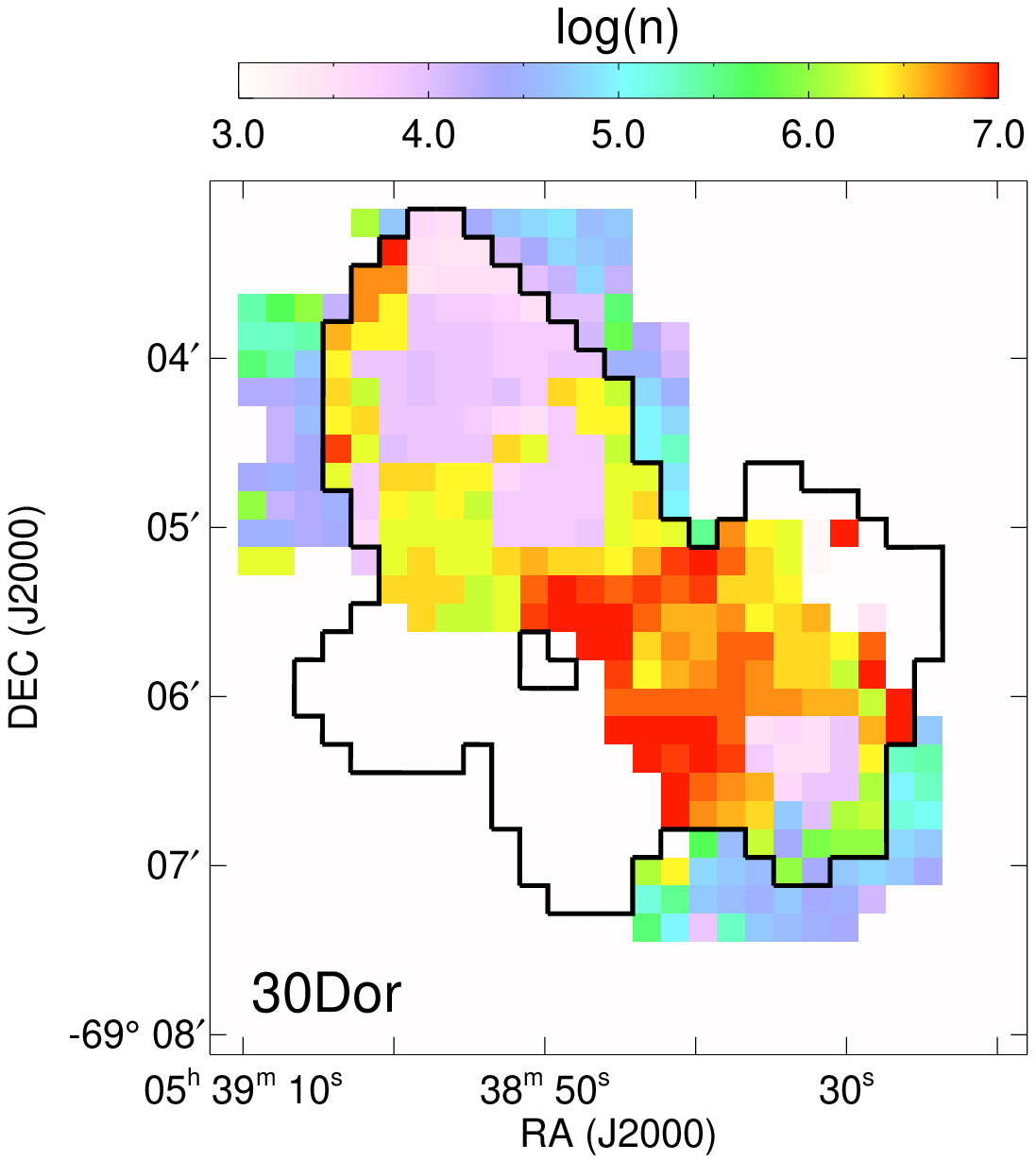}
\includegraphics[bb=100 0 384 355,width=0.22\hsize,clip]{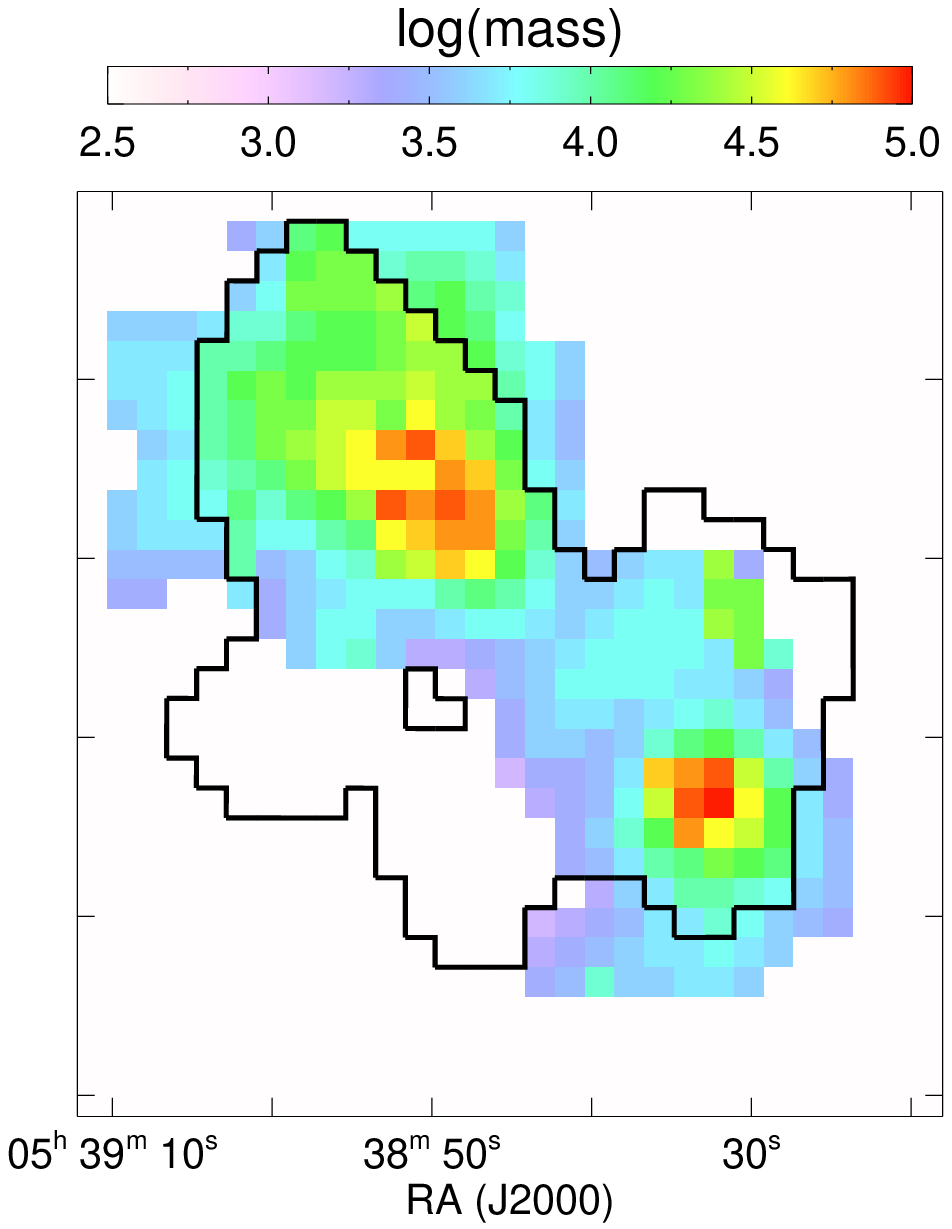}
\includegraphics[bb=100 0 384 355,width=0.22\hsize,clip]{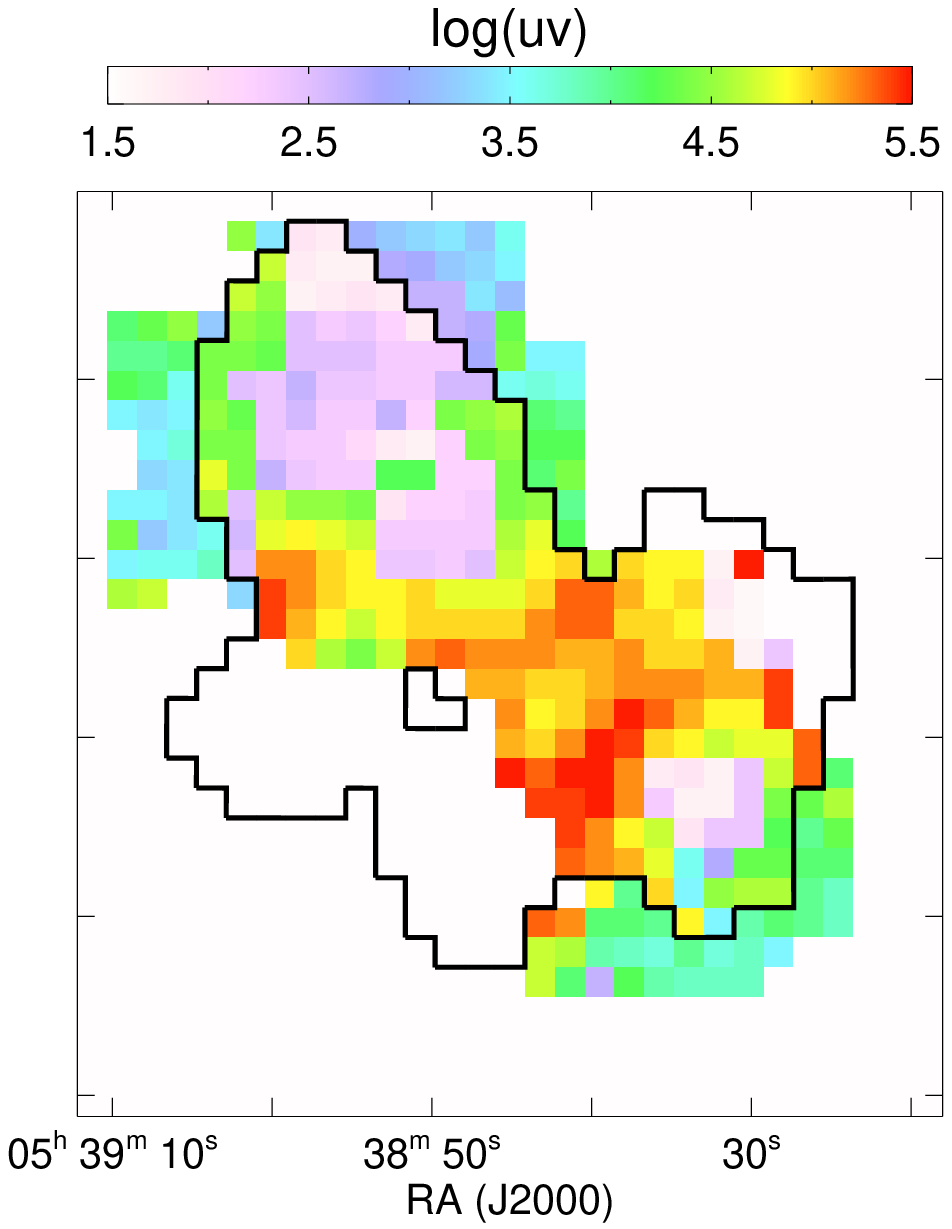}
\includegraphics[bb=100 0 384 355,width=0.22\hsize,clip]{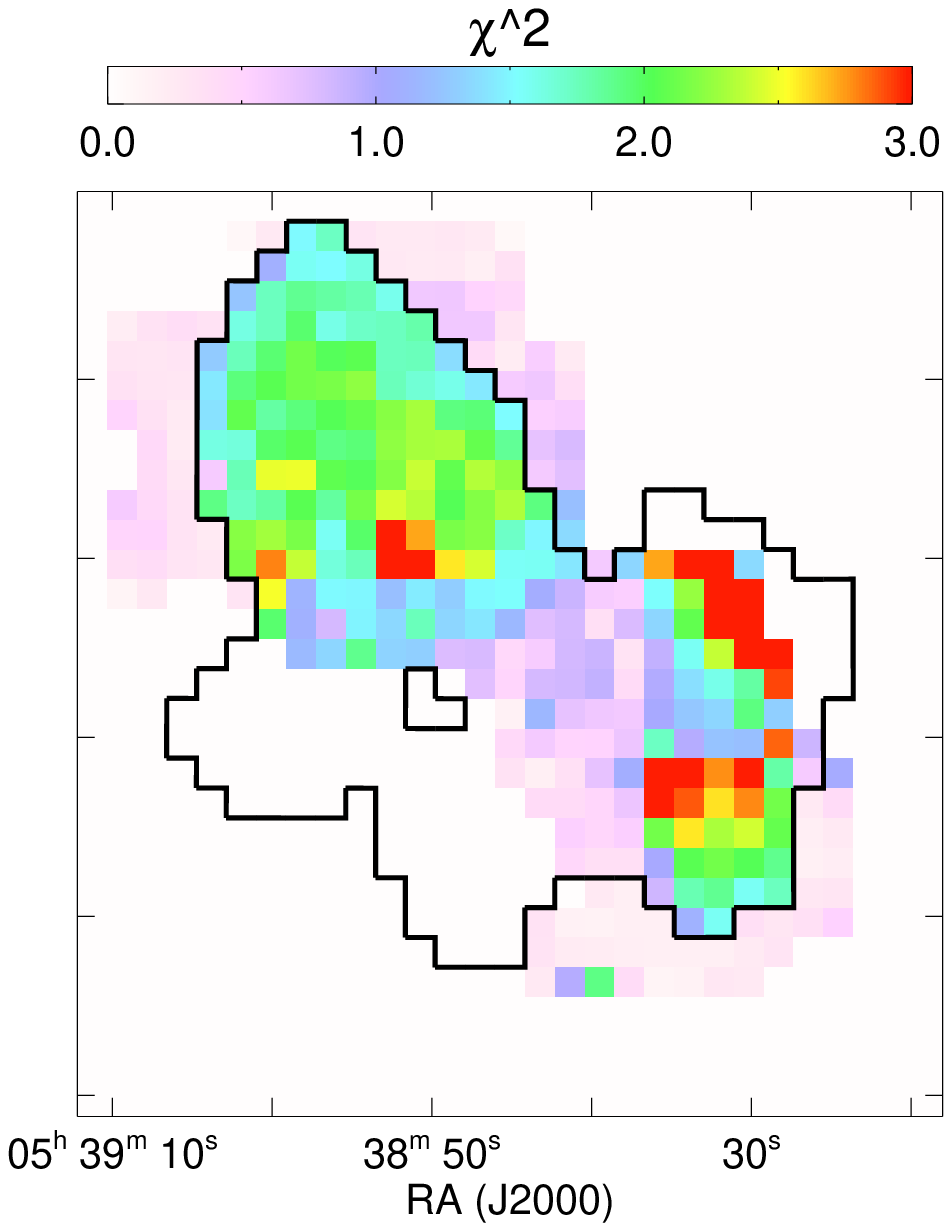}
\caption{Same as Fig.~\ref{figure:nmuv_N159} but for 30~Dor.}
\label{figure:nmuv_30Dor}
\end{figure*}

\subsubsection{Comparison of different inputs and model assumptions}\label{subsec:PDR_compare_assumptions}

Here we discuss the effect of different model assumptions and inputs on the obtained physical properties. First, we compare the results between the model with an upper mass limit of $10^3$ and $10^1 M_\odot$. In the model, a smaller clump has a higher mean density (see the mass-size relation in Sect.~\ref{sec:PDRintro}), and removing clumps with the mass range of $10^1$--$10^3$ increases the mean ensemble density by a factor 3.7. The derived densities using an upper mass limit of $10^1$ in the fit are typically higher than the results with an upper mass limit of $10^3$ by a factor of 1--4. On the other hand, the choice of the lower mass limit between $10^{-3}$ and $10^{-2}$ does not make a significant difference. This is because clumps with a mass of $10^{-3}$ are dominated by \cplus\ and do not contribute much to the CO and \ci\ emissions in most of our density and UV field strength parameter space; for example with $n=10^4$ and $\chi=10^2$, a clump with $m=10^{-3}$ gives line ratios of \cii/CO(3-2)$\sim 5\times 10^5$ and \cii/\ci\transl$\sim 7\times 10^4$.

We also ran the KOSMA-$\tau$ model with the smallest number of very small grains with the extinction curve of LMC average (model 26) and LMC 2 (model 29) in \citet{WD2001}. The extinction curve in the LMC 2 shows a weaker 2175\AA\ bump than that of the LMC average. Thus the model 29 has a size distribution with the fewest small grains in \citet{WD2001}. The smaller amount of small grains results in a lower heating efficiency, and consequently a lower temperature, which changes the continuum SED shape. The most affected emission line is \oi\ because the \oi\ lines have the highest level energies. When we compare the fitting result of model 26 and model 28 (same extinction curve, but a different amount of very small grains), the derived density and UV field of model 26 are higher than those of model 28, because it has a lower heating efficiency and requires a larger density and/or stronger UV field to produce the same amount of the line emission. The derived mass is not significantly different, because the mass is well constrained by the continuum emission, which does not depend significantly on small grains. The density and UV field derived with the model 29 is even slightly higher, but the difference between model 26 and model 29 (different extinction curve) is not as significant as the difference between model 28 and 26.

As input intensities, we examine the following cases; (1) integrated intensities over the whole velocity range, (2) a sum of the Gaussians defined by the CO velocity profiles together with the integrated intensity of \oi\ emissions, (3) same as (2) but scale the PACS \oi\ intensities using the fraction of the sum of Gaussians to the integrated intensity for the \cii. The default input used above is case (2). Case (2) corresponds to the assumption that the \oi\ emission has the same line profile as CO lines, and case (3) corresponds to the assumption that the \oi\ emission has the same line profile as \cii. The pointed observations at a few positions in N159 and 30Dor (Sect.~\ref{subsec:result_oi_profile}) show that the real case is somewhere between those two assumptions. The obtained mass is stable against the choice of the input intensities. The obtained density and mass shows a significant scatter ($\leq 1$ dex) but no systematic trend. 

\subsubsection{Spatial distribution of density, mass, and UV field}

Figures~\ref{figure:nmuv_N159}--\ref{figure:nmuv_30Dor} shows the spatial distribution of density, mass, and UV field in our four regions obtained by the model fit using all line and continuum emissions. Areas where at least one of the \oi\ lines is detected are indicated in the figures, since the fitting with and without \oi\ emission makes a difference to the obtained range of density and the UV field strength (see Sect.~\ref{subsec:PDRfit_result}). On the other hand, the derived mass distribution is stable against different fitting and model assumptions, tracing the CO clouds. 

When we only look at areas where \oi\ emission detected, the UV field and mass distribution appear anti-correlated, indicating that the areas outside of giant molecular clouds are more excited by OB stars that are not embedded by molecular clouds. Surprisingly, the distribution of the density is quite similar to that of the UV field. This may mean that only dense clumps survive in higher UV field regions. However it contradicts the interferometry observations, where high density tracers (HCO$^+$ and HCN) are detected toward the center of molecular clouds in N159 \citep{Seale2012} and the detection of NH$_3$ toward N159~W \citep{OttJ2010}. In N160, there are patches with almost 2 orders of magnitude stronger UV field than surrounding pixels, which are not fully consistent with the positions of early type stars (Fig.~\ref{figure:N160_cii_stars}). Also, the derived density at these positions are $10^6$--$10^7$~\cc, which suggests a very high pressure. It is unlikely that this jump in the UV field is real. Instead we think that it is due to the degeneracy problem between a high $n$ -- high $\chi$ solution and a low $n$ -- low $\chi$ solution (see Sect.~\ref{subsec:PDRfit_result}). In 30~Dor, the derived UV field strength close to R136 is somewhat too high because it is slightly higher than the radiation field derived by the star radiation in the plane of R136 \citep{Chevance2016}.

\subsubsection{Discussion}

Overall, the results of the PDR model fit suggest that a model with one clump ensemble component cannot reproduce all of the line and continuum emissions consistently at most of the observed positions. A natural next step would be to consider two clump ensembles with different physical properties. With the velocity-resolved spectra, it is well justified; instead of using the sum of Gaussians that conform to the CO line profiles, we would assign each velocity component to a separate clump ensemble. They should be fitted simultaneously, with free parameters to distribute the continuum emission to each component. We did not pursue this strategy in this paper since we do not have enough data points to fit additional free parameters. A very useful future addition to the present data would be velocity-resolved high-$J$ CO spectra and mid- (to high-$J$) \thco\ spectra, to better constrain the density and UV field, and investigate the origin of low-$J$ and high-$J$ CO emissions.

The results of the PDR model fit using different tracers provide a warning in interpreting the ``best'' fit of the model. When excluding one line or continuum emission from the fit, the physical properties of the best fit model vary significantly, and can jump by a few orders of magnitude in some cases. This clearly indicates that we should use as many  tracers as possible to obtain a consistent picture. 

\section{Summary} \label{sec:summary}

We mapped 3--13 arcmin$^2$ areas in 30~Dor, N158, N160, and N159 in \cii\ 158\um\ with GREAT on board SOFIA, as well as CO(2-1) to (6-5), \thco(2-1) and (3-2), \ci\ \transl\ and \ci\ \transu\ with APEX. We also observed the velocity-resolved \oi\ 145\um\ and 63\um\ at selected positions in N159 and 30~Dor for the first time. The results are as follows.

\begin{itemize}
\item In all four star-forming regions, the line profiles of the CO, \thco, and \ci\ emission are similar while \cii\  typically has a wider line profile or an additional velocity component. On average, 30\% of the emission in \cii\ cannot be reproduced by the CO-defined line profile. This fraction is lower toward molecular clouds.
\item The \oi\ line profiles match those of CO at some positions, but are more similar to the \cii\ profiles at other positions. This indicates that we cannot simply assume that the velocity components of the \oi\ emission are the same as either CO or \cii.
\item Using the \hi\ and \cii\ line profiles, we estimated that contribution of atomic gas to the \cii\ emission is 15\% or less. The thermal pressure that is required to emit the \cii\ associated with the atomic gas is $4\times 10^3$--$10^5$~K\,\cc, which is higher than the standard Galactic ISM value and consistent with previous studies. For some positions in 30~Dor, the \hi\ absorption coincides with a velocity component in \cii, which may be because of a cold molecular cloud with a mixture of atomic hydrogen.
\item We interpret the different line profiles among CO, \cii, and \oi\ in the LMC as contributions from spatially separated clouds and/or clouds in different phases, which give different line ratios depending on their physical properties.
\item We investigate channel maps in individual regions and distinguish clouds.
\item We derived the column density of CO, \czero, and \cplus\ and compared them to those of SMC regions measured by \citet{Requena-Torres2016}. We do not see a clear correlation between $N$(\cplus)/$N$(CO) and metallicity. The trend cannot be fully explained by a different UV field strength and metallicity.
\item We modeled the line and continuum emission using the latest KOSMA-$\tau$ PDR model, which treats the dust-related physics consistently and computes the dust continuum SED as well as the line emission. At most positions, a single clump ensemble does not satisfactorily reproduce all the observed emission lines and continuum. Toward the CO peak at N159~W, we propose a scenario in which the CO, \cii\ and \oi\ 63\um\ emission are affected by mutual shielding between clumps.
\item We show that the best fit model results depend sensitively on which combination of continuum and emission lines are used in the fit.  
\end{itemize}

\begin{acknowledgements}
This work is based in part on observations made with the NASA/DLR Stratospheric Observatory for Infrared Astronomy (SOFIA). SOFIA is jointly operated by the Universities Space Research Association, Inc. (USRA), under NASA contract NAS2-97001, and the Deutsches SOFIA Institut (DSI) under DLR contract 50 OK 0901 to the University of Stuttgart. This work is carried out within the Collaborative Research Centre 956, sub-project A3 and C1, funded by the Deutsche Forschungsgemeinschaft (DFG). We thank M.-Y. Lee for fruitful discussions and providing the data of Herschel/SPIRE at N159~W. We thank R. Braun for providing the opacity corrected \hi\ column density in the LMC.
\end{acknowledgements}

\begin{appendix}

\section{Gain correction using the total raw count}\label{app:gain_correction}

\begin{figure}
\centering
\includegraphics[width=0.95\hsize]{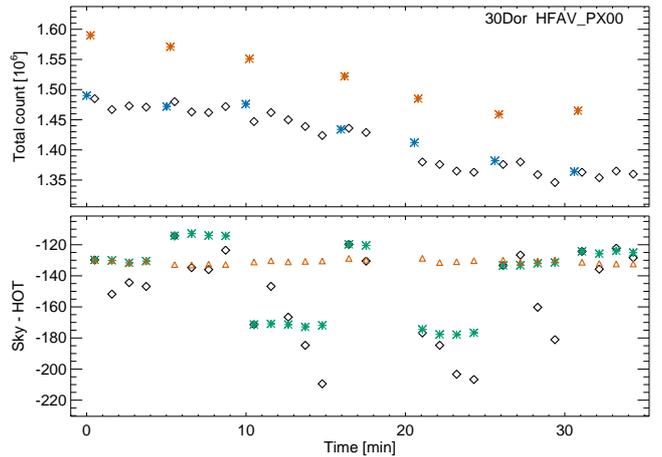}
\caption{(Upper panel) The total count averaged over 5000 channels with a good sensitivity and outside of the source emission versus time in the 30~Dor leg for the pixel HFAV\_PX00. The red and blue points indicate hot-load and cold-load measurements, respectively, and black points are the sky observations. (Lower panel) The sky count minus hot count divided by the gain, averaged over the same channels. See \citet{Guan2012} for the detailed formula. Black diamonds show the case without gain correction, green asterisks are after the gain correction within observations that share the same load measurement, red triangles represent the data after the final grain correction (see text).}
\label{figure:gaindrift}
\end{figure}

In the HFA data, we often see that the total raw count is drifting (Fig.~\ref{figure:gaindrift}). Since the raw count of the load measurements follow the same drifting trend in general, this is most likely determined by a gain drift of the receiver rather than a change of the sky transmission. The lower panel of Fig.~\ref{figure:gaindrift} shows the sky$-$hot count divided by the gain \citep[See][for the detailed formula]{Guan2012}. When the gain is calculated from the last load measurement before the sky observations, the sky$-$hot shows a strong drift between two load measurements (black points), which leads a corresponding variation of the fitted atmospheric transmission. The first method to correct it is to scale the raw count of individual observations to the one right after each load measurement, assuming that the total count variation between two load measurements is caused by the gain drift. The resultant sky$-$hot is presented as green points in the lower panel of Fig.~\ref{figure:gaindrift}. The drift between two load measurements is corrected, but a new load measurement often does not provide the same level of sky$-$hot; it is jumping around. For our paper, relative intensities between different measurements are more important than the absolute intensity, because (1) we would like to achieve a good S/N by averaging measurements over 10--20 min of observing time and (2) we would like to compare four raster points in 30~Dor which are spread among the whole leg. Therefore, we took a somewhat radical approach: we scaled all hot measurement to the first hot measurement, all cold measurement to the first cold measurement, and all sky measurement to the first sky measurement for  each pixel of each source. The sky$-$hot after this gain correction is shown as red triangles in Fig.~\ref{figure:gaindrift} lower panel. This assumes that the sky opacity is constant over the observing time of one source, and any change in the total raw count is caused by a gain variation. This correction implies that all measurements have the same atmospheric opacity correction, but the absolute correction is uncertain since the choice of the measurement to be scaled to is arbitrary. We note that the lower panel of Fig.~\ref{figure:gaindrift} indicates that taking the first measurement is not a bad choice because it has an intermediate gain value. This method is justified by following reasons. The variation in sky$-$hot among different load measurement blocks (green points) is not synchronized among different pixels. If this variation is because of the variation of the sky opacity, all pixels should show the same trend. The second reason is that for similar observations of another source (NGC2024, Graf et al. in prep.), strong baseline structure is suppressed significantly by our adopted scaling method. 

In terms of the absolute intensity scale, we estimate the uncertainty by the variation of sky$-$hot among different load measurement blocks (green points), which is translated to the variation of the atmospheric transmission. Among the data used in this paper, the uncertainty of the absolute intensity is $\leq 66$\% for HFAV\_PX00 and $\leq 33$\% for other HFAV pixels. We did the same analysis for LFA channels, although they were much more stable. The maximum uncertainty of the absolute intensity for LFA pixels is 20\%.

\section{Integrated intensity maps and selected spectra in N160, N158, and 30~Dor}\label{app:map_and_spectra}

Here we show the integrated intensity maps and spectra at selected positions of N160, N158, and 30~Dor as shown in Paper I for N159.

\begin{figure*}
\centering
\includegraphics[bb=30 20 414 340,width=0.51\hsize,clip]{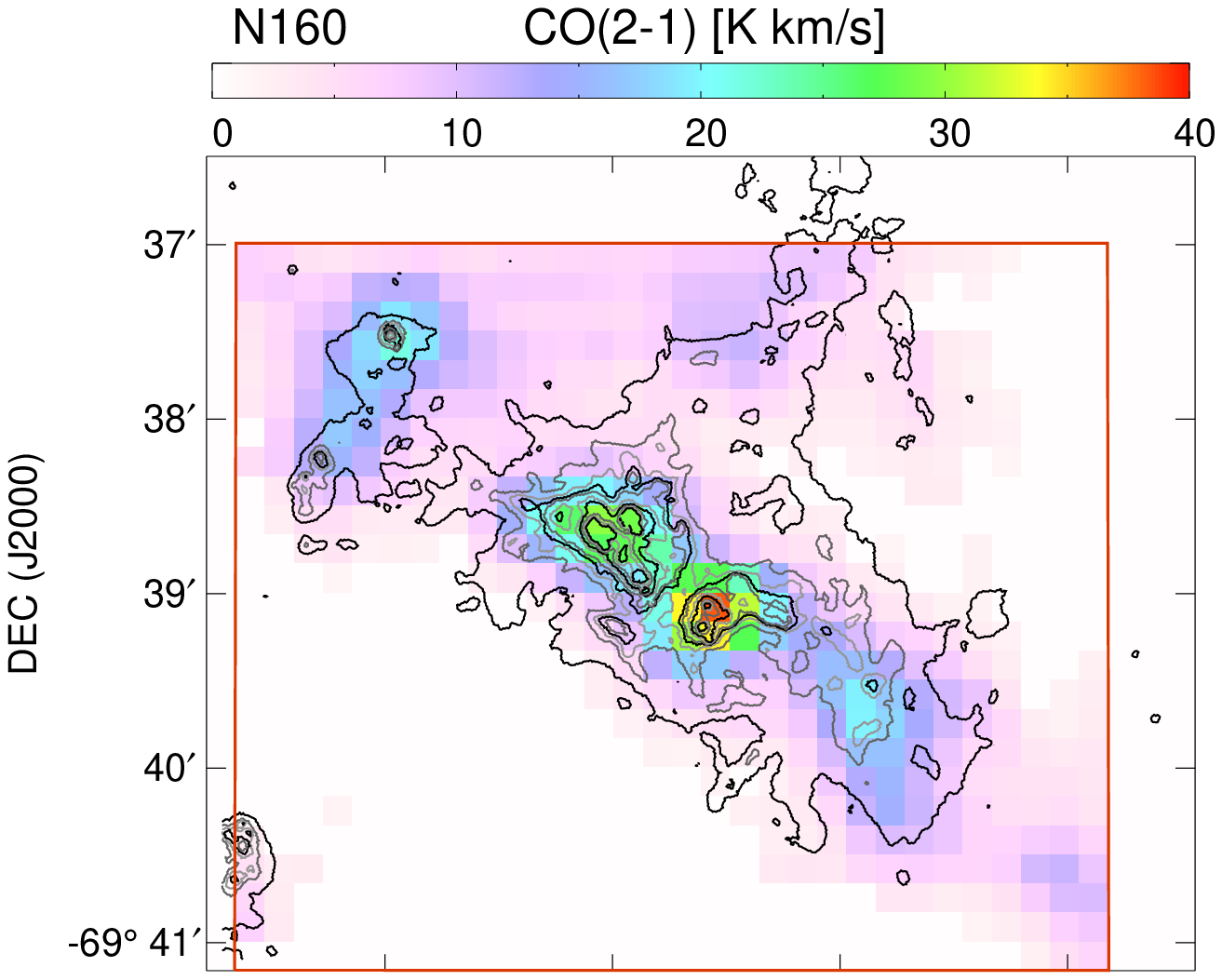}
\includegraphics[bb=60 20 414 340,width=0.47\hsize,clip]{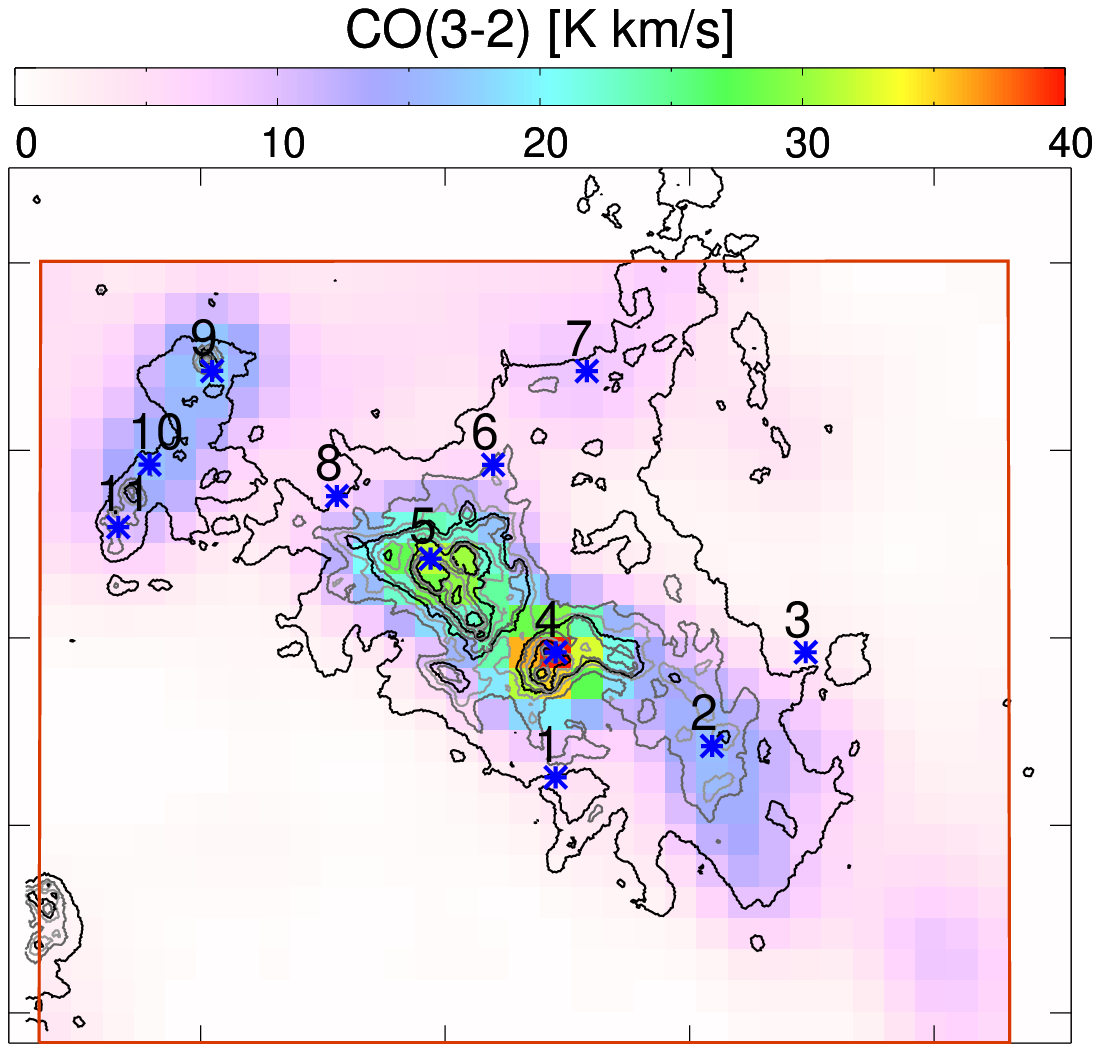}
\includegraphics[bb=30 20 414 340,width=0.51\hsize,clip]{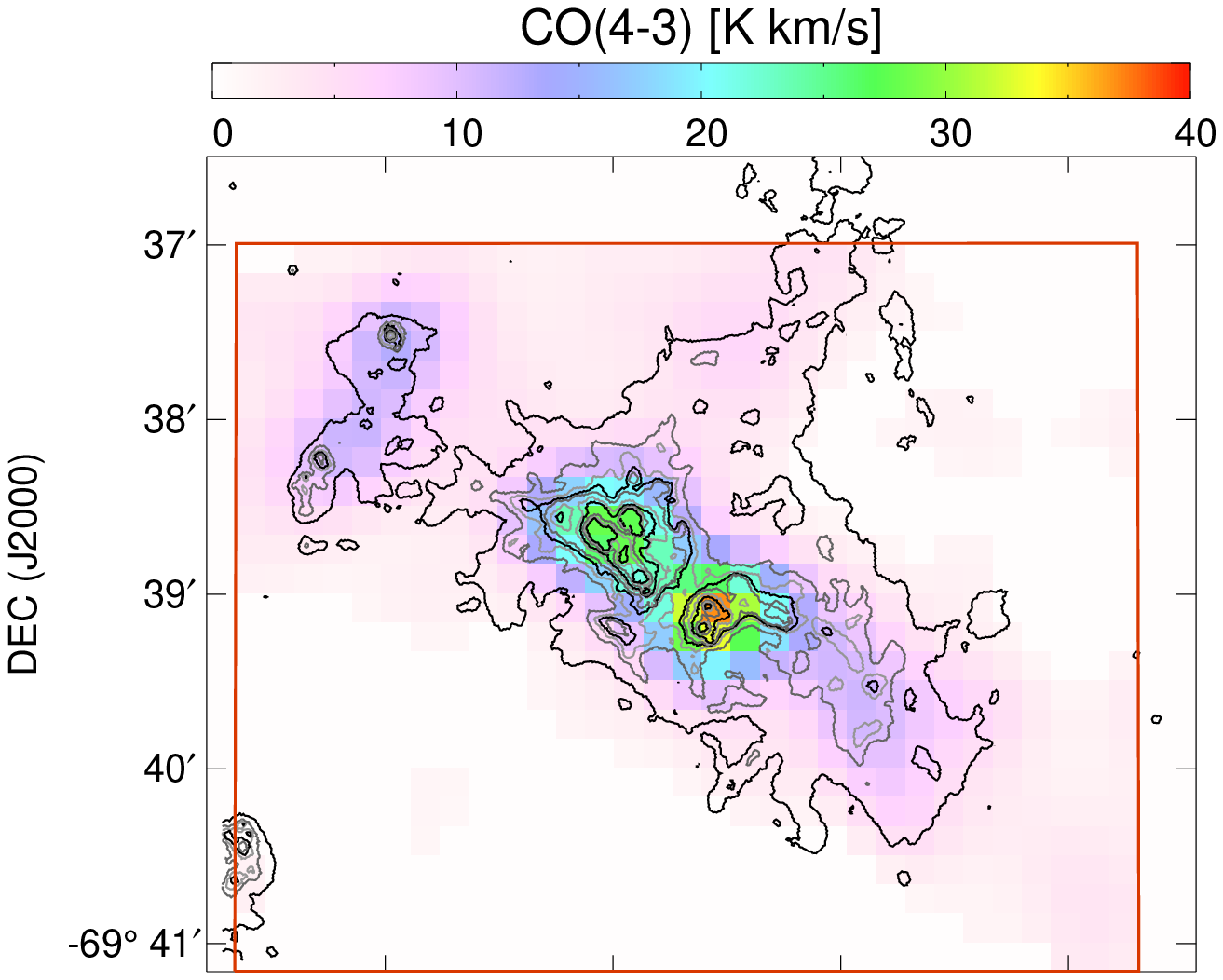}
\includegraphics[bb=60 20 414 340,width=0.47\hsize,clip]{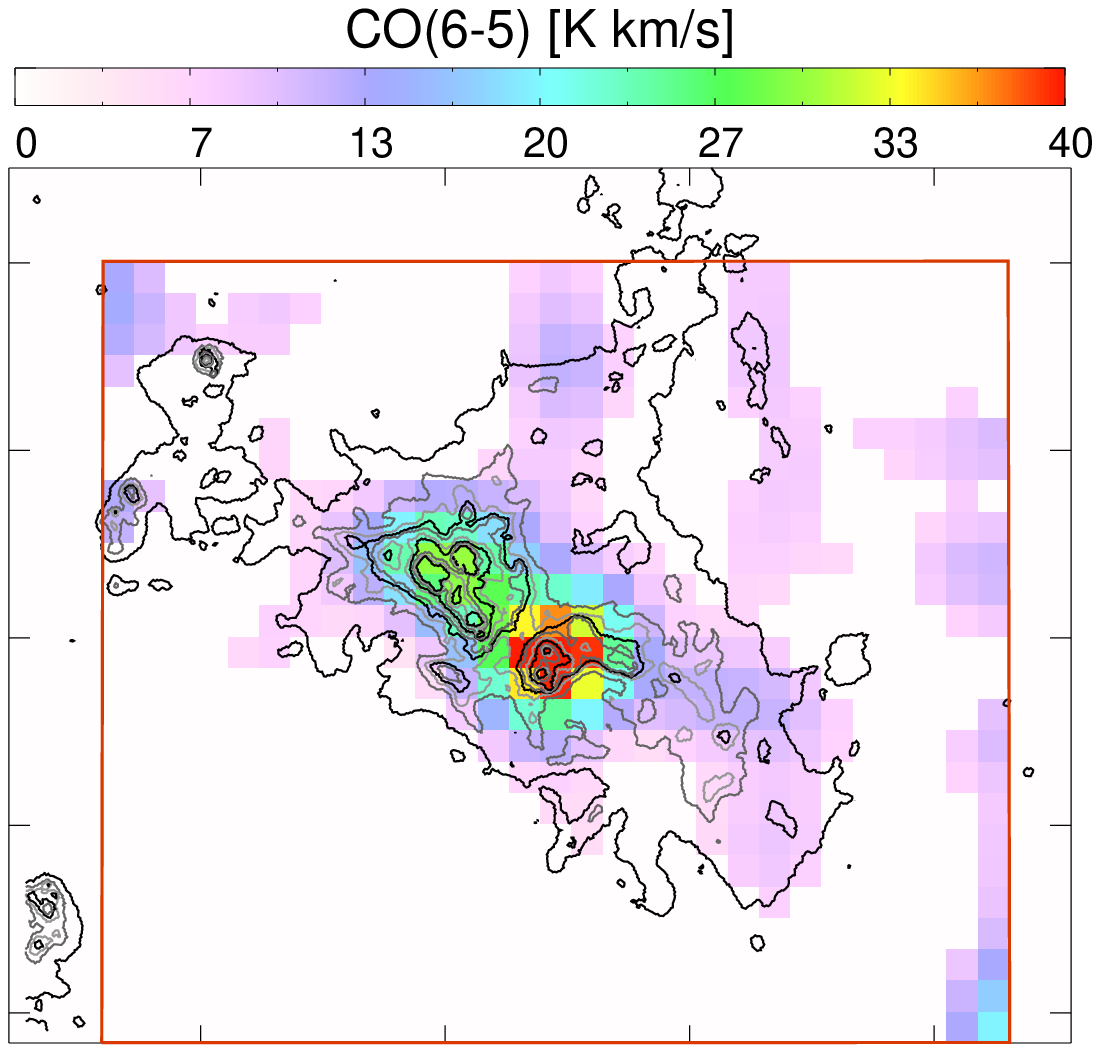}
\includegraphics[bb=30 0 414 340,width=0.51\hsize,clip]{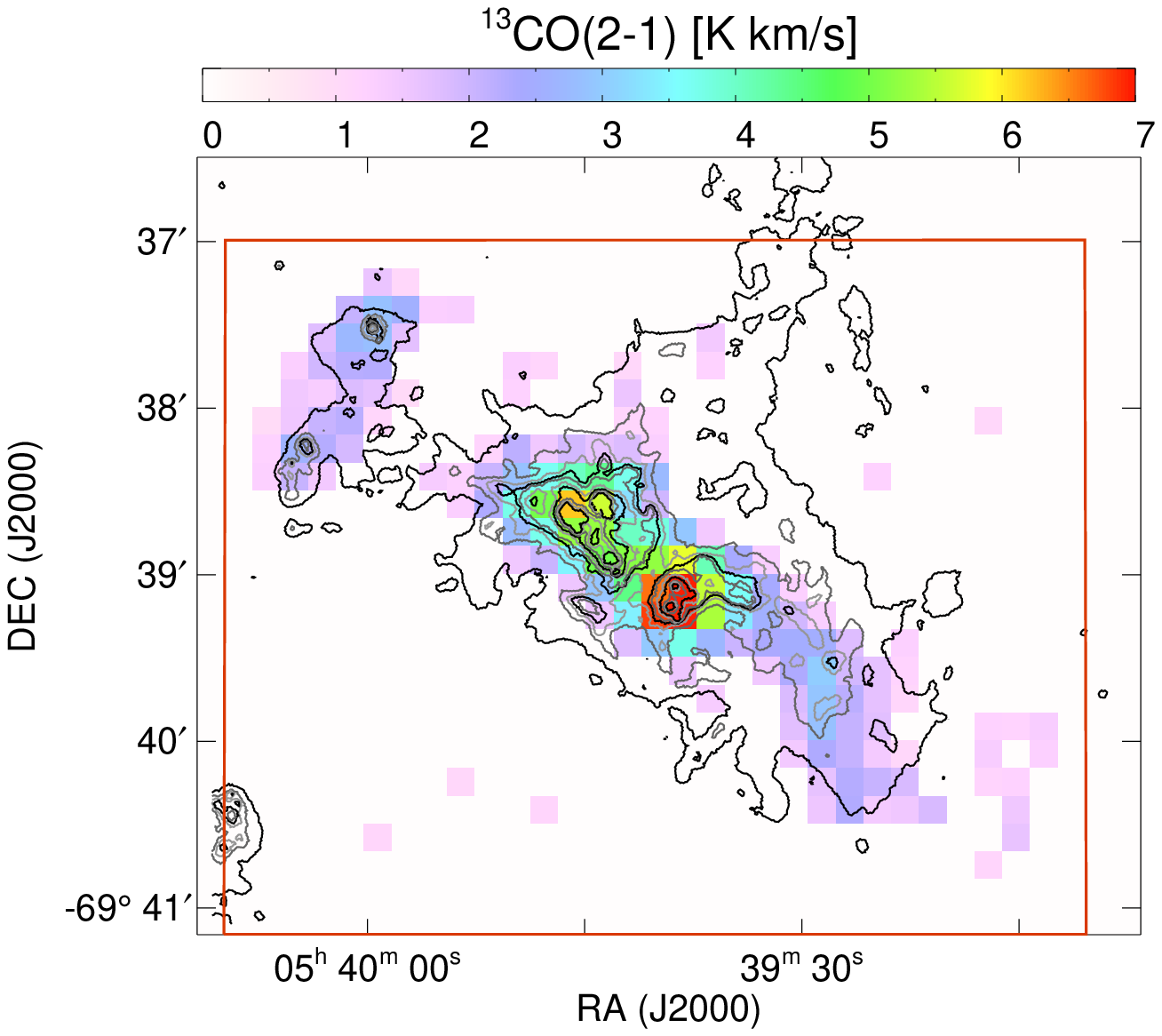}
\includegraphics[bb=60 0 414 340,width=0.47\hsize,clip]{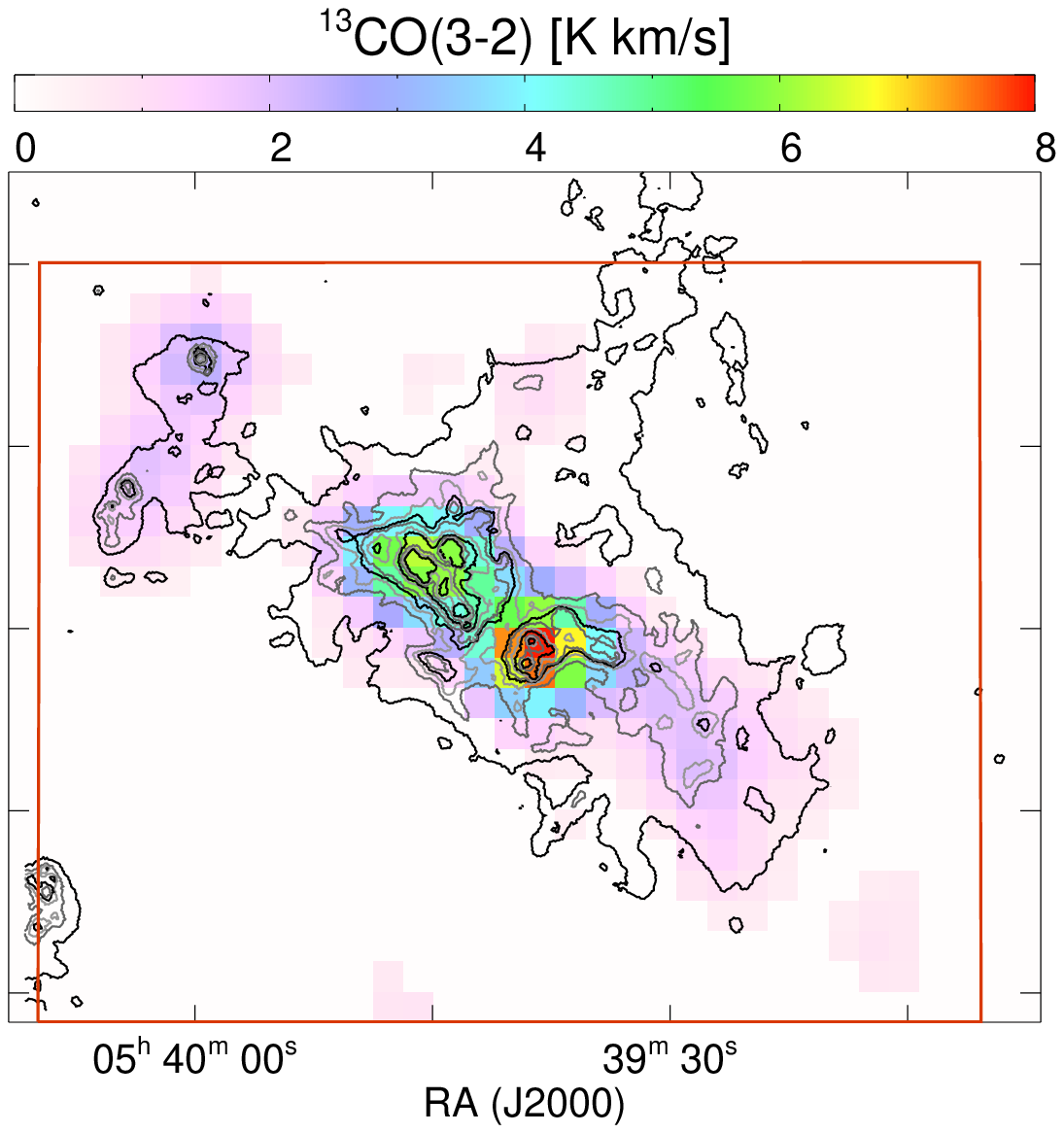}
\caption{Integrated intensity maps (colors, 30\arcsec\ resolution) overlaid with contours of the IRAC 8\um\ emission in N160. The red lines outline the observed area. Blue asterisks in CO(3-2) and \cii\ maps mark the positions where the spectra shown in Fig.~\ref{figure:selected_spectra_N160} have been extracted.}
\label{figure:integmap_N160}
\end{figure*}

\addtocounter{figure}{-1}

\begin{figure*}
\centering
\includegraphics[bb=30 0 414 340,width=0.51\hsize,clip]{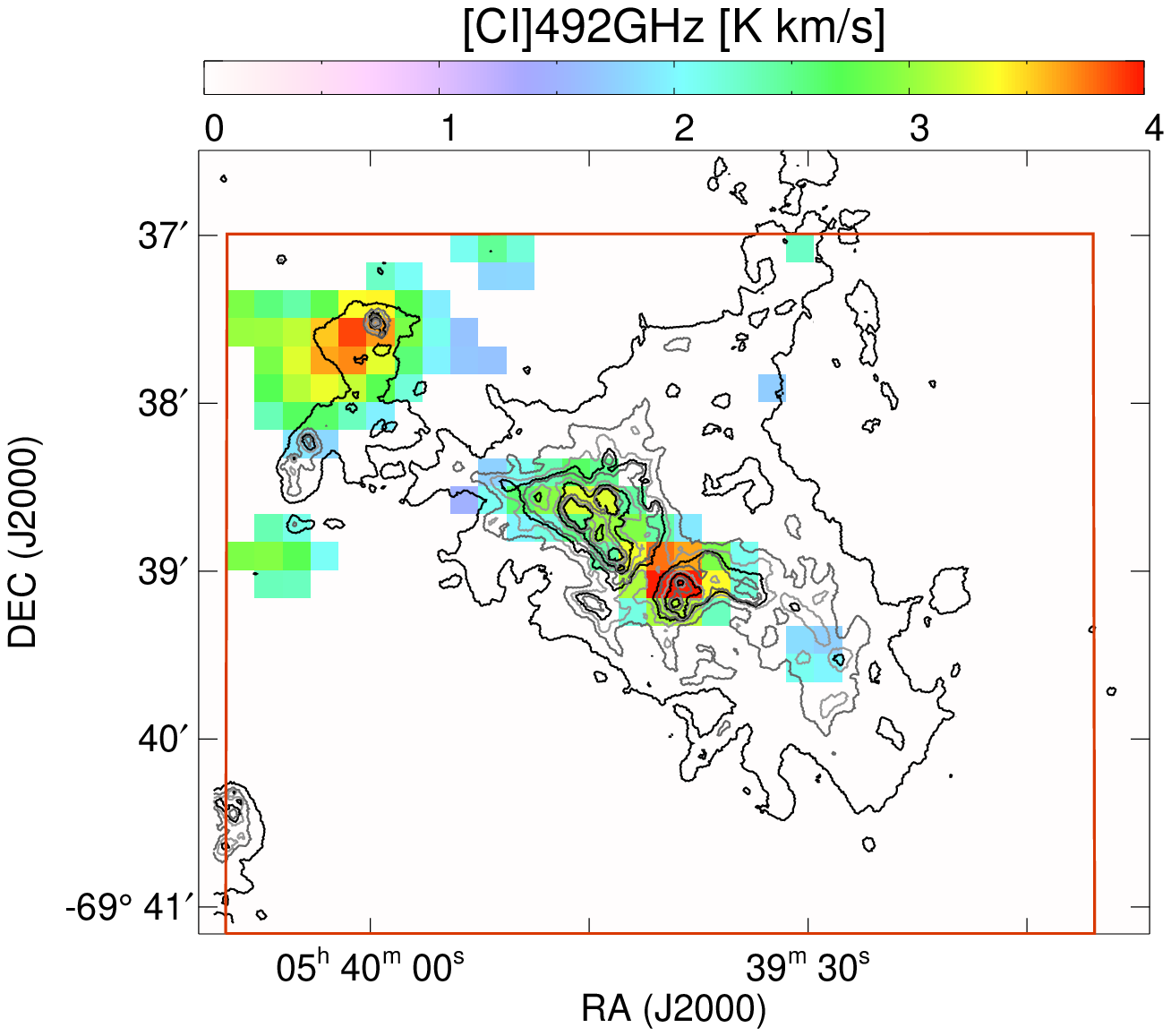}
\includegraphics[bb=60 0 414 340,width=0.47\hsize,clip]{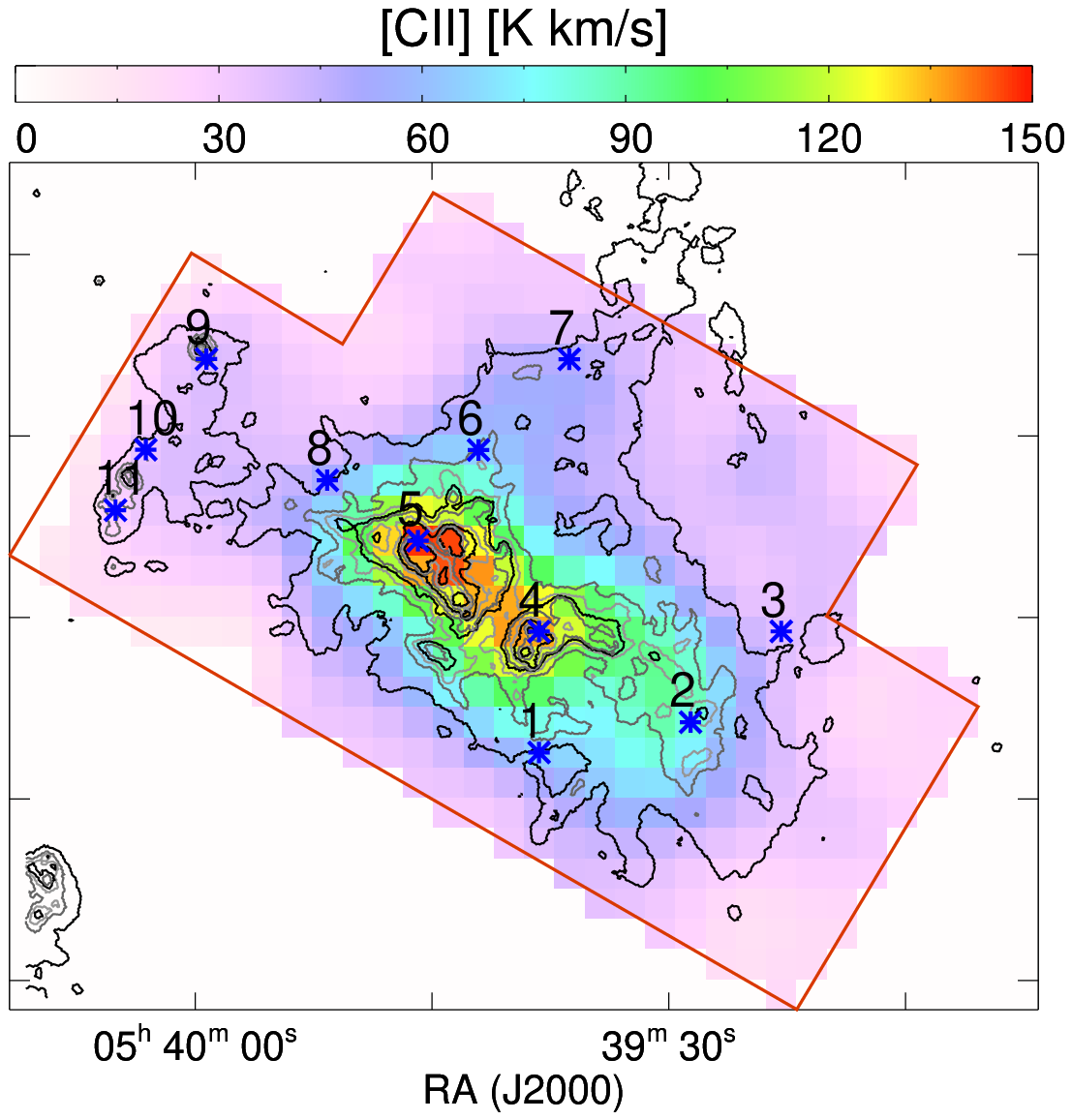}
\caption{\it{(continued)}}
\end{figure*}

\begin{figure*}
\centering
\includegraphics{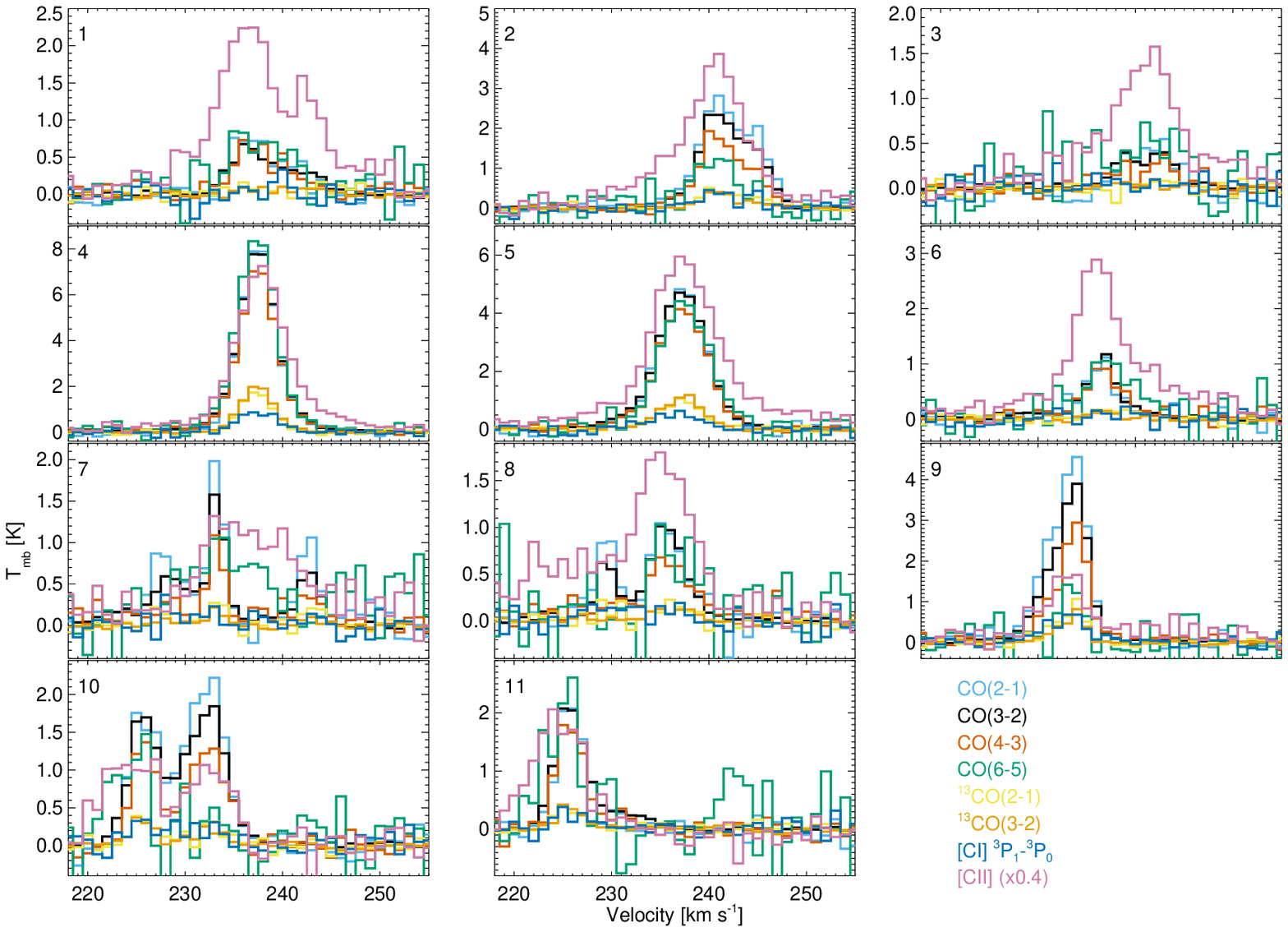}
\caption{Spectra at selected positions in N160, marked in the CO(3-2) and \cii\ panels of Fig.~\ref{figure:integmap_N160}.}
\label{figure:selected_spectra_N160}
\end{figure*}

\begin{figure*}
\centering
\includegraphics[bb=0 20 404 340,width=0.48\hsize,clip]{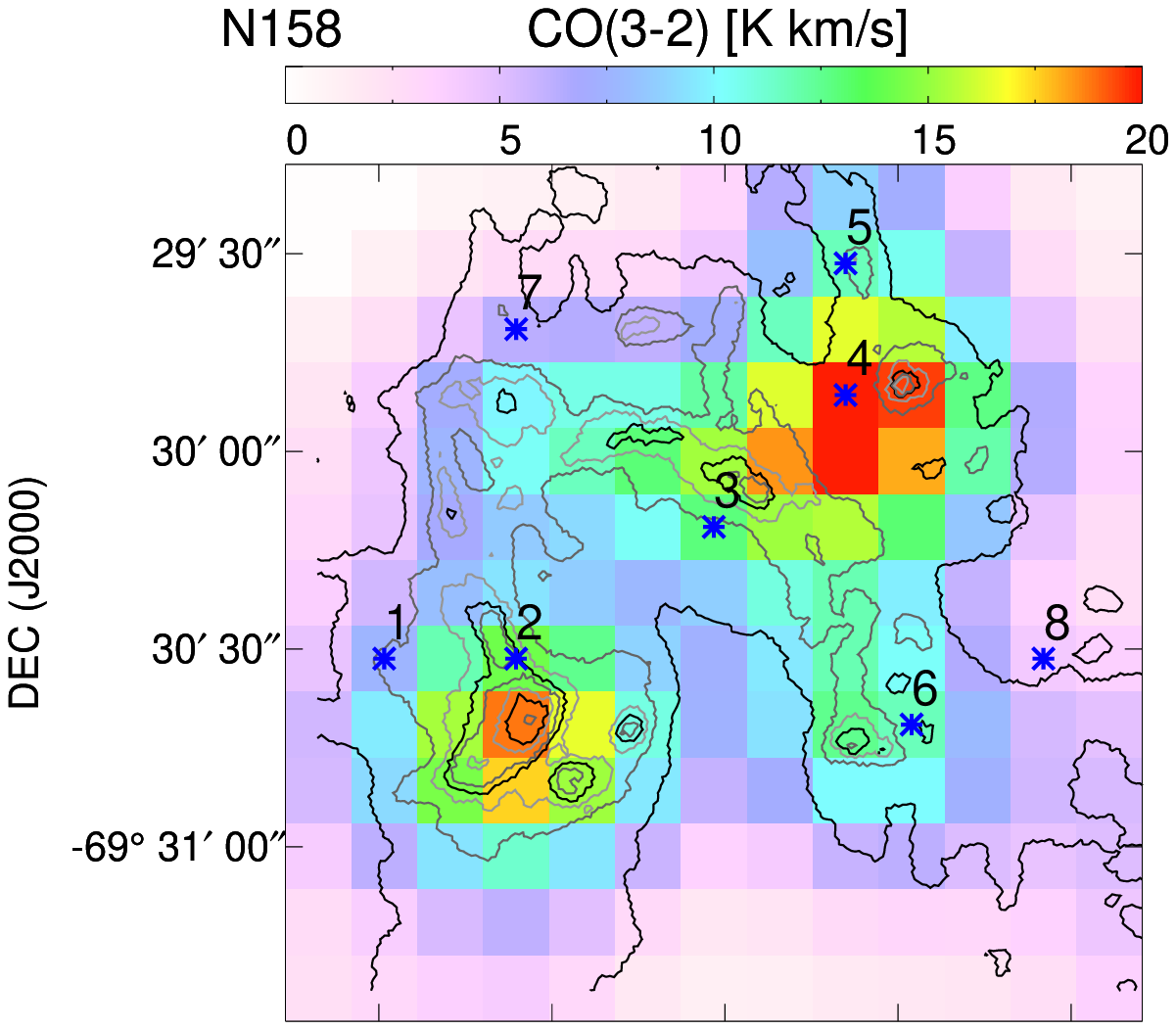}
\includegraphics[bb=70 20 474 340,width=0.48\hsize,clip]{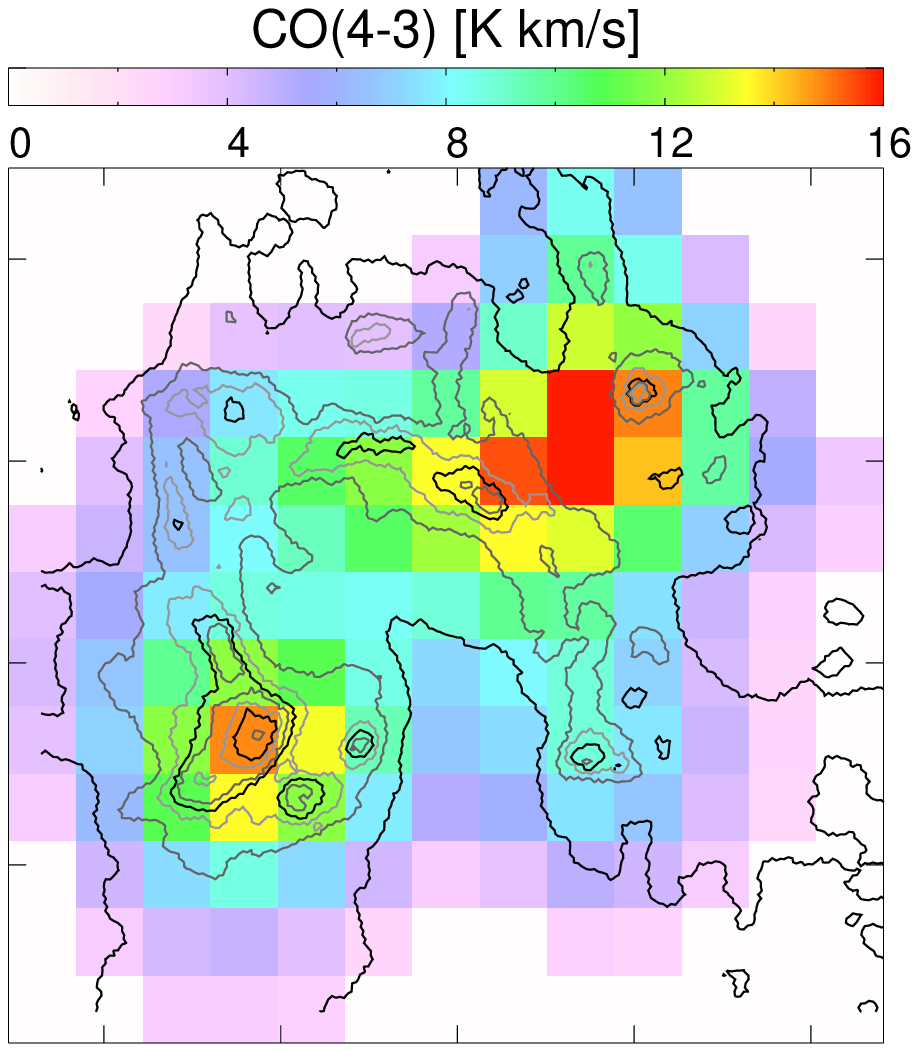}
\includegraphics[bb=0 0 404 340,width=0.48\hsize,clip]{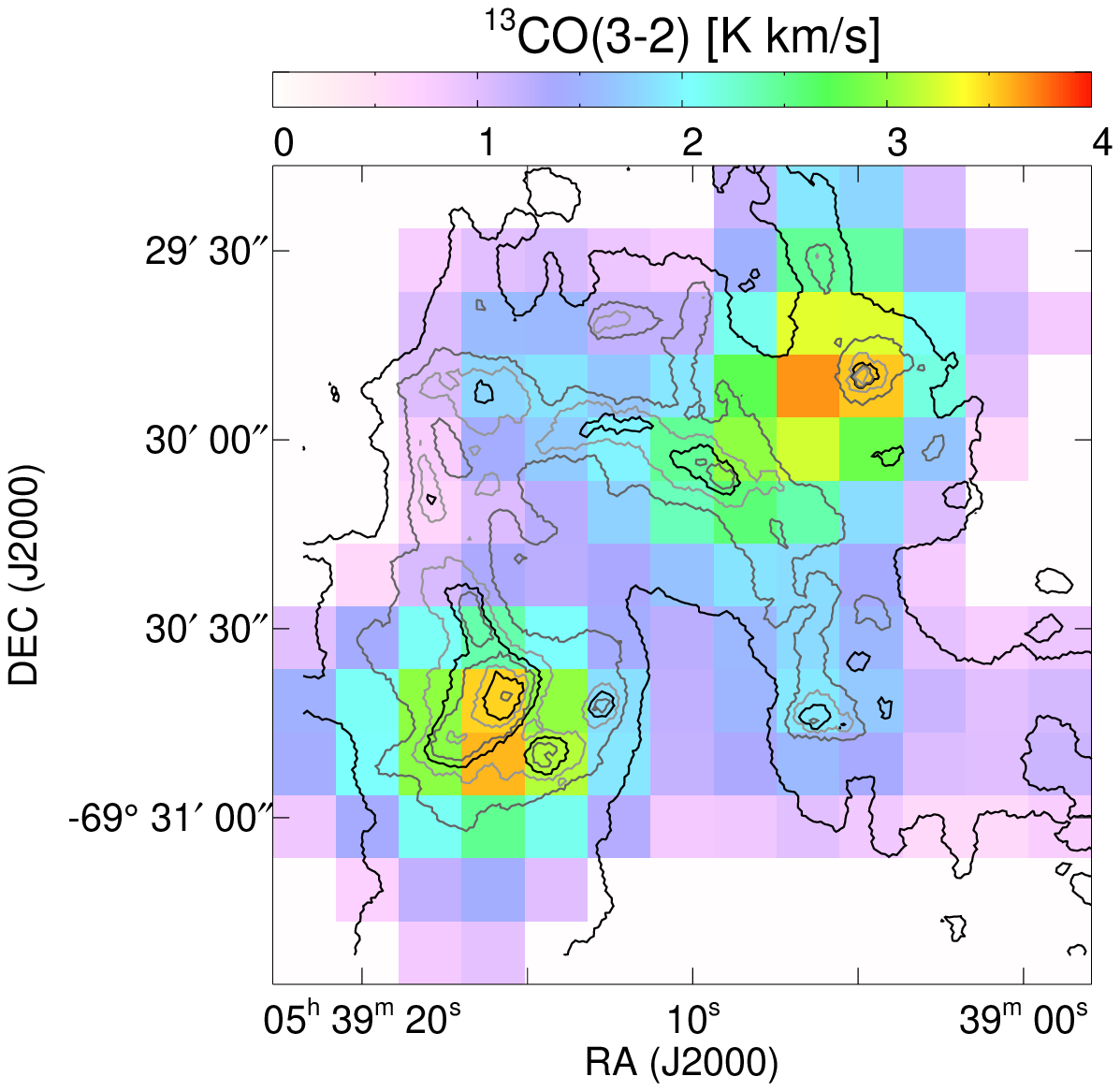}
\includegraphics[bb=70 0 474 340,width=0.48\hsize,clip]{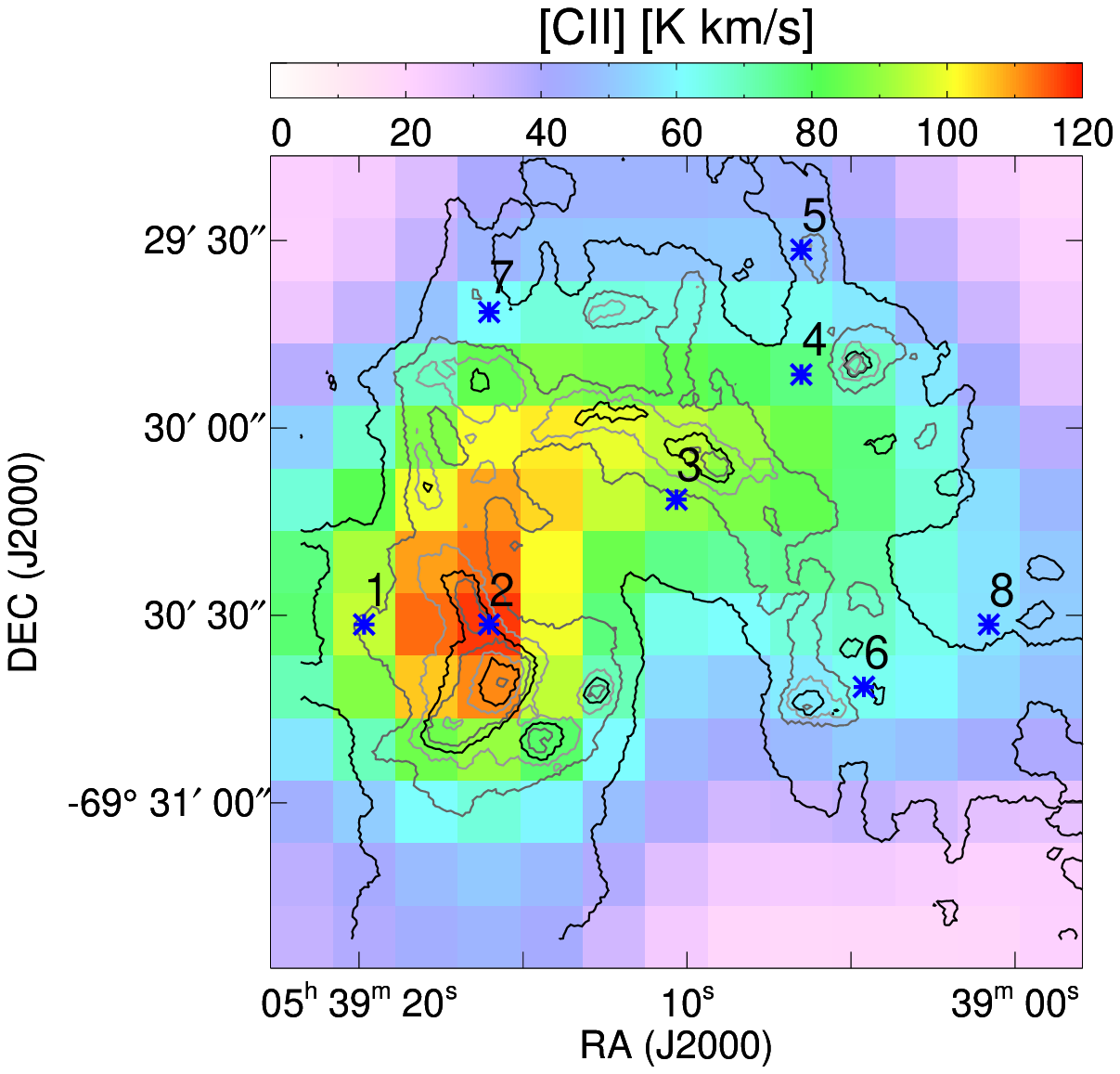}
\caption{Integrated intensity maps (colors, 30\arcsec\ resolution) overlaid with contours of the IRAC 8\um\ emission in N158. The \ci\transl\ is not shown because it is detected only at a few positions around the two CO peaks. Blue asterisks in CO(3-2) and \cii\ maps mark the positions mark the positions where the spectra shown in Fig.~\ref{figure:selected_spectra_N158} have been extracted.}
\label{figure:integmap_N158}
\end{figure*}

\begin{figure*}
\centering
\includegraphics{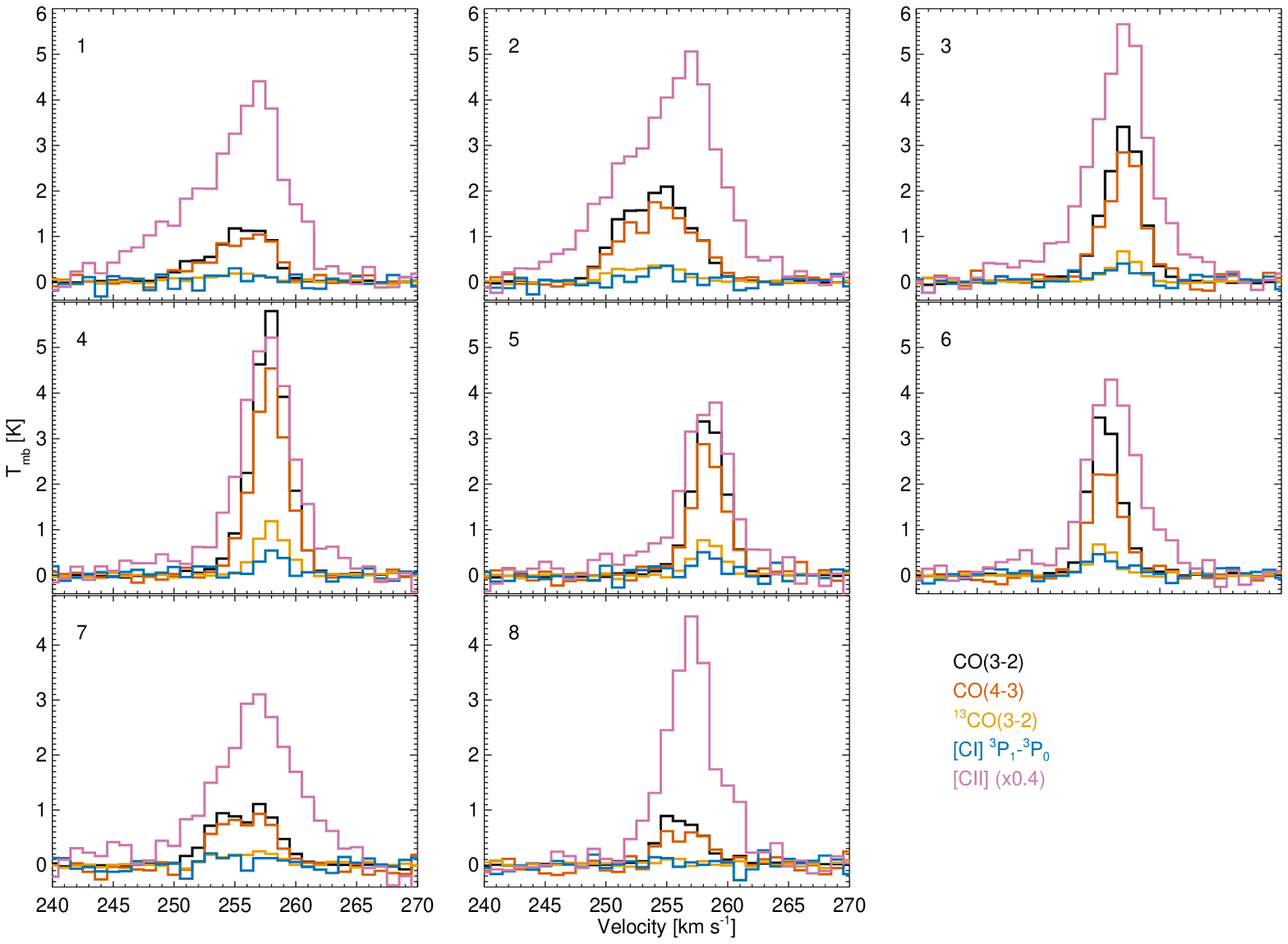}
\caption{Spectra at selected positions in N158, marked in the CO(3-2) and \cii\ panels of Fig.~\ref{figure:integmap_N158}.}
\label{figure:selected_spectra_N158}
\end{figure*}

\begin{figure*}
\centering
\includegraphics[bb=0 20 404 340,width=0.48\hsize,clip]{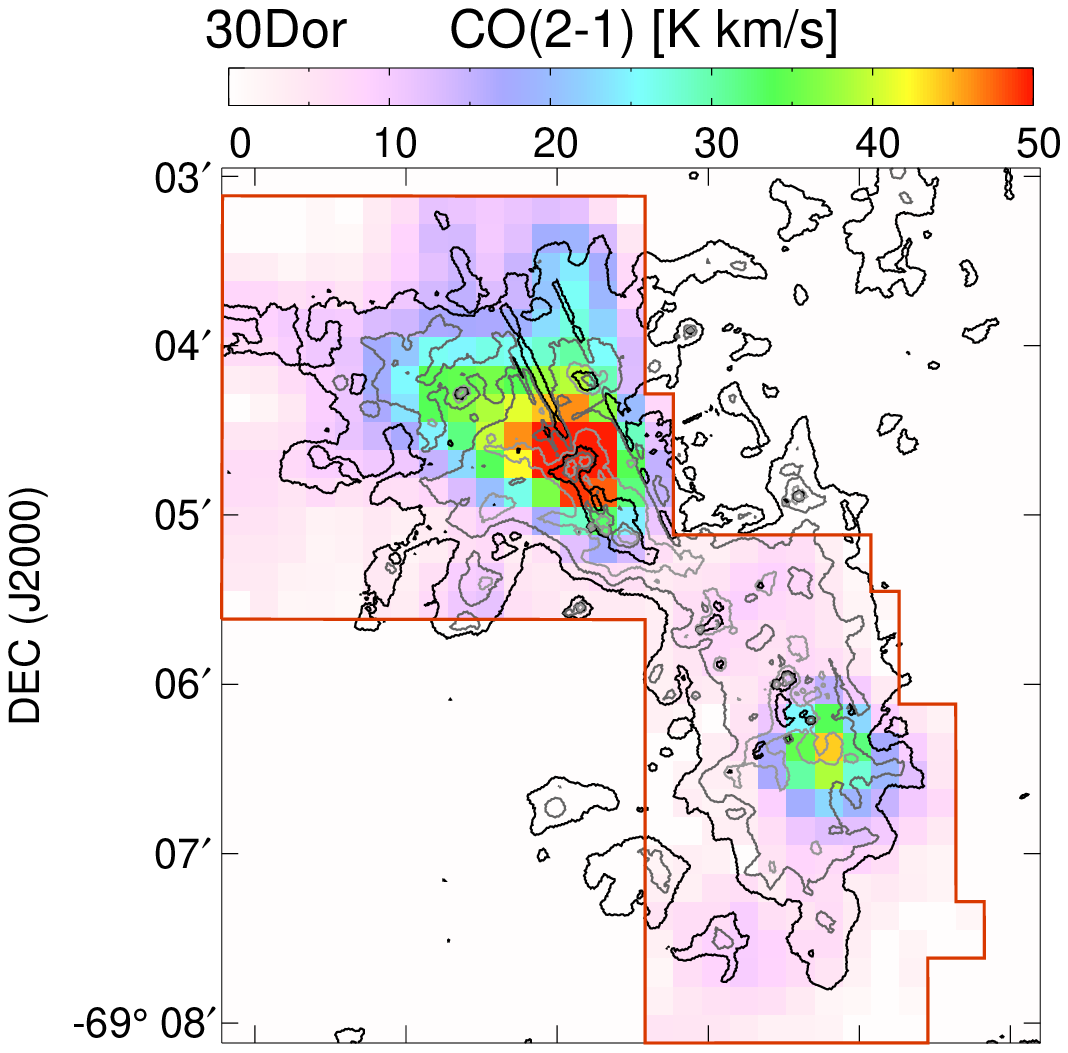}
\includegraphics[bb=70 20 474 340,width=0.48\hsize,clip]{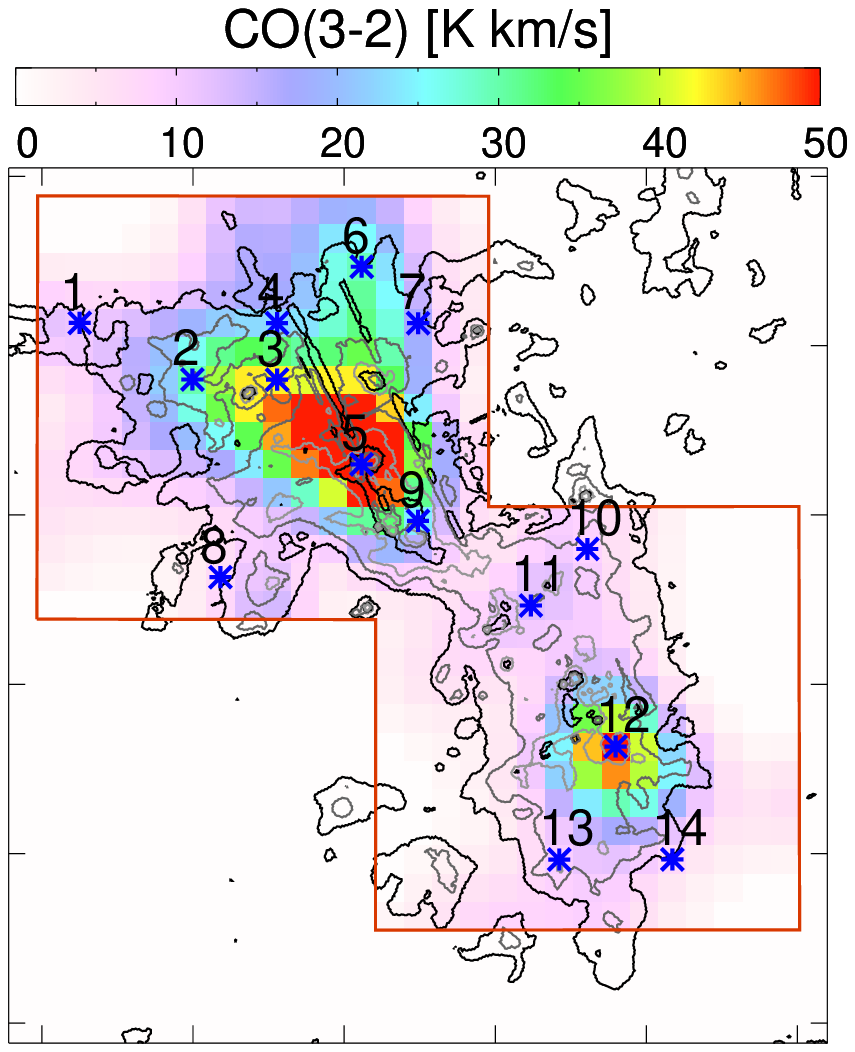}
\includegraphics[bb=0 20 404 340,width=0.48\hsize,clip]{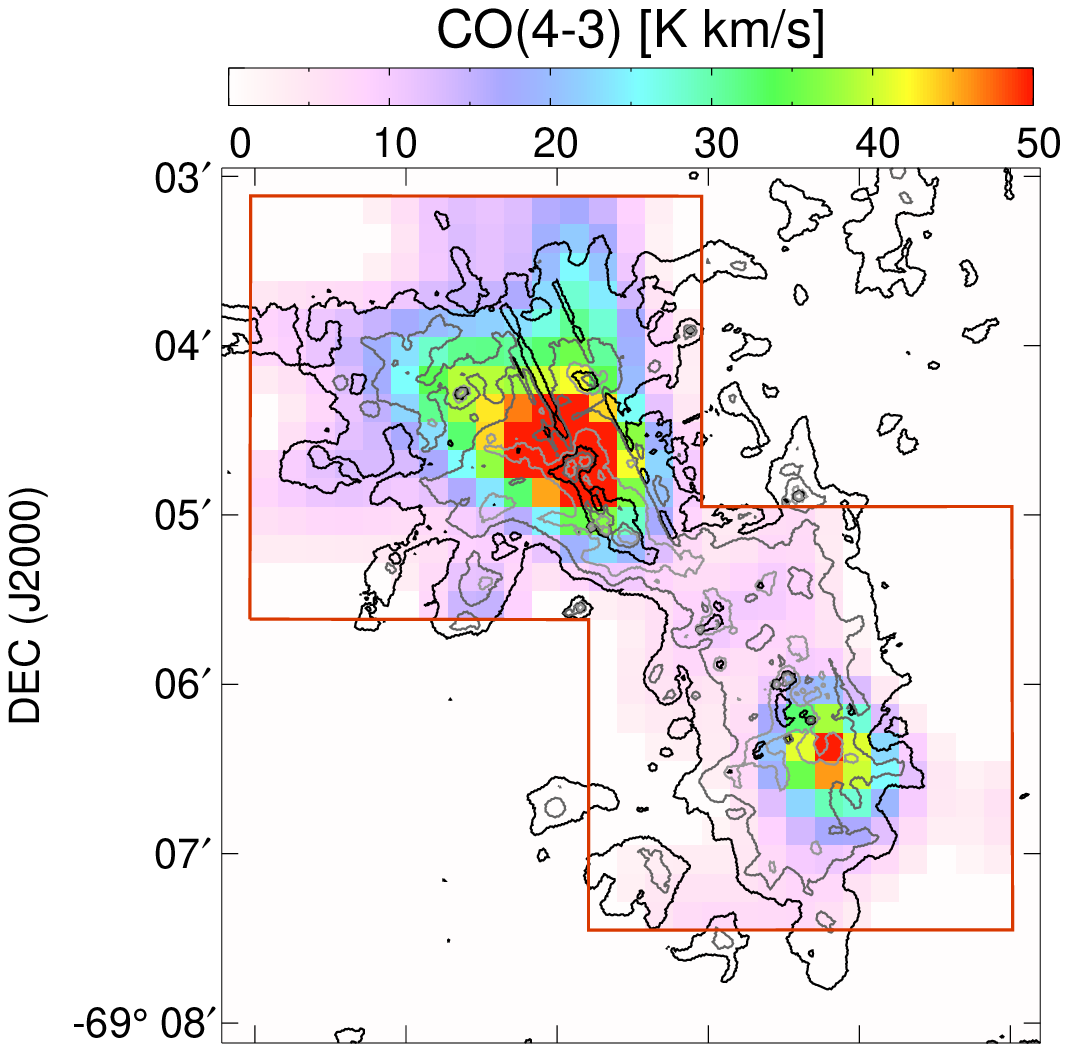}
\includegraphics[bb=70 20 474 340,width=0.48\hsize,clip]{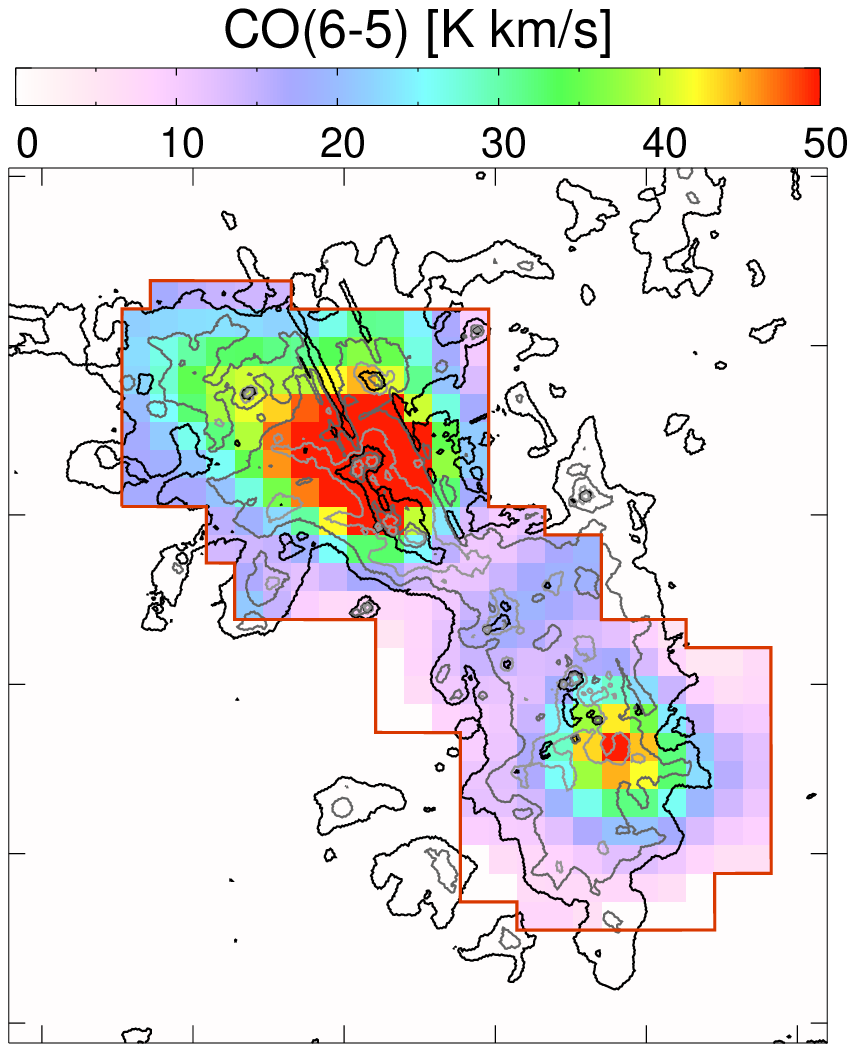}
\includegraphics[bb=0 20 404 340,width=0.48\hsize,clip]{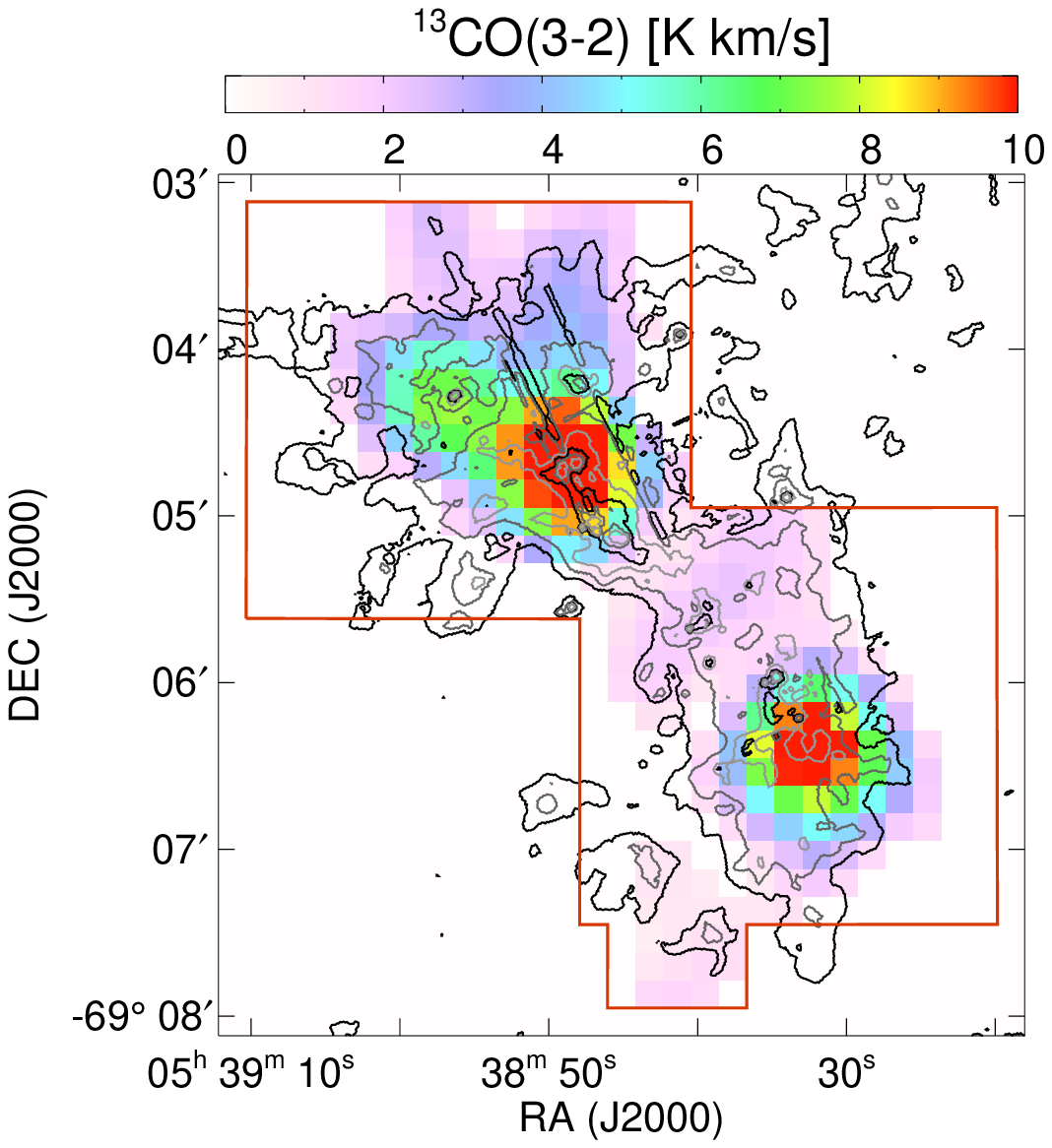}
\includegraphics[bb=70 20 474 340,width=0.48\hsize,clip]{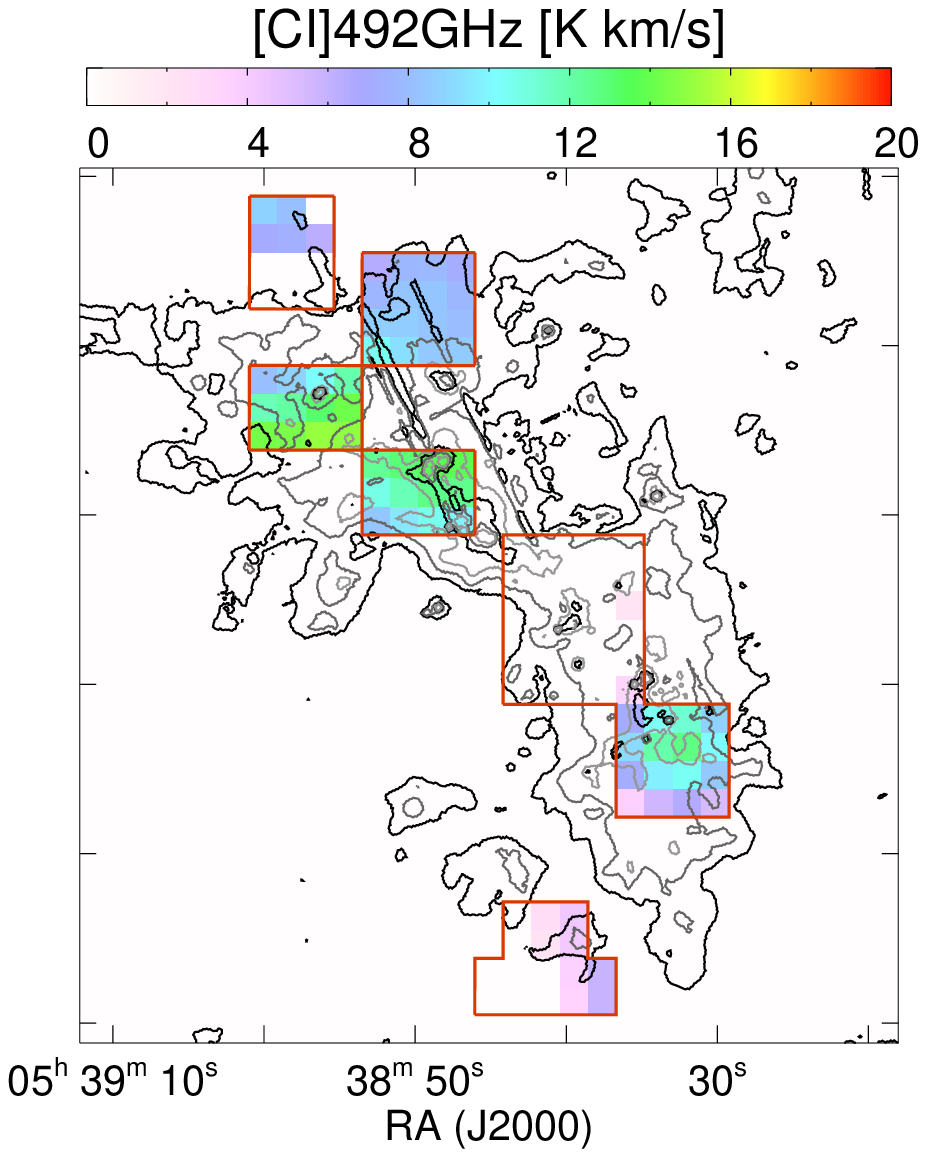}
\caption{Integrated intensity maps (colors, 30\arcsec\ resolution) overlaid with contours of the IRAC 8\um\ emission in 30~Dor. The red lines outline the observed area. Blue asterisks in CO(3-2) and \cii\ maps mark the positions where the spectra shown in Fig.~\ref{figure:selected_spectra_30Dor} have been extracted.}
\label{figure:integmap_30Dor}
\end{figure*}

\addtocounter{figure}{-1}

\begin{figure*}
\centering
\includegraphics[bb=0 20 404 340,width=0.48\hsize,clip]{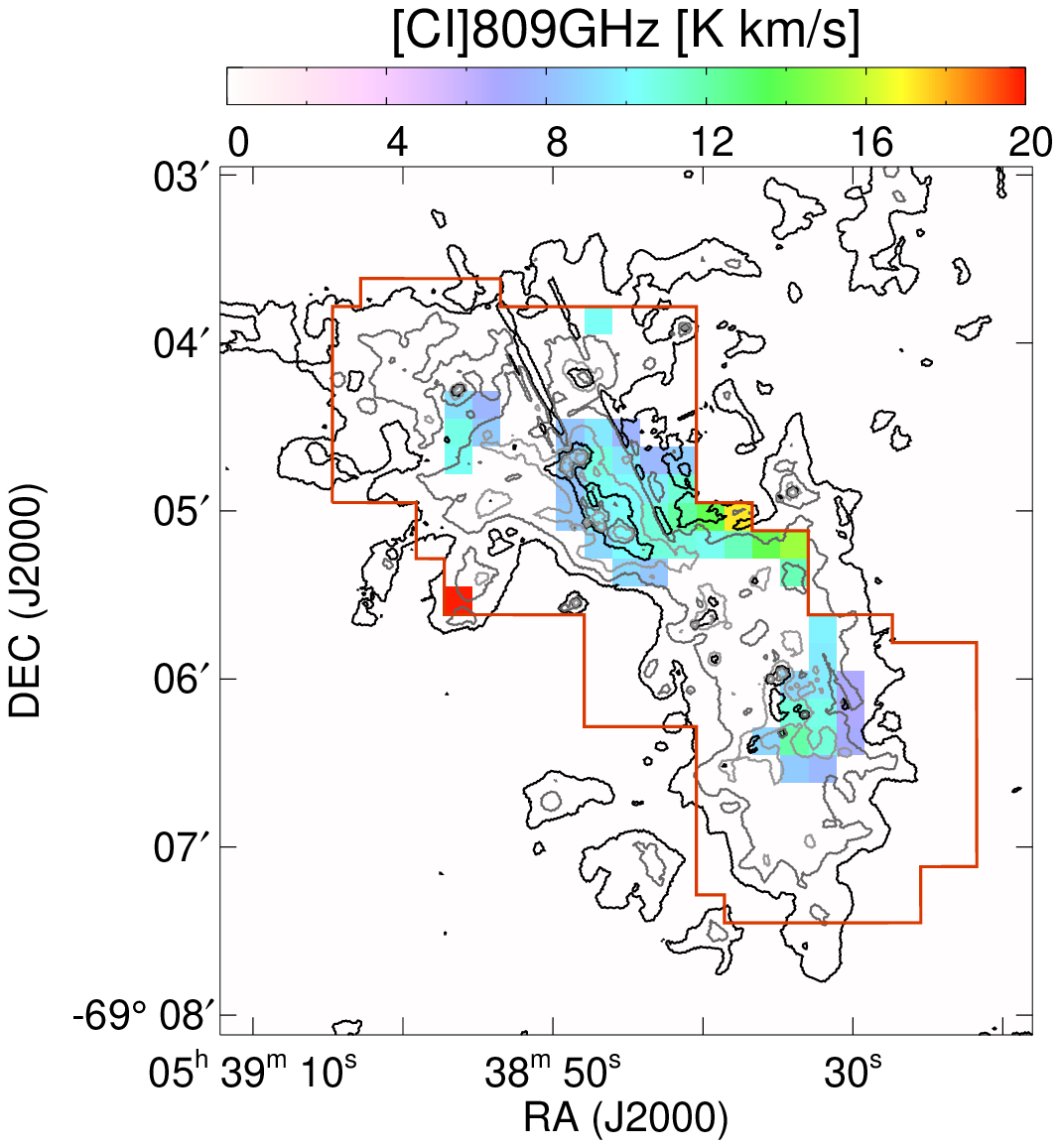}
\includegraphics[bb=70 20 474 340,width=0.48\hsize,clip]{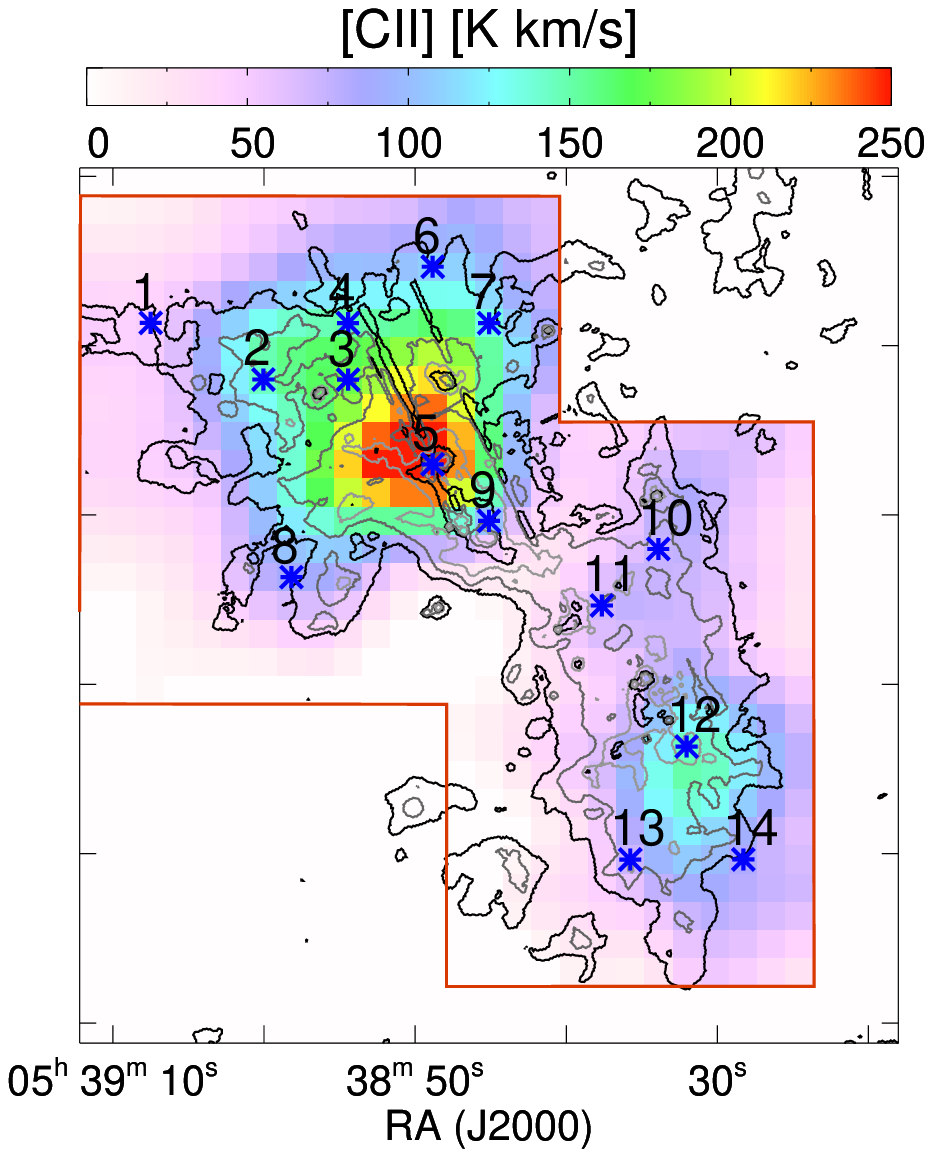}
\caption{\it{(continued)}}
\end{figure*}

\begin{figure*}
\centering
\includegraphics{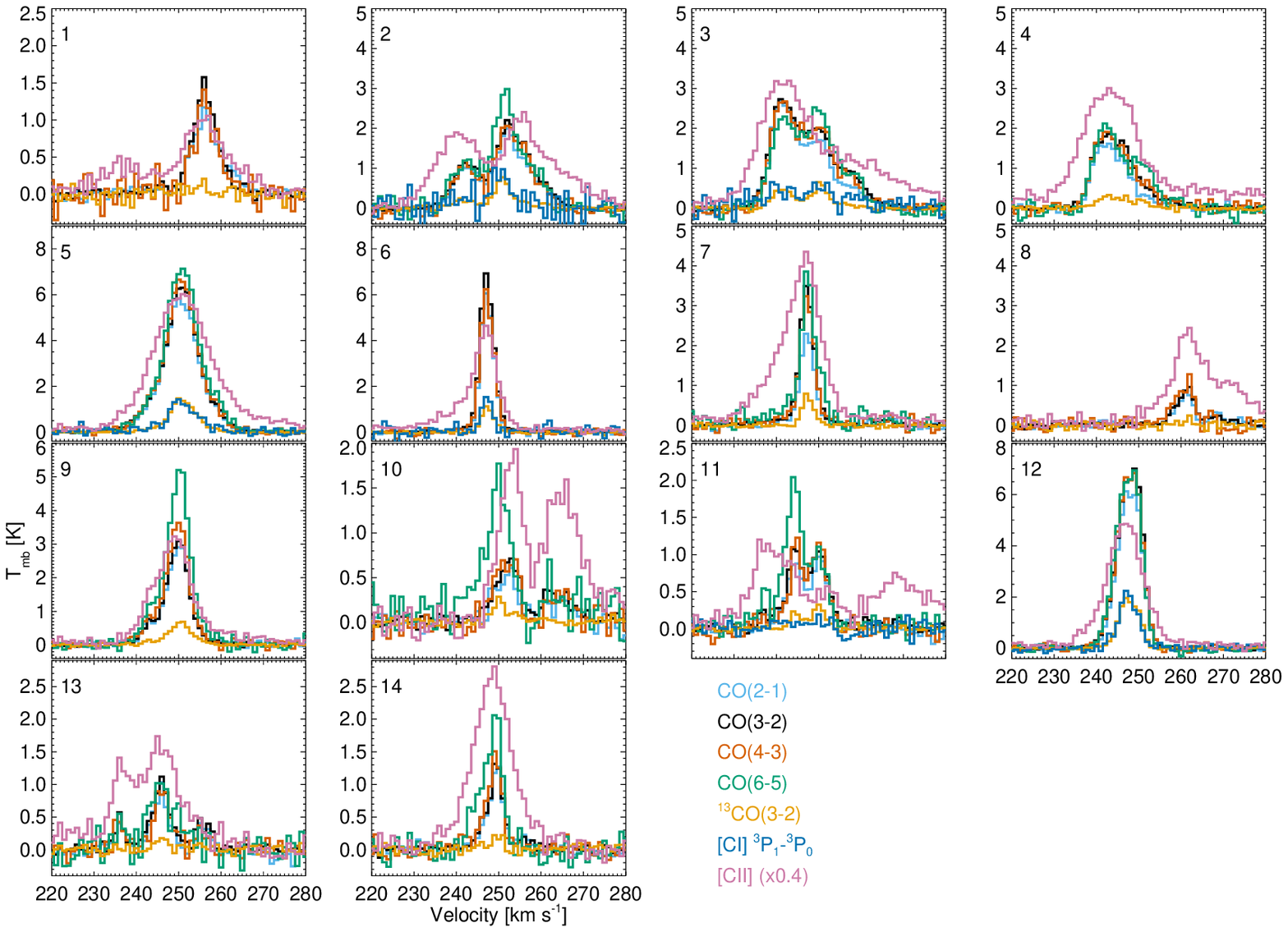}
\caption{Spectra at selected positions in 30~Dor, marked in the CO(3-2) and \cii\ panels of Fig.~\ref{figure:integmap_30Dor}.}
\label{figure:selected_spectra_30Dor}
\end{figure*}

\end{appendix}


\begin{thebibliography}{98}
\expandafter\ifx\csname natexlab\endcsname\relax\def\natexlab#1{#1}\fi

\bibitem[{{Andree-Labsch} {et~al.}(2017){Andree-Labsch}, {Ossenkopf-Okada}, \&
  {R{\"o}llig}}]{Andree-Labsch2017}
{Andree-Labsch}, S., {Ossenkopf-Okada}, V., \& {R{\"o}llig}, M. 2017, \aap,
  598, A2

\bibitem[{{Bisbas} {et~al.}(2018){Bisbas}, {Tan}, {Csengeri}, {Wu}, {Lim},
  {Caselli}, {G{\"u}sten}, {Ricken}, \& {Riquelme}}]{Bisbas2018}
{Bisbas}, T.~G., {Tan}, J.~C., {Csengeri}, T., {et~al.} 2018, \mnras, 478, L54

\bibitem[{{Bolatto} {et~al.}(1999){Bolatto}, {Jackson}, \&
  {Ingalls}}]{Bolatto1999}
{Bolatto}, A.~D., {Jackson}, J.~M., \& {Ingalls}, J.~G. 1999, \apj, 513, 275

\bibitem[{{Bolatto} {et~al.}(2000{\natexlab{a}}){Bolatto}, {Jackson}, {Israel},
  {Zhang}, \& {Kim}}]{Bolatto2000b}
{Bolatto}, A.~D., {Jackson}, J.~M., {Israel}, F.~P., {Zhang}, X., \& {Kim}, S.
  2000{\natexlab{a}}, \apj, 545, 234

\bibitem[{{Bolatto} {et~al.}(2000{\natexlab{b}}){Bolatto}, {Jackson},
  {Kraemer}, \& {Zhang}}]{Bolatto2000a}
{Bolatto}, A.~D., {Jackson}, J.~M., {Kraemer}, K.~E., \& {Zhang}, X.
  2000{\natexlab{b}}, \apjl, 541, L17

\bibitem[{{Bolatto} {et~al.}(2013){Bolatto}, {Wolfire}, \&
  {Leroy}}]{Bolatto2013}
{Bolatto}, A.~D., {Wolfire}, M., \& {Leroy}, A.~K. 2013, \araa, 51, 207

\bibitem[{{Braun}(2012)}]{Braun2012}
{Braun}, R. 2012, \apj, 749, 87

\bibitem[{{Braun} {et~al.}(2009){Braun}, {Thilker}, {Walterbos}, \&
  {Corbelli}}]{Braun2009}
{Braun}, R., {Thilker}, D.~A., {Walterbos}, R.~A.~M., \& {Corbelli}, E. 2009,
  \apj, 695, 937

\bibitem[{{Bron}(2014)}]{Bron2014}
{Bron}, E. 2014, PhD thesis, LERMA, Observatoire de Paris, PSL Research
  University, CNRS, Sorbonne Universit{\'e}s, UPMC Univ.~Paris 06, F-92190,
  Meudon, France <EMAIL>emeric.bron@obspm.fr</EMAIL>

\bibitem[{{Carlhoff}(2013)}]{Carlhoff2013}
{Carlhoff}, P. 2013, PhD thesis, Universit\"{a}t zu K\"{o}ln

\bibitem[{{Chen} {et~al.}(2010){Chen}, {Indebetouw}, {Chu}, {Gruendl},
  {Testor}, {Heitsch}, {Seale}, {Meixner}, \& {Sewilo}}]{Chen2010}
{Chen}, C.-H.~R., {Indebetouw}, R., {Chu}, Y.-H., {et~al.} 2010, \apj, 721,
  1206

\bibitem[{{Chevance} {et~al.}(2016){Chevance}, {Madden}, {Lebouteiller},
  {Godard}, {Cormier}, {Galliano}, {Hony}, {Indebetouw}, {Le Bourlot}, {Lee},
  {Le Petit}, {Pellegrini}, {Roueff}, \& {Wu}}]{Chevance2016}
{Chevance}, M., {Madden}, S.~C., {Lebouteiller}, V., {et~al.} 2016, \aap, 590,
  A36

\bibitem[{{Cormier} {et~al.}(2015){Cormier}, {Madden}, {Lebouteiller}, {Abel},
  {Hony}, {Galliano}, {R{\'e}my-Ruyer}, {Bigiel}, {Baes}, {Boselli},
  {Chevance}, {Cooray}, {De Looze}, {Doublier}, {Galametz}, {Hughes},
  {Karczewski}, {Lee}, {Lu}, \& {Spinoglio}}]{Cormier2015}
{Cormier}, D., {Madden}, S.~C., {Lebouteiller}, V., {et~al.} 2015, \aap, 578,
  A53

\bibitem[{{Crowther} \& {Dessart}(1998)}]{CrowtherDessart1998}
{Crowther}, P.~A. \& {Dessart}, L. 1998, \mnras, 296, 622

\bibitem[{{Cubick} {et~al.}(2008){Cubick}, {Stutzki}, {Ossenkopf}, {Kramer}, \&
  {R{\"o}llig}}]{Cubick2008}
{Cubick}, M., {Stutzki}, J., {Ossenkopf}, V., {Kramer}, C., \& {R{\"o}llig}, M.
  2008, \aap, 488, 623

\bibitem[{{de Boer} {et~al.}(1998){de Boer}, {Braun}, {Vallenari}, \&
  {Mebold}}]{deBoer1998}
{de Boer}, K.~S., {Braun}, J.~M., {Vallenari}, A., \& {Mebold}, U. 1998, \aap,
  329, L49

\bibitem[{{Dedes} {et~al.}(2010){Dedes}, {R{\"o}llig}, {Mookerjea}, {Okada},
  {Ossenkopf}, {Bruderer}, {Benz}, {Melchior}, {Kramer}, {Gerin}, {G{\"u}sten},
  {Akyilmaz}, {Berne}, {Boulanger}, {de Lange}, {Dubbeldam}, {France},
  {Fuente}, {Goicoechea}, {Harris}, {Huisman}, {Jellema}, {Joblin}, {Klein},
  {Le Petit}, {Lord}, {Martin}, {Martin-Pintado}, {Neufeld}, {Philipp},
  {Phillips}, {Pilleri}, {Rizzo}, {Salez}, {Schieder}, {Simon}, {Siebertz},
  {Stutzki}, {van der Tak}, {Teyssier}, \& {Yorke}}]{Dedes2010}
{Dedes}, C., {R{\"o}llig}, M., {Mookerjea}, B., {et~al.} 2010, \aap, 521, L24

\bibitem[{{Draine}(2011)}]{Draine2011a}
{Draine}, B.~T. 2011, {Physics of the Interstellar and Intergalactic Medium}

\bibitem[{{Fahrion} {et~al.}(2017){Fahrion}, {Cormier}, {Bigiel}, {Hony},
  {Abel}, {Cigan}, {Csengeri}, {Graf}, {Lebouteiller}, {Madden}, {Wu}, \&
  {Young}}]{Fahrion2017}
{Fahrion}, K., {Cormier}, D., {Bigiel}, F., {et~al.} 2017, \aap, 599, A9

\bibitem[{{Fari{\~n}a} {et~al.}(2009){Fari{\~n}a}, {Bosch}, {Morrell},
  {Barb{\'a}}, \& {Walborn}}]{Farina2009}
{Fari{\~n}a}, C., {Bosch}, G.~L., {Morrell}, N.~I., {Barb{\'a}}, R.~H., \&
  {Walborn}, N.~R. 2009, \aj, 138, 510

\bibitem[{{Fleener} {et~al.}(2010){Fleener}, {Payne}, {Chu}, {Chen}, \&
  {Gruendl}}]{Fleener2010}
{Fleener}, C.~E., {Payne}, J.~T., {Chu}, Y.-H., {Chen}, C.-H.~R., \& {Gruendl},
  R.~A. 2010, \aj, 139, 158

\bibitem[{{Fukui} {et~al.}(2015){Fukui}, {Harada}, {Tokuda}, {Morioka},
  {Onishi}, {Torii}, {Ohama}, {Hattori}, {Nayak}, {Meixner}, {Sewi{\l}o},
  {Indebetouw}, {Kawamura}, {Saigo}, {Yamamoto}, {Tachihara}, {Minamidani},
  {Inoue}, {Madden}, {Galametz}, {Lebouteiller}, {Mizuno}, \&
  {Chen}}]{Fukui2015}
{Fukui}, Y., {Harada}, R., {Tokuda}, K., {et~al.} 2015, \apjl, 807, L4

\bibitem[{{Galametz} {et~al.}(2013){Galametz}, {Hony}, {Galliano}, {Madden},
  {Albrecht}, {Bot}, {Cormier}, {Engelbracht}, {Fukui}, {Israel}, {Kawamura},
  {Lebouteiller}, {Li}, {Meixner}, {Misselt}, {Montiel}, {Okumura}, {Panuzzo},
  {Roman-Duval}, {Rubio}, {Sauvage}, {Seale}, {Sewi{\l}o}, \& {van
  Loon}}]{Galametz2013}
{Galametz}, M., {Hony}, S., {Galliano}, F., {et~al.} 2013, \mnras, 431, 1596

\bibitem[{{Garnett}(1999)}]{Garnett1999}
{Garnett}, D.~R. 1999, in IAU Symposium, Vol. 190, New Views of the Magellanic
  Clouds, ed. Y.-H. {Chu}, N.~{Suntzeff}, J.~{Hesser}, \& D.~{Bohlender}, 266

\bibitem[{{Gibson} {et~al.}(2005){Gibson}, {Taylor}, {Higgs}, {Brunt}, \&
  {Dewdney}}]{Gibson2005}
{Gibson}, S.~J., {Taylor}, A.~R., {Higgs}, L.~A., {Brunt}, C.~M., \& {Dewdney},
  P.~E. 2005, \apj, 626, 195

\bibitem[{{Goldsmith} {et~al.}(2012){Goldsmith}, {Langer}, {Pineda}, \&
  {Velusamy}}]{Goldsmith2012}
{Goldsmith}, P.~F., {Langer}, W.~D., {Pineda}, J.~L., \& {Velusamy}, T. 2012,
  \apjs, 203, 13

\bibitem[{{Gordon} {et~al.}(2017){Gordon}, {Jones}, {Gehrz}, \&
  {Helton}}]{Gordon2017}
{Gordon}, M.~S., {Jones}, T.~J., {Gehrz}, R.~D., \& {Helton}, L.~A. 2017, \apj,
  834, 122

\bibitem[{{Guan} {et~al.}(2012){Guan}, {Stutzki}, {Graf}, {G{\"u}sten},
  {Okada}, {Requena-Torres}, {Simon}, \& {Wiesemeyer}}]{Guan2012}
{Guan}, X., {Stutzki}, J., {Graf}, U.~U., {et~al.} 2012, \aap, 542, L4

\bibitem[{{G{\"u}sten} {et~al.}(2006){G{\"u}sten}, {Nyman}, {Schilke},
  {Menten}, {Cesarsky}, \& {Booth}}]{Guesten2006}
{G{\"u}sten}, R., {Nyman}, L.~{\AA}., {Schilke}, P., {et~al.} 2006, \aap, 454,
  L13

\bibitem[{{Heikkil{\"a}} {et~al.}(1999){Heikkil{\"a}}, {Johansson}, \&
  {Olofsson}}]{Heikkila1999}
{Heikkil{\"a}}, A., {Johansson}, L.~E.~B., \& {Olofsson}, H. 1999, \aap, 344,
  817

\bibitem[{{Henize}(1956)}]{Henize1956}
{Henize}, K.~G. 1956, \apjs, 2, 315

\bibitem[{{Heydari-Malayeri} {et~al.}(2002){Heydari-Malayeri}, {Charmandaris},
  {Deharveng}, {Meynadier}, {Rosa}, {Schaerer}, \&
  {Zinnecker}}]{HeydariMalayeri2002}
{Heydari-Malayeri}, M., {Charmandaris}, V., {Deharveng}, L., {et~al.} 2002,
  \aap, 381, 941

\bibitem[{{Heydari-Malayeri} \& {Testor}(1986)}]{HeydariMalayeriTestor1986}
{Heydari-Malayeri}, M. \& {Testor}, G. 1986, \aap, 162, 180

\bibitem[{{Heyminck} {et~al.}(2012){Heyminck}, {Graf}, {G{\"u}sten}, {Stutzki},
  {H{\"u}bers}, \& {Hartogh}}]{Heyminck2012}
{Heyminck}, S., {Graf}, U.~U., {G{\"u}sten}, R., {et~al.} 2012, \aap, 542, L1

\bibitem[{{Indebetouw} {et~al.}(2013){Indebetouw}, {Brogan}, {Chen}, {Leroy},
  {Johnson}, {Muller}, {Madden}, {Cormier}, {Galliano}, {Hughes}, {Hunter},
  {Kawamura}, {Kepley}, {Lebouteiller}, {Meixner}, {Oliveira}, {Onishi}, \&
  {Vasyunina}}]{Indebetouw2013}
{Indebetouw}, R., {Brogan}, C., {Chen}, C.-H.~R., {et~al.} 2013, \apj, 774, 73

\bibitem[{{Indebetouw} {et~al.}(2004){Indebetouw}, {Johnson}, \&
  {Conti}}]{Indebetouw2004}
{Indebetouw}, R., {Johnson}, K.~E., \& {Conti}, P. 2004, \aj, 128, 2206

\bibitem[{{Israel} \& {Maloney}(2011)}]{Israel2011}
{Israel}, F.~P. \& {Maloney}, P.~R. 2011, \aap, 531, A19

\bibitem[{{Israel} {et~al.}(1996){Israel}, {Maloney}, {Geis}, {Herrmann},
  {Madden}, {Poglitsch}, \& {Stacey}}]{Israel1996}
{Israel}, F.~P., {Maloney}, P.~R., {Geis}, N., {et~al.} 1996, \apj, 465, 738

\bibitem[{{Johansson} {et~al.}(1998){Johansson}, {Greve}, {Booth}, {Boulanger},
  {Garay}, {de Graauw}, {Israel}, {Kutner}, {Lequeux}, {Murphy}, {Nyman}, \&
  {Rubio}}]{Johansson1998}
{Johansson}, L.~E.~B., {Greve}, A., {Booth}, R.~S., {et~al.} 1998, \aap, 331,
  857

\bibitem[{{Kalari} {et~al.}(2018){Kalari}, {Rubio}, {Elmegreen}, {Guzm{\'a}n},
  {Zinnecker}, \& {Herrera}}]{Kalari2018}
{Kalari}, V.~M., {Rubio}, M., {Elmegreen}, B.~G., {et~al.} 2018, \apj, 852, 71

\bibitem[{{Kasemann} {et~al.}(2006){Kasemann}, {G{\"u}sten}, {Heyminck},
  {Klein}, {Klein}, {Philipp}, {Korn}, {Schneider}, {Henseler}, {Baryshev}, \&
  {Klapwijk}}]{Kasemann2006}
{Kasemann}, C., {G{\"u}sten}, R., {Heyminck}, S., {et~al.} 2006, in Society of
  Photo-Optical Instrumentation Engineers (SPIE) Conference Series, Vol. 6275,
  Society of Photo-Optical Instrumentation Engineers (SPIE) Conference Series

\bibitem[{{Kavars} {et~al.}(2005){Kavars}, {Dickey}, {McClure-Griffiths},
  {Gaensler}, \& {Green}}]{Kavars2005}
{Kavars}, D.~W., {Dickey}, J.~M., {McClure-Griffiths}, N.~M., {Gaensler},
  B.~M., \& {Green}, A.~J. 2005, \apj, 626, 887

\bibitem[{{Kawada} {et~al.}(2011){Kawada}, {Takahashi}, {Yasuda}, {Kiriyama},
  {Mori}, {Mouri}, {Kaneda}, {Okada}, {Takahashi}, \& {Murakami}}]{Kawada2011}
{Kawada}, M., {Takahashi}, A., {Yasuda}, A., {et~al.} 2011, \pasj, 63, 903

\bibitem[{{Kim} {et~al.}(2003){Kim}, {Staveley-Smith}, {Dopita}, {Sault},
  {Freeman}, {Lee}, \& {Chu}}]{Kim2003}
{Kim}, S., {Staveley-Smith}, L., {Dopita}, M.~A., {et~al.} 2003, \apjs, 148,
  473

\bibitem[{{Klaassen} {et~al.}(2005){Klaassen}, {Plume}, {Gibson}, {Taylor}, \&
  {Brunt}}]{Klaassen2005}
{Klaassen}, P.~D., {Plume}, R., {Gibson}, S.~J., {Taylor}, A.~R., \& {Brunt},
  C.~M. 2005, \apj, 631, 1001

\bibitem[{{Klein} {et~al.}(2014){Klein}, {Ciechanowicz}, {Leinz}, {Heyminck},
  {G{\"u}sten}, {Kasemann}, {Wunsch}, {Maier}, \& {Sekimoto}}]{Klein2014}
{Klein}, T., {Ciechanowicz}, M., {Leinz}, C., {et~al.} 2014, Terahertz Science
  and Technology, IEEE Transactions on, 4, 588

\bibitem[{{Lazendic} {et~al.}(2002){Lazendic}, {Whiteoak}, {Klamer},
  {Harbison}, \& {Kuiper}}]{Lazendic2002}
{Lazendic}, J.~S., {Whiteoak}, J.~B., {Klamer}, I., {Harbison}, P.~D., \&
  {Kuiper}, T.~B.~H. 2002, \mnras, 331, 969

\bibitem[{{Lee} {et~al.}(2016){Lee}, {Madden}, {Lebouteiller}, {Gusdorf},
  {Godard}, {Wu}, {Galametz}, {Cormier}, {Le Petit}, {Roueff}, {Bron},
  {Carlson}, {Chevance}, {Fukui}, {Galliano}, {Hony}, {Hughes}, {Indebetouw},
  {Israel}, {Kawamura}, {Le Bourlot}, {Lesaffre}, {Meixner}, {Muller}, {Nayak},
  {Onishi}, {Roman-Duval}, \& {Sewi{\l}o}}]{Lee2016}
{Lee}, M.-Y., {Madden}, S.~C., {Lebouteiller}, V., {et~al.} 2016, \aap, 596,
  A85

\bibitem[{{Lefloch} \& {Lazareff}(1994)}]{Lefloch1994}
{Lefloch}, B. \& {Lazareff}, B. 1994, \aap, 289, 559

\bibitem[{{Leurini} {et~al.}(2015){Leurini}, {Wyrowski}, {Wiesemeyer},
  {Gusdorf}, {G{\"u}sten}, {Menten}, {Gerin}, {Levrier}, {H{\"u}bers},
  {Jacobs}, {Ricken}, \& {Richter}}]{Leurini2015}
{Leurini}, S., {Wyrowski}, F., {Wiesemeyer}, H., {et~al.} 2015, \aap, 584, A70

\bibitem[{{Li} \& {Goldsmith}(2003)}]{LiGoldsmith2003}
{Li}, D. \& {Goldsmith}, P.~F. 2003, \apj, 585, 823

\bibitem[{{Lucke} \& {Hodge}(1970)}]{LuckeHodge1970}
{Lucke}, P.~B. \& {Hodge}, P.~W. 1970, \aj, 75, 171

\bibitem[{{Madden} {et~al.}(2011){Madden}, {Galametz}, {Cormier},
  {Lebouteiller}, {Galliano}, {Hony}, {R{\'e}my}, {Sauvage}, {Contursi},
  {Sturm}, {Poglitsch}, {Pohlen}, {Smith}, {Bendo}, \&
  {O'Halloran}}]{Madden2011}
{Madden}, S.~C., {Galametz}, M., {Cormier}, D., {et~al.} 2011, in EAS
  Publications Series, Vol.~52, EAS Publications Series, ed. M.~{R{\"o}llig},
  R.~{Simon}, V.~{Ossenkopf}, \& J.~{Stutzki}, 95--101

\bibitem[{{Madden} {et~al.}(1997){Madden}, {Poglitsch}, {Geis}, {Stacey}, \&
  {Townes}}]{Madden1997}
{Madden}, S.~C., {Poglitsch}, A., {Geis}, N., {Stacey}, G.~J., \& {Townes},
  C.~H. 1997, \apj, 483, 200

\bibitem[{{Mart{\'{\i}}n-Hern{\'a}ndez}
  {et~al.}(2005){Mart{\'{\i}}n-Hern{\'a}ndez}, {Vermeij}, \& {van der
  Hulst}}]{MartinHernandez2005}
{Mart{\'{\i}}n-Hern{\'a}ndez}, N.~L., {Vermeij}, R., \& {van der Hulst}, J.~M.
  2005, \aap, 433, 205

\bibitem[{{Meixner} {et~al.}(2006){Meixner}, {Gordon}, {Indebetouw}, {Hora},
  {Whitney}, {Blum}, {Reach}, {Bernard}, {Meade}, {Babler}, {Engelbracht},
  {For}, {Misselt}, {Vijh}, {Leitherer}, {Cohen}, {Churchwell}, {Boulanger},
  {Frogel}, {Fukui}, {Gallagher}, {Gorjian}, {Harris}, {Kelly}, {Kawamura},
  {Kim}, {Latter}, {Madden}, {Markwick-Kemper}, {Mizuno}, {Mizuno}, {Mould},
  {Nota}, {Oey}, {Olsen}, {Onishi}, {Paladini}, {Panagia}, {Perez-Gonzalez},
  {Shibai}, {Sato}, {Smith}, {Staveley-Smith}, {Tielens}, {Ueta}, {van Dyk},
  {Volk}, {Werner}, \& {Zaritsky}}]{Meixner2006}
{Meixner}, M., {Gordon}, K.~D., {Indebetouw}, R., {et~al.} 2006, \aj, 132, 2268

\bibitem[{{Meixner} {et~al.}(2013){Meixner}, {Panuzzo}, {Roman-Duval},
  {Engelbracht}, {Babler}, {Seale}, {Hony}, {Montiel}, {Sauvage}, {Gordon},
  {Misselt}, {Okumura}, {Chanial}, {Beck}, {Bernard}, {Bolatto}, {Bot},
  {Boyer}, {Carlson}, {Clayton}, {Chen}, {Cormier}, {Fukui}, {Galametz},
  {Galliano}, {Hora}, {Hughes}, {Indebetouw}, {Israel}, {Kawamura}, {Kemper},
  {Kim}, {Kwon}, {Lebouteiller}, {Li}, {Long}, {Madden}, {Matsuura}, {Muller},
  {Oliveira}, {Onishi}, {Otsuka}, {Paradis}, {Poglitsch}, {Reach},
  {Robitaille}, {Rubio}, {Sargent}, {Sewi{\l}o}, {Skibba}, {Smith},
  {Srinivasan}, {Tielens}, {van Loon}, \& {Whitney}}]{Meixner2013}
{Meixner}, M., {Panuzzo}, P., {Roman-Duval}, J., {et~al.} 2013, \aj, 146, 62

\bibitem[{{Mills} \& {Turtle}(1984)}]{Mills1984}
{Mills}, B.~Y. \& {Turtle}, A.~J. 1984, in IAU Symposium, Vol. 108, Structure
  and Evolution of the Magellanic Clouds, ed. S.~{van den Bergh} \& K.~S.~D.
  {de Boer}, 283--290

\bibitem[{{Misselt} {et~al.}(1999){Misselt}, {Clayton}, \&
  {Gordon}}]{Misselt1999}
{Misselt}, K.~A., {Clayton}, G.~C., \& {Gordon}, K.~D. 1999, \apj, 515, 128

\bibitem[{{Mochizuki} {et~al.}(1994){Mochizuki}, {Nakagawa}, {Doi}, {Yui},
  {Okuda}, {Shibai}, {Yui}, {Nishimura}, \& {Low}}]{Mochizuki1994}
{Mochizuki}, K., {Nakagawa}, T., {Doi}, Y., {et~al.} 1994, \apjl, 430, L37

\bibitem[{{Mookerjea} {et~al.}(2012){Mookerjea}, {Ossenkopf}, {Ricken},
  {G{\"u}sten}, {Graf}, {Jacobs}, {Kramer}, {Simon}, \&
  {Stutzki}}]{Mookerjea2012}
{Mookerjea}, B., {Ossenkopf}, V., {Ricken}, O., {et~al.} 2012, \aap, 542, L17

\bibitem[{{Nakajima} {et~al.}(2005){Nakajima}, {Kato}, {Nagata}, {Tamura},
  {Sato}, {Sugitani}, {Nagashima}, {Nagayama}, {Iwata}, {Ita}, {Tanabe},
  {Kurita}, {Nakaya}, \& {Baba}}]{Nakajima2005}
{Nakajima}, Y., {Kato}, D., {Nagata}, T., {et~al.} 2005, \aj, 129, 776

\bibitem[{{Okada} {et~al.}(2012){Okada}, {G{\"u}sten}, {Requena-Torres},
  {R{\"o}llig}, {Hartogh}, {H{\"u}bers}, {Klein}, {Ricken}, {Simon}, \&
  {Stutzki}}]{Okada2012}
{Okada}, Y., {G{\"u}sten}, R., {Requena-Torres}, M.~A., {et~al.} 2012, \aap,
  542, L10

\bibitem[{{Okada} {et~al.}(2003){Okada}, {Onaka}, {Shibai}, \&
  {Doi}}]{Okada2003}
{Okada}, Y., {Onaka}, T., {Shibai}, H., \& {Doi}, Y. 2003, \aap, 412, 199

\bibitem[{{Okada} {et~al.}(2015){Okada}, {Requena-Torres}, {G{\"u}sten},
  {Stutzki}, {Wiesemeyer}, {P{\"u}tz}, \& {Ricken}}]{Okada2015}
{Okada}, Y., {Requena-Torres}, M.~A., {G{\"u}sten}, R., {et~al.} 2015, \aap,
  580, A54

\bibitem[{{Ossenkopf} {et~al.}(2015){Ossenkopf}, {Koumpia}, {Okada},
  {Mookerjea}, {van der Tak}, {Simon}, {P{\"u}tz}, \&
  {G{\"u}sten}}]{Ossenkopf2015}
{Ossenkopf}, V., {Koumpia}, E., {Okada}, Y., {et~al.} 2015, \aap, 580, A83

\bibitem[{{Ott} {et~al.}(2010){Ott}, {Henkel}, {Staveley-Smith}, \&
  {Wei{\ss}}}]{OttJ2010}
{Ott}, J., {Henkel}, C., {Staveley-Smith}, L., \& {Wei{\ss}}, A. 2010, \apj,
  710, 105

\bibitem[{{P{\'e}rez-Beaupuits} {et~al.}(2012){P{\'e}rez-Beaupuits},
  {Wiesemeyer}, {Ossenkopf}, {Stutzki}, {G{\"u}sten}, {Simon}, {H{\"u}bers}, \&
  {Ricken}}]{PerezBeaupuits2012}
{P{\'e}rez-Beaupuits}, J.~P., {Wiesemeyer}, H., {Ossenkopf}, V., {et~al.} 2012,
  \aap, 542, L13

\bibitem[{{Pilbratt} {et~al.}(2010){Pilbratt}, {Riedinger}, {Passvogel},
  {Crone}, {Doyle}, {Gageur}, {Heras}, {Jewell}, {Metcalfe}, {Ott}, \&
  {Schmidt}}]{Pilbratt2010}
{Pilbratt}, G.~L., {Riedinger}, J.~R., {Passvogel}, T., {et~al.} 2010, \aap,
  518, L1+

\bibitem[{{Pilleri} {et~al.}(2012){Pilleri}, {Fuente}, {Cernicharo},
  {Ossenkopf}, {Bern{\'e}}, {Gerin}, {Pety}, {Goicoechea}, {Rizzo},
  {Montillaud}, {Gonz{\'a}lez-Garc{\'{\i}}a}, {Joblin}, {Le Bourlot}, {Le
  Petit}, \& {Kramer}}]{Pilleri2012b}
{Pilleri}, P., {Fuente}, A., {Cernicharo}, J., {et~al.} 2012, \aap, 544, A110

\bibitem[{{Pineda} {et~al.}(2014){Pineda}, {Langer}, \&
  {Goldsmith}}]{Pineda2014}
{Pineda}, J.~L., {Langer}, W.~D., \& {Goldsmith}, P.~F. 2014, \aap, 570, A121

\bibitem[{{Pineda} {et~al.}(2017){Pineda}, {Langer}, {Goldsmith}, {Horiuchi},
  {Kuiper}, {Muller}, {Hughes}, {Ott}, {Requena-Torres}, {Velusamy}, \&
  {Wong}}]{Pineda2017}
{Pineda}, J.~L., {Langer}, W.~D., {Goldsmith}, P.~F., {et~al.} 2017, \apj, 839,
  107

\bibitem[{{Pineda} {et~al.}(2012){Pineda}, {Mizuno}, {R{\"o}llig}, {Stutzki},
  {Kramer}, {Klein}, {Rubio}, {Kawamura}, {Minamidani}, {Benz}, {Burton},
  {Fukui}, {Koo}, \& {Onishi}}]{Pineda2012}
{Pineda}, J.~L., {Mizuno}, N., {R{\"o}llig}, M., {et~al.} 2012, \aap, 544, A84

\bibitem[{{Pineda} {et~al.}(2008){Pineda}, {Mizuno}, {Stutzki}, {Cubick},
  {Aravena}, {Bensch}, {Bertoldi}, {Bronfman}, {Fujishita}, {Graf},
  {Hitschfeld}, {Honingh}, {Jakob}, {Jacobs}, {Kawamura}, {Klein}, {Kramer},
  {May}, {Miller}, {Mizuno}, {M{\"u}ller}, {Onishi}, {Ossenkopf}, {Rabanus},
  {R{\"o}llig}, {Rubio}, {Sasago}, {Schieder}, {Simon}, {Sun}, {Volgenau},
  {Yamamoto}, \& {Fukui}}]{Pineda2008}
{Pineda}, J.~L., {Mizuno}, N., {Stutzki}, J., {et~al.} 2008, \aap, 482, 197

\bibitem[{{Poglitsch} {et~al.}(1995){Poglitsch}, {Krabbe}, {Madden}, {Nikola},
  {Geis}, {Johansson}, {Stacey}, \& {Sternberg}}]{Poglitsch1995}
{Poglitsch}, A., {Krabbe}, A., {Madden}, S.~C., {et~al.} 1995, \apj, 454, 293

\bibitem[{{Poglitsch} {et~al.}(2010){Poglitsch}, {Waelkens}, {Geis},
  {Feuchtgruber}, {Vandenbussche}, {Rodriguez}, {Krause}, {Renotte}, {van
  Hoof}, {Saraceno}, {Cepa}, {Kerschbaum}, {Agn{\`e}se}, {Ali}, {Altieri},
  {Andreani}, {Augueres}, {Balog}, {Barl}, {Bauer}, {Belbachir}, {Benedettini},
  {Billot}, {Boulade}, {Bischof}, {Blommaert}, {Callut}, {Cara}, {Cerulli},
  {Cesarsky}, {Contursi}, {Creten}, {De Meester}, {Doublier}, {Doumayrou},
  {Duband}, {Exter}, {Genzel}, {Gillis}, {Gr{\"o}zinger}, {Henning},
  {Herreros}, {Huygen}, {Inguscio}, {Jakob}, {Jamar}, {Jean}, {de Jong},
  {Katterloher}, {Kiss}, {Klaas}, {Lemke}, {Lutz}, {Madden}, {Marquet},
  {Martignac}, {Mazy}, {Merken}, {Montfort}, {Morbidelli}, {M{\"u}ller},
  {Nielbock}, {Okumura}, {Orfei}, {Ottensamer}, {Pezzuto}, {Popesso},
  {Putzeys}, {Regibo}, {Reveret}, {Royer}, {Sauvage}, {Schreiber}, {Stegmaier},
  {Schmitt}, {Schubert}, {Sturm}, {Thiel}, {Tofani}, {Vavrek}, {Wetzstein},
  {Wieprecht}, \& {Wiezorrek}}]{Poglitsch2010}
{Poglitsch}, A., {Waelkens}, C., {Geis}, N., {et~al.} 2010, \aap, 518, L2+

\bibitem[{{Requena-Torres} {et~al.}(2016){Requena-Torres}, {Israel}, {Okada},
  {G{\"u}sten}, {Stutzki}, {Risacher}, {Simon}, \&
  {Zinnecker}}]{Requena-Torres2016}
{Requena-Torres}, M.~A., {Israel}, F.~P., {Okada}, Y., {et~al.} 2016, \aap,
  589, A28

\bibitem[{{Risacher} {et~al.}(2016){Risacher}, {G{\"u}sten}, {Stutzki},
  {H{\"u}bers}, {Bell}, {Buchbender}, {B{\"u}chel}, {Csengeri}, {Graf},
  {Heyminck}, {Higgins}, {Honingh}, {Jacobs}, {Klein}, {Okada}, {Parikka},
  {P{\"u}tz}, {Reyes}, {Ricken}, {Riquelme}, {Simon}, \&
  {Wiesemeyer}}]{Risacher2016}
{Risacher}, C., {G{\"u}sten}, R., {Stutzki}, J., {et~al.} 2016, \aap, 595, A34

\bibitem[{{R{\"o}llig} {et~al.}(2006){R{\"o}llig}, {Ossenkopf}, {Jeyakumar},
  {Stutzki}, \& {Sternberg}}]{Roellig2006}
{R{\"o}llig}, M., {Ossenkopf}, V., {Jeyakumar}, S., {Stutzki}, J., \&
  {Sternberg}, A. 2006, \aap, 451, 917

\bibitem[{{R{\"o}llig} {et~al.}(2016){R{\"o}llig}, {Simon}, {G{\"u}sten},
  {Stutzki}, {Israel}, \& {Jacobs}}]{Roellig2016}
{R{\"o}llig}, M., {Simon}, R., {G{\"u}sten}, R., {et~al.} 2016, \aap, 591, A33

\bibitem[{{R{\"o}llig} {et~al.}(2013){R{\"o}llig}, {Szczerba}, {Ossenkopf}, \&
  {Gl{\"u}ck}}]{Roellig2013a}
{R{\"o}llig}, M., {Szczerba}, R., {Ossenkopf}, V., \& {Gl{\"u}ck}, C. 2013,
  \aap, 549, A85

\bibitem[{{Saigo} {et~al.}(2017){Saigo}, {Onishi}, {Nayak}, {Meixner},
  {Tokuda}, {Harada}, {Morioka}, {Sewi{\l}o}, {Indebetouw}, {Torii},
  {Kawamura}, {Ohama}, {Hattori}, {Yamamoto}, {Tachihara}, {Minamidani},
  {Inoue}, {Madden}, {Galametz}, {Lebouteiller}, {Chen}, {Mizuno}, \&
  {Fukui}}]{Saigo2017}
{Saigo}, K., {Onishi}, T., {Nayak}, O., {et~al.} 2017, \apj, 835, 108

\bibitem[{{Sandell} {et~al.}(2015){Sandell}, {Mookerjea}, {G{\"u}sten},
  {Requena-Torres}, {Riquelme}, \& {Okada}}]{Sandell2015}
{Sandell}, G., {Mookerjea}, B., {G{\"u}sten}, R., {et~al.} 2015, \aap, 578, A41

\bibitem[{{Schneider} {et~al.}(2012){Schneider}, {G{\"u}sten}, {Tremblin},
  {Hennemann}, {Minier}, {Hill}, {Comer{\'o}n}, {Requena-Torres}, {Kraemer},
  {Simon}, {R{\"o}llig}, {Stutzki}, {Djupvik}, {Zinnecker}, {Marston},
  {Csengeri}, {Cormier}, {Lebouteiller}, {Audit}, {Motte}, {Bontemps},
  {Sandell}, {Allen}, {Megeath}, \& {Gutermuth}}]{Schneider2012}
{Schneider}, N., {G{\"u}sten}, R., {Tremblin}, P., {et~al.} 2012, \aap, 542,
  L18

\bibitem[{{Seale} {et~al.}(2012){Seale}, {Looney}, {Wong}, {Ott}, {Klein}, \&
  {Pineda}}]{Seale2012}
{Seale}, J.~P., {Looney}, L.~W., {Wong}, T., {et~al.} 2012, \apj, 751, 42

\bibitem[{{Simon} {et~al.}(2012){Simon}, {Schneider}, {Stutzki}, {G{\"u}sten},
  {Graf}, {Hartogh}, {Guan}, {Staguhn}, \& {Benford}}]{Simon2012}
{Simon}, R., {Schneider}, N., {Stutzki}, J., {et~al.} 2012, \aap, 542, L12

\bibitem[{{Sofia} {et~al.}(2004){Sofia}, {Lauroesch}, {Meyer}, \&
  {Cartledge}}]{Sofia2004}
{Sofia}, U.~J., {Lauroesch}, J.~T., {Meyer}, D.~M., \& {Cartledge}, S.~I.~B.
  2004, \apj, 605, 272

\bibitem[{{Sternberg} \& {Dalgarno}(1995)}]{Sternberg1995}
{Sternberg}, A. \& {Dalgarno}, A. 1995, \apjs, 99, 565

\bibitem[{{St{\"o}rzer} {et~al.}(1996){St{\"o}rzer}, {Stutzki}, \&
  {Sternberg}}]{Stoerzer1996}
{St{\"o}rzer}, H., {Stutzki}, J., \& {Sternberg}, A. 1996, \aap, 310, 592

\bibitem[{{Tang} {et~al.}(2016){Tang}, {Li}, {Heiles}, {Wang}, {Pan}, \&
  {Wang}}]{Tang2016}
{Tang}, N., {Li}, D., {Heiles}, C., {et~al.} 2016, \aap, 593, A42

\bibitem[{{Testor} \& {Niemela}(1998)}]{TestorNiemela1998}
{Testor}, G. \& {Niemela}, V. 1998, \aaps, 130, 527

\bibitem[{{Tielens} \& {Hollenbach}(1985)}]{TH85I}
{Tielens}, A.~G.~G.~M. \& {Hollenbach}, D. 1985, \apj, 291, 722

\bibitem[{{Vassilev} {et~al.}(2008){Vassilev}, {Meledin}, {Lapkin}, {Belitsky},
  {Nystr{\"o}m}, {Henke}, {Pavolotsky}, {Monje}, {Risacher}, {Olberg},
  {Strandberg}, {Sundin}, {Fredrixon}, {Ferm}, {Desmaris}, {Dochev},
  {Pantaleev}, {Bergman}, \& {Olofsson}}]{Vassilev2008}
{Vassilev}, V., {Meledin}, D., {Lapkin}, I., {et~al.} 2008, \aap, 490, 1157

\bibitem[{{Wang} {et~al.}(2009){Wang}, {Chin}, {Henkel}, {Whiteoak}, \&
  {Cunningham}}]{Wang2009}
{Wang}, M., {Chin}, Y.-N., {Henkel}, C., {Whiteoak}, J.~B., \& {Cunningham}, M.
  2009, \apj, 690, 580

\bibitem[{{Weingartner} \& {Draine}(2001)}]{WD2001}
{Weingartner}, J.~C. \& {Draine}, B.~T. 2001, \apj, 548, 296

\bibitem[{{Welty} {et~al.}(2016){Welty}, {Lauroesch}, {Wong}, \&
  {York}}]{Welty2016}
{Welty}, D.~E., {Lauroesch}, J.~T., {Wong}, T., \& {York}, D.~G. 2016, \apj,
  821, 118

\bibitem[{{Yeh} {et~al.}(2015){Yeh}, {Seaquist}, {Matzner}, \&
  {Pellegrini}}]{Yeh2015}
{Yeh}, S.~C.~C., {Seaquist}, E.~R., {Matzner}, C.~D., \& {Pellegrini}, E.~W.
  2015, \apj, 807, 117

\bibitem[{{Young} {et~al.}(2012){Young}, {Becklin}, {Marcum}, {Roellig}, {De
  Buizer}, {Herter}, {G{\"u}sten}, {Dunham}, {Temi}, {Andersson}, {Backman},
  {Burgdorf}, {Caroff}, {Casey}, {Davidson}, {Erickson}, {Gehrz}, {Harper},
  {Harvey}, {Helton}, {Horner}, {Howard}, {Klein}, {Krabbe}, {McLean}, {Meyer},
  {Miles}, {Morris}, {Reach}, {Rho}, {Richter}, {Roeser}, {Sandell}, {Sankrit},
  {Savage}, {Smith}, {Shuping}, {Vacca}, {Vaillancourt}, {Wolf}, \&
  {Zinnecker}}]{Young2012}
{Young}, E.~T., {Becklin}, E.~E., {Marcum}, P.~M., {et~al.} 2012, \apjl, 749,
  L17

\end{thebibliography}
\end{document}